\documentclass[10pt,floatfix,amssymb,amsmath,prd,aps,twocolumn,preprintnumbers,superscriptaddress,nofootinbib]{revtex4-2}
\usepackage[normalem]{ulem} 
\usepackage{hyperref}
\hypersetup{
	colorlinks=true,
    linktocpage=true,
	linkcolor=blue,
	citecolor=red,   
	urlcolor=blue,
	pdftitle={Heavy quark mass effects in charged-current deep-inelastic scattering at approximate NNLO in the Aivazis-Collins-Olness-Tung scheme},
	pdfauthor={Risse, P. et al.}
}
\usepackage{cleveref}
\usepackage{bm}
\usepackage{ulem}
\usepackage{tabularx}
\usepackage{multirow}
\usepackage{siunitx}
\usepackage{graphicx}
\usepackage{xcolor}
\usepackage[caption=false]{subfig}
 
\usepackage{pbox}
\usepackage{listings}
\definecolor{codegreen}{rgb}{0,0.6,0}
\definecolor{codegray}{rgb}{0.5,0.5,0.5}
\definecolor{codepurple}{rgb}{0.58,0,0.82}
\definecolor{backcolour}{rgb}{0.94,0.94,0.88}
\lstdefinestyle{mystyle}{
    backgroundcolor=\color{backcolour},  
    basicstyle=\ttfamily\lst@ifdisplaystyle\footnotesize\fi,
    commentstyle=\color{codegreen},
    keepspaces=true,                 
    keywordstyle=\color{blue},       
    numbers=left,                    
    numbersep=5pt,                   
    numberstyle=\tiny\color{codegray},
    rulecolor=\color{black},         
    stepnumber=2,                    
    morekeywords={mkdir,cmake,make},
    frame=tb,
    breaklines=true,
    escapeinside={(*@}{@*)},
}
\usepackage{xifthen}
\usepackage{xparse}

\usepackage{scrextend}

\makeatletter 

\def\fstrut{\vrule height 10pt depth 4pt width 0pt}

\let\oldcite\cite
\renewcommand{\cite}[1]{\mbox{\oldcite{#1}}}    
\renewcommand\onecolumngrid{
\do@columngrid{one}{\@ne}%
\def\set@footnotewidth{\onecolumngrid}
\def\footnoterule{\kern-6pt\hrule width 1.5in\kern6pt}%
}

\renewcommand\twocolumngrid{
        \def\footnoterule{
        \dimen@\skip\footins\divide\dimen@\thr@@
        \kern-\dimen@\hrule width.5in\kern\dimen@}
        \do@columngrid{mlt}{\tw@}
}%

\makeatother    

\newcommand{\aSACOTchi}{\mbox{\textit{aSACOT-\texorpdfstring{$\chi$}{χ}}}}
\newcommand{\SACOTchi}{\mbox{\textit{SACOT-\texorpdfstring{$\chi$}{χ}}}}
\newcommand{\ZMVFNS}{\mbox{\textit{ZM-VFNS}}}
\newcommand{\xfitter}{\mbox{\texttt{xFitter}}}
\newcommand{\FONLL}[1][]{\ifthenelse{\equal{#1}{}}
	{\textit{FONLL}}
    {\textit{FONLL-#1}}}

\newcommand{\order}[1]{\ensuremath{\mathcal{O}\left(#1\right)}}
\newcommand{\alphas}{\ensuremath{\alpha_s}}

\newcommand{\review}[1]{{#1}}

\newcommand{\apfelxx}{\mbox{\texttt{APFEL++}}}
\newcommand{\newimplemented}{{\color{red}$\bm\checkmark$}}
\newcommand{\alreadyimplemented}{{\color{black}$\bm\checkmark$}}
\newcommand{\ncteq}{\mbox{\sl nCTEQ}}
\newcommand{\myIntFrac}[4]{\int_{#1}^{#2}\!\frac{\mathrm{d}#3}{#4}\,}

\newcommand{\ACOT}{ACOT}
\newcommand{\SACOT}[1]{\mbox{aSACOT-$\chi(n=#1)$}}
\newcommand{\SACOTnlo}{\mbox{SACOT-$\chi$ NLO}}
\newcommand{\ZMNLO}{\mbox{ZM NLO}}
\newcommand{\ZMNNLO}{\mbox{ZM NNLO}}
\newcommand{\ZMonlyNNLO}{\SACOTnlo{} + \ZMNNLO}

\newcommand{\numhighl}[1]{\textcolor{darkgray}{\textsl{#1}}}

\NewDocumentCommand{\F}{O{} O{} m}{\ifthenelse{\isempty{#1}}
	{\ensuremath{F_{#3}^{#2}}}
	{\ensuremath{F_{#3,#1}^{#2}}}}
\NewDocumentCommand{\coef}{O{} O{} m}{\ifthenelse{\isempty{#2}}
	{\ensuremath{C_{#1}^{#3}}}
	{\ensuremath{C_{#1}^{#3,#2}}}}
\NewDocumentCommand{\coeftilde}{O{} O{} m}{\ifthenelse{\isempty{#2}}
	{\ensuremath{\tilde{C}_{#1}^{#3}}}
	{\ensuremath{\tilde{C}_{#1}^{#3,#2}}}}
\newcommand{\anti}[1]{\ensuremath{\overline{#1}}}
\newcommand{\sumDtype}{\ensuremath{\sum_{D}^{d,s,b}}}
\newcommand{\sumUtype}{\ensuremath{\sum_{U}^{u,c,t}}}
\newcommand{\sumbothtype}{\ensuremath{\sumUtype\,\sumDtype}}

\newcommand{\orcid}[1]{\,\href{https://orcid.org/#1}{\includegraphics[width=9pt]{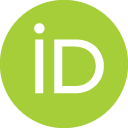}}\,}
\newcommand{\orcidVB}{0000-0003-0148-0272} 
\newcommand{\orcidTJ}{0000-0002-1334-7607} 
\newcommand{\orcidKK}{0000-0003-1412-447X} 
\newcommand{\orcidAK}{0000-0002-4090-0084} 
\newcommand{\orcidFO}{0000-0001-6799-2436} 
\newcommand{\orcidIS}{0000-0003-0373-474X} 
\newcommand{\orcidPR}{0000-0002-8570-5506} 
\newcommand{\muenster}{\affiliation{Institut f{\"u}r Theoretische Physik, Universit{\"a}t M{\"u}nster,
Wilhelm-Klemm-Stra{\ss}e 9, D-48149 M{\"u}nster, Germany}}
\newcommand{\smu}{\affiliation{Department of Physics, Southern Methodist University, Dallas, TX 75275-0175, U.S.A.}}
\newcommand{\jlab}{\affiliation{Theory Center, Jefferson Lab, Newport News, VA 23606, U.S.A.}}
\newcommand{\saclay}{\affiliation{IRFU, CEA, Université Paris-Saclay, 91191 Gif-sur-Yvette, France}}
\newcommand{\krakow}{\affiliation{Institute of Nuclear Physics Polish Academy of Sciences, PL-31342 Krakow, Poland}}
\newcommand{\grenoble}{\affiliation{Laboratoire de Physique Subatomique et de Cosmologie, Université Grenoble-Alpes, 
\\CNRS/IN2P3, 53 avenue des Martyrs, 38026 Grenoble, France}
}

\begin{document} 


\preprint{MS-TP-25-07}
\preprint{SMU-PHY-25-01}
\preprint{JLAB-THY-25-4279}
\preprint{IFJPAN-IV-2025-9}

\title{Heavy quark mass effects in charged-current deep-inelastic scattering\linebreak
	at approximate NNLO in the Aivazis-Collins-Olness-Tung scheme}

\author{P. Risse\orcid{\orcidPR}}
\email{prisse@smu.edu}
\muenster\smu\jlab
\author{V. Bertone\orcid{\orcidVB}} \saclay
\author{T. Je\v{z}o\orcid{\orcidTJ}} \muenster
\author{K. Kova\v{r}\'{i}k\orcid{\orcidKK}} \muenster

\author{A. Kusina\orcid{\orcidAK}} \krakow
\author{F. I. Olness\orcid{\orcidFO}} \smu
\author{I. Schienbein\orcid{\orcidIS}} \grenoble

\begin{abstract}
\vspace{0.5cm}
The approximate SACOT-$\chi$ scheme for heavy quark production in deep-inelastic scattering was initially formulated for the neutral current structure functions $F_2$ and $F_L$.
We extend this approach to the charged current case
(also including $F_3$), and thereby complete the definitions for the most relevant inclusive structure functions.
Furthermore, we implement these structure functions in the open-source code \apfelxx{} which provides fast numerical evaluations over a wide kinematic range; this addition to the \apfelxx{} code is publicly available, with details provided in the appendix.
This  SACOT-$\chi$ implementation enables
detailed numerical insights on the mass dependence of the structure functions and cross sections in the $(x,Q^2)$-plane for both neutral and charged current processes. 
We consider kinematic regions relevant for the experimental measurements from 
fixed-target $\nu$DIS experiments (NuTeV, CCFR and Chorus) and HERA, and also projections for the upcoming EIC.
In particular, the $\nu$DIS experiments reveal a surprisingly strong dependence on the mass effects, offering valuable insights that may help resolve long-standing challenges in accurately describing these datasets.
\null \vspace{3.5cm}
\end{abstract}

\date{\today}

\maketitle

\tableofcontents
\newpage
\section{Introduction}
\label{sec:intro}
As we near the fifteen-year mark in the operation of the Large Hadron Collider (LHC), a wide range of measurements at the LHC have achieved very high precision. Any prediction of an observable measured at the LHC requires detailed information on the structure of the proton. Therefore, any precise theoretical prediction is crucially dependent on an accurate extraction of the proton structure from experimental data.
Nowadays, the parton distribution functions (PDFs), which parametrize the structure of the proton in a simplified collinear picture, are determined using predominantly theoretical calculations at next-to-next-to-leading order (NNLO), when available. The goal is to determine the structure of the proton with an accuracy of 1\% or better.

The cornerstone of any determination of PDFs are the double-differential cross-section data for inclusive deep-inelastic scattering (DIS). The most precise data on proton DIS come from the combination of measurements from H1 and ZEUS experiments at the HERA collider~\cite{H1:2012xnw,ZEUS:2014wft,H1:2015ubc,H1:2018flt}. To best utilize the precision of the data, one uses the most accurate theoretical predictions for the corresponding $F_2$, $F_3$ and $F_L$ DIS structure functions to interpret the data in terms of PDFs.

Current state-of-the-art descriptions of the DIS structure functions are given using General-Mass Variable-Flavor-Number Schemes (GM-VFNS)~\cite{Aivazis:1993kh,Aivazis:1993pi,Kretzer:1998ju,Tung:2001mv,Kniehl:2011bk,Thorne:1997ga,Thorne:2006qt,Cacciari:1998it,Forte:2010ta,Hoche:2019ncc,Gauld:2021zmq,Guzzi:2011ew,Gao:2021fle,Guzzi:2024can,Barontini:2024xgu} at NNLO. 
Variable-Flavor-Number (VFN) schemes dynamically modify the number of active quarks depending on the kinematic region to ensure optimal precision across different energy scales. 
While the Zero-Mass Variable-Flavor-Number Scheme (ZM-VFNS)~\cite{Collins:1986mp} neglects the quark-mass effects in the matrix elements, the GM-VFNS approach extends and improves on the ZM-VFNS by incorporating relevant quark-mass effects both in kinematics and in the underlying matrix elements.
However, at NNLO, fully accounting for quark-mass effects becomes highly complex, leading to results that are cumbersome to implement efficiently.
Several procedures have been designed that leverage the simpler ZM matrix elements, but apply the quark-mass effects in the kinematics~\cite{Thorne:2008xf,Nadolsky:2009ge,Stavreva:2012bs}. These sometimes called ``intermediate mass''-schemes yield results closer to the fully massive calculation~\cite{Xie:2019eoe} than the ZM matrix elements, but retain their numerical simplicity.

The present analysis aims for an efficient implementation of the DIS structure functions up to NNLO. For that we first utilize (and extend) the ``intermediate mass''-scheme that was developed in  Ref.~\cite{Stavreva:2012bs} in the context of the \ACOT{}-scheme~\cite{Aivazis:1993kh,Aivazis:1993pi,Kretzer:1998ju} for the NNLO contribution. 
Secondly, we effectively factorize out the PDF dependence of the calculation and store the PDF-independent result in interpolation tables as implemented in the open-source code  \apfelxx{}~\cite{Bertone:2017gds,APFELppGithub,Bertone:2013vaa}. 
In this framework, the precomputed tables, which are computationally the most expensive ingredient, are easily combined with the PDFs to yield the final physical prediction; 
using the tables, rather than a full recalculation, can speed up the calculation by orders of magnitude.

The simplified NNLO calculation of the structure functions in the \ACOT-scheme, here called \aSACOTchi{} (for approximate, Simplified \ACOT-$\chi$), was developed for the neutral current structure functions $\F{2}$ and $\F{L}$. In this paper we extend the approach to cover the neutral current $\F{3}$, and also the complete set of charged current structure functions. Thereby, the \aSACOTchi{} scheme is presented for the most relevant inclusive DIS structure functions.\footnote{Note that there are also the structure functions $\F{4}$ and $\F{5}$. Their contribution is suppressed by the mass of the lepton and therefore less studied. A calculation of $\F{4}$ and $\F{5}$ in the ACOT scheme at NLO is currently underway~\cite{Spezzano:2025todo}.}
The calculation requires the \SACOTchi{} scheme up to NLO. Thus, the additions to \apfelxx{}, accompanying this paper, include \SACOTchi{} (NLO) and \aSACOTchi{} (NNLO) for all nine structure functions; $F_{2,L,3}$ for neutral-current, and charged-current neutrino and anti-neutrino DIS.

Implementing all nine structure functions not only enables us to assess the impact of the \SACOTchi{} scheme, but also allows to apply the scheme to cross-section predictions.
With an emphasis on charged current predictions, we investigate the impact on current experimental measurements including HERA~II~\cite{H1:2015ubc} and neutrino DIS~\cite{NuTeV:2005wsg,CCFRNuTeV:2000qwc,Yang:2001rm,CHORUS:2008vjb}, as well as projected measurements at the upcoming Electron Ion Collider (EIC)~\cite{AbdulKhalek:2021gbh,Khalek:2021ulf}. As the kinematic coverage of the neutrino DIS data sets is in the region where the heavy-quark masses impact the predictions most, we place special emphasis on analyzing these predictions.

The outline of our article is as follows. 
In~\cref{sec:theoretical_methods}, we present the main elements of the \aSACOTchi{} scheme, and the treatment of the heavy-quark masses.  
In~\cref{sec:results}, we outline the efficient numerical implementation, and demonstrate the impact for a variety of experimental measurements including fixed-target $\nu$DIS (CCFR, NuTeV, CHORUS), HERA, and the upcoming EIC.
In~\cref{sec:conclusions}, we recap the key results of this investigation. 

We provide a set of appendices that further detail the implementation and application of the $\aSACOTchi{}$ scheme. 
In~\cref{sec:Details_of_the_numerical_implementation}, we present the details of the numerical implementation; this is an essential step to efficiently include NNLO corrections. 
In~\cref{sec:feynman}, we display the Feynman diagrams, including the new channels that are present at NNLO. 
In~\cref{sec:nScaling}, we explore the details of the $n$-scaling and use this to estimate the theoretical uncertainty. 
In~\cref{sec:LibraryDocumentation},  we provide the key components required to use the $\aSACOTchi{}$ implementation within the $\apfelxx{}$ framework; the code can be downloaded from the GitHub repository~\cite{APFELppGithub}.

\section{Theoretical methods}
\label{sec:theoretical_methods}

This section introduces the main concepts of \aSACOTchi{} for neutral current interactions, and we also extend this to charged current interactions. For a detailed introduction to the scheme, we refer the reader to the original formulation in Ref.~\cite{Stavreva:2012bs}. 

The \aSACOTchi{} scheme can be summarized as
\begin{align}
    \aSACOTchi{} &\equiv \notag\\
    \SACOTchi{}&\left[\order{\alphas^{0+1}}\right] + \text{ZM-VFNS}\left[\order{\alphas^{2}}\right]\bigg|_{\chi(n)}.
\end{align}
That is, we apply the exact  scheme at LO and NLO in $\alphas{}$, and at NNLO we use the ZM-VFNS Wilson coefficients, but restrict the (lower) integration bound in the convolution to the generalized $\chi(n)$-scaling variable (see below). 
\review{%
%
As a reminder, in the \SACOTchi{} scheme, the dynamic mass of an incoming heavy quark---or of a heavy quark appearing in an internal line with an on-shell cut---is set to zero.
This procedure is not an approximation but due to an internal freedom in the formulation of the ACOT scheme, as shown in Ref.~\cite{Kramer:2000hn}. 
The only approximation made here is to set the heavy quark mass to zero in the NNLO terms (\textit{e.g.}\ for incoming gluons) except for the integration boundaries (implemented via the $\chi$-variable).}
%

%
In Ref.~\cite{Stavreva:2012bs} it has been shown that this treatment captures the dominant mass effects.\footnote{To be more precise, it has been shown that, up to $\order{\alphas{}}$, the phase-space mass is the dominant contribution (cf. Fig.~6 and~7 of Ref.~\cite{Stavreva:2012bs}). We are working under the assumption that this observation holds for higher orders as well.}
In technical terms, we separate the phase-space mass from the dynamic mass. 
The phase-space mass is the kinematic mass that constrains the effective phase space, and we retain these contributions.
The dynamic mass is the mass appearing in the hard-scattering cross section $\hat{\sigma}(m)$; 
this is neglected for those higher-order terms where the massless Wilson coefficients are implemented.

Thus, the convolution between a generic Wilson coefficient $\coef{\lambda}$ and a PDF $f(x,Q^2)$ is given by
\begin{equation}
    [\coef{\lambda}\otimes f](x,Q^2) = \myIntFrac{\chi(n)}{1}{z}{z} \coef{\lambda}\left(z,Q^2\right)f\left(\frac{\chi}{z},Q^2\right)\,.
    \label{eq:basic_mellin_convolution}
\end{equation}
Following the original formulation, we choose the generalized $\chi(n)$-scaling variable to be
\begin{equation}
    \chi(n) = x\left(1+\frac{(n\,m_H)^2}{Q^2}\right) \,,
    \label{eq:n_scaling_variable}
\end{equation}
where $x$ is the Bjorken-$x$ (the lower integration limit in the massless treatment) and $m_H$ is the mass of the heavy quark. For the following discussion, $n=\{0,1,2,3\}$ is a scaling factor, which can be interpreted as having $n=0,1,2,3$ heavy quarks produced, and can be used to measure the impact of heavy quark contributions on the physical predictions. The special case $n=0$ replicates the massless case. 
In our numerical implementation $n$ can be chosen freely.
See \cref{sec:nScaling} for details on the $n$-scaling.

The key ingredient of the \aSACOTchi{} scheme is the decomposition of the structure functions into individual flavor contributions both for the initial and final states. This allows one to apply the appropriate $\chi(n)$-rescaling factor with $m_H$ set to the mass of the participating heavy quark.
This has to be done case by case for the individual structure functions, which we describe in the following sections starting with the neutral current example.

\review{Note that for this paper we assume the charm, bottom and top quarks are massive and the remaining quarks are massless. The numerical implementation however is general, such that the massive/massless quarks can be chosen freely.}

\subsection{Definition of \aSACOTchi{} in neutral current DIS}

This section reviews the core definitions of the neutral current \aSACOTchi{} in a manner updated to coincide with the present numerical implementation, and to prepare the relevant equations for extension to charged current interactions. The last subsection adapts these concepts to define $\F{3}$ in the  \aSACOTchi{} scheme.

For neutral current interactions at NNLO, a structure function $\F{\lambda}$ with $\lambda \in \{2,L\}$ is constructed as
\begin{equation}
    \F{\lambda} = \sum_{k=0}^6\sum_{l=1}^6 \F[k][l]{\lambda}\,,
    \label{eq:NC_structure_function_decomposition_top_level}
\end{equation}
where the index $k$ represents initial-state partons with $k=0$ denoting the gluon and $k=1,2,3,\dots$ denoting $d,u,s,\dots$ quarks and anti-quarks. The index $l$ runs over the final-state (anti-)quarks and does not include the gluon as they do not impose phase space constraints.

In the following, we discuss the partonic decomposition and the resulting $\chi(n)$-prescription starting from the massless expressions. Generally a structure function in the ZM-VFNS can be calculated from
\begin{equation}
    x^{-1}\F{\lambda} = \coef[ns]{\lambda}\otimes q_{ns} + \langle a^2\rangle\left(\coef[s]{\lambda}\otimes q_s + \coef[g]{\lambda}\otimes g\right)\,,
\end{equation}
where the \textit{non-singlet} (\textit{ns}) coefficient $\coef[ns]{\lambda}$ couples to the \textit{ns}-combination
\begin{equation}
    q_{ns} = \sum_{i=1}^{n_f} \left(a^2_i-\langle a^2\rangle\right)q_i^+ = \sum_{i=1}^{n_f} \left(a^2_i-\langle a^2\rangle\right)(q_i + \anti{q}_i)\,,
\end{equation}
the \textit{singlet} (\textit{s}) coefficient $\coef[s]{\lambda}$ to the \textit{s}-combination
\begin{equation}
    q_s = \sum_{i=1}^{n_f}q^+_i
\end{equation}
and the gluon coefficient $\coef[g]{\lambda}$ to the gluon distribution. $a_i$ are the full electroweak couplings\footnote{The full electroweak couplings include the $\gamma$-couplings, the $\gamma/Z$ interference and the $Z$-couplings, see e.g.~Eqs.~(B9)--(B11) of Ref.~\cite{Stavreva:2012bs}.} 
and the $n_f$-average over the squares $\langle a^2 \rangle$ is defined by
\begin{equation}
    \langle a^2 \rangle \equiv \langle a^2 \rangle^{(n_f)} = \frac{1}{n_f}\sum_{i=1}^{n_f} a_i^2\,.
\end{equation}
Isolating the contribution to $\F{\lambda}$ from an incoming quark of flavor $k$ we find
\begin{equation}
    x^{-1}\F[k]{\lambda} = \left[a^2_k \coef[ns]{\lambda}+\langle a^2\rangle\coef[ps]{\lambda}\right]\otimes q^+_k
\end{equation}
with the \textit{purely singlet} (\textit{ps}) coefficient $\coef[ps]{\lambda}$ given by
\begin{equation}
    \coef[ps]{\lambda} = \coef[s]{\lambda}-\coef[ns]{\lambda}\,.
\end{equation}
The gluon-initiated contribution is instead given by
\begin{equation}
    x^{-1}\F[0]{\lambda} = \langle a^2\rangle \coef[g]{\lambda} \otimes g\,.
\end{equation}

As a next step, we disentangle the final-state flavors. As all contributions from the individual final-state flavors are the same in the massless scheme, explicit factors of $n_f$ appear inside the Wilson coefficients, where the contributions have been summed up. At NNLO, we find the relations
\begin{subequations}
\begin{align}
    \coef[ns]{\lambda} &= \quad\,\,\coeftilde[ns,A]{\lambda} + n_f\,\coeftilde[ns,B]{\lambda} \\
    \coef[ps]{\lambda} &= n_f\,\coeftilde[ps]{\lambda}\\
    \coef[g]{\lambda}\, &= n_f\,\coeftilde[g]{\lambda}\,,
\end{align}%
\label{eq:coef_nf_dependence}%
\end{subequations}
where the \textit{ns}-coefficient splits into a contribution not proportional to $n_f$ (index $A$) and a contribution proportional to $n_f$ (index $B$). The \textit{ps}- and the gluon coefficient are entirely proportional to $n_f$. 

Finally, we can perform the final-state flavor decomposition. Starting with the initial parton being a quark of flavor $k$, we find
\begin{align}
    x^{-1}\F[k][l]{\lambda} = \left\{a^2_k \left[\coeftilde[ns,A]{\lambda}\delta_{kl} + \coeftilde[ns,B]{\lambda}\right]+\,a^2_l\,\coeftilde[ps]{\lambda}\right\}\otimes_{kl} q^+_k\,,
\end{align}
where the Kronecker $\delta$ is necessary to avoid double counting. For the case of the initial parton being a gluon, we find
\begin{equation}
    x^{-1}\F[0][j]{\lambda} = a^2_j\,\coeftilde[g]{\lambda}\otimes_{j} g\,.
\end{equation}
Note that we have indexed the Mellin convolutions, $\otimes_{kl}$ and $\otimes_{j}$, with the participating flavor indices. This notation makes it explicit that the convolution (\textit{cf.} \cref{eq:basic_mellin_convolution}) has to be performed with the lower integration bound set to 
\begin{equation}
    \chi_{kl}(n) = x\left(1+\frac{[n\max(m_k,m_l)]^2}{Q^2}\right)
\end{equation}
in the quark-initiated case and to 
\begin{equation}
    \chi_{j}(n) = x\left(1+\frac{[n\,m_j]^2}{Q^2}\right)
\end{equation}
in the gluon-initiated case. Summing over all initial- and final-state partons as prescribed in \cref{eq:NC_structure_function_decomposition_top_level} yields the total structure function in the \aSACOTchi{} scheme.

\subsubsection{Extension of \aSACOTchi{} to \texorpdfstring{$\F{3}$}{F3}}

The original formulation of the \aSACOTchi{} scheme was restricted to $\F{2}$ and $\F{L}$ only. However, the parton decomposition as described in the steps above can be repeated for $\F{3}$. Only a few differences have to be taken into account:
\begin{itemize}
    \item instead of the total distributions, the convolution has to be performed with the difference distributions: $q^+_k \rightarrow q^-_k = (q_k-\anti{q}_k)$
    \item the full electroweak charges have to be replaced by the parity-violating couplings: $a^2_k\rightarrow\tilde{a}^2_k$
    \item the \textit{ns}-coefficients $\coef[ns]{\lambda}$ have to be replaced with the appropriate coefficients for $\F{3}$: $\coef[ns]{\lambda}\rightarrow\coef[ns]{3}$
\end{itemize}
Applying these changes together with the fact that both gluon and \textit{ps}-coefficients vanish in the massless scheme, one finds 
\begin{subequations}
\begin{align}
    \F[k][l]{3} &= \tilde{a}^2_k\left[\coeftilde[ns,A]{3}\delta_{kl}+\coeftilde[ns,B]{3}\right]\otimes_{kl}q^-_k \\
    \F[0][l]{3} &= 0\,.
\end{align}
\end{subequations}

\subsection{Extension of \aSACOTchi{} to charged current}
\label{sec:extension_of_aSACOT-chi_to_CC}

In this section, we extend the \aSACOTchi{} scheme to charged current interactions. The general principle of dissecting the massless Wilson coefficients into the individual flavor contributions remains the same. However, as we have a flavor change at the electroweak vertex, the flavor structure becomes more involved and we have to keep track of an additional flavor $j$:
\begin{equation}
    \F[]{\lambda} = \sum_{k=0}^{6}\sum_{l=1}^{6}\sum_{j=1}^{6} \F[k][j,l]{\lambda}\,.
\end{equation}
As an example, consider one of the diagrams of \cref{fig:NNLO_ns_CC_nf} for a $\nu_{\mu} \rightarrow \mu^-$ interaction (i.e.~with a $W^+$ boson): The incoming flavor $k$ has to be of down-type ($k\in\{d,s,b\}$) and the outgoing flavor $j$ of up-type ($j\in\{u,c,t\}$). However, the radiated flavor $l$ can be of any type. Therefore, we have to consider cases where, e.g., $k=b,\,j=t$ and $l=c$ and keep track of three masses simultaneously. The $\chi(n)$-scaling variable is generalized to 
\begin{equation}
    \chi_{klj}(n) = x\left(1+\frac{[n\max(m_k,m_l,m_j)]^2}{Q^2}\right)
\end{equation}
for convolutions with three flavor indices denoted by~$\otimes_{klj}$.

As in the neutral current case, we start the flavor decomposition from the formulation in the massless scheme: the structure functions for an incoming neutrino $\F{\lambda}(\nu)$ and an incoming anti-neutrino $\F{\lambda}(\anti{\nu})$ are conveniently computed using the linear combinations
\begin{align}
	\F[][+]{\lambda} &= \frac{1}{2}\left[\F{\lambda}(\nu) + \F{\lambda}(\anti{\nu})\right] \notag\\
	&= \sumbothtype|V_{UD}|^2\left(\coef[+,ns]{\lambda} + \coef[ps]{\lambda}\right)\otimes\left( D^+ + U^+\right) \notag\\
    &\quad+ \left[\sumbothtype|V_{UD}|^2\right]4\coef[g]{\lambda}\otimes g
\end{align}
\begin{align}
	\F[][-]{\lambda} &= \frac{1}{2}\left[\F{\lambda}(\nu) - \F{\lambda}(\anti{\nu})\right] \notag\\
	&= \sumbothtype|V_{UD}|^2 \coef[-,ns]{\lambda}\otimes\left(D^- - U^-\right) \,,
\end{align}
where the sums run over up-/down-type quarks. Note that the \textit{ps}- and gluon coefficients do not contribute to the ``$-$''-combination. To disentangle the final-state flavors, we notice that the massless Wilson coefficients have the same $n_f$-dependence as in the neutral current case given in \cref{eq:coef_nf_dependence}. In the following, we investigate the flavor structure of the \textit{ns}-, \textit{ps}- and gluon coefficients separately.

{\bfseries \textit{ns}-coefficients:}
The relevant Feynman diagrams are given in \cref{fig:NNLO_ns_CC}. Following the flavor notation of these diagrams, we note that $k$ and $j$ are connected: if $k$ is of down-type, $j$ has to be up-type. The third flavor $l$ is independent analogously to the neutral current relations. Thus, we arrive at the decomposition:

\begin{widetext}
\begin{subequations}
\begin{equation}
	\F[k][+,ns,j,l]{\lambda} = 
	\begin{cases}
		\phantom{-}|V_{kj}|^2 \left[\coeftilde[+,ns,A]{\lambda}\delta_{kl} + \coeftilde[+,ns,B]{\lambda}\right]\otimes_{kjl} q^+_k & \text{if } k\in U,\, j\in D \\
		
		\phantom{-}|V_{jk}|^2 \left[\coeftilde[+,ns,A]{\lambda}\delta_{kl} + \coeftilde[+,ns,B]{\lambda}\right]\otimes_{kjl} q^+_k & \text{if }  k\in D,\, j\in U \\
		\phantom{-}0 & \text{else,} 
	\end{cases}
    \label{eq:CC_F2L_ns_flavor_dissection_P}
\end{equation}
\begin{equation}
	\F[k][-,ns,j,l]{\lambda} = 
	\begin{cases}
		-|V_{kj}|^2 \left[\coeftilde[-,ns,A]{\lambda}\delta_{kl} + \coeftilde[-,ns,B]{\lambda}\right]\otimes_{kjl} q^-_k & \text{if } k\in U,\, j\in D \\
		
		\phantom{-}|V_{jk}|^2 \left[\coeftilde[-,ns,A]{\lambda}\delta_{kl} + \coeftilde[-,ns,B]{\lambda}\right]\otimes_{kjl} q^-_k & \text{if } k\in D,\, j\in U \\
		\phantom{-}0 & \text{else.} 
	\end{cases}
    \label{eq:CC_F2L_ns_flavor_dissection_M}
\end{equation}%
\end{subequations}
\end{widetext}
Note the extra minus sign in $\F[k][-,ns,j,l]{\lambda}$ when $k$ is of up-type.

{\bfseries \textit{ps}-coefficient:} From the graphs in \cref{fig:NNLO_ps_CC} one can see that in these contributions the flavor indices $j$ and $l$ are connected. The incoming flavor $k$ is independent. Thus we find
\begin{equation}
	\F[k][+,ps,j,l]{\lambda} =
	\begin{cases}
		|V_{jl}|^2\,\coeftilde[ps]{\lambda} \otimes_{kjl} q_k^+ &\text{if } j\in U,\, l\in D\\
		|V_{lj}|^2\,\coeftilde[ps]{\lambda} \otimes_{kjl} q_k^+ &\text{if } j\in D,\, l\in U\\
		0 & \text{else.}
	\end{cases} 
\end{equation}

{\bfseries gluon coefficient:} In the gluon-initiated contributions, shown in \cref{fig:NNLO_g_CC}, $j$ and $l$ are connected yielding
\begin{equation}
	\F[0][+,j,l]{\lambda} = 
	\begin{cases}
		4\,|V_{jl}|^2\,\coeftilde[g]{\lambda}\otimes_{jl} g &\text{if } j\in U,\, l\in D\\
		4\,|V_{lj}|^2\,\coeftilde[g]{\lambda}\otimes_{jl} g &\text{if } j\in D,\, l\in U\\
		0 & \text{else.} 
	\end{cases}
\end{equation}

{\bfseries Final formula:} The final structure functions are constructed by summing over the three classes of coefficients, which yields
\begin{subequations}
\begin{align}
	\F[][+]{\lambda} &= \sum_{k=0}^{6}\sum_{j=1}^{6}\sum_{l=1}^{6} \left[ \F[k][+,ns,j,l]{\lambda} + \F[k][ps,j,l]{\lambda} + \F[k][g,j,l]{\lambda} \right]\\
	\F[][-]{\lambda} &= \sum_{k=0}^{6}\sum_{j=1}^{6}\sum_{l=1}^{6}  \,\,\F[k][-,ns,j,l]{\lambda}\,.
\end{align}
\end{subequations}

\subsubsection{\aSACOTchi{} for charged current \texorpdfstring{$\F{3}$}{F3}}

The same derivation can be made for $\F{3}$ with the replacement $\anti{q}_k \rightarrow -\anti{q}_k$. Both the \textit{ps}-coefficient and the gluon-initiated coefficient are zero, such that the massless result reads
\begin{subequations}
\begin{align}
    \F[][+]{3} &= \sumbothtype|V_{UD}|^2 \coef[+,ns]{3}\otimes\left( D^- + U^-\right) \\
    \F[][-]{3} &= \sumbothtype|V_{UD}|^2 \coef[-,ns]{3}\otimes\left(D^+ - U^+\right) 
\end{align}
\end{subequations}
However, since the flavor structure is analogous to $\F{2,L}$, we can recycle \cref{eq:CC_F2L_ns_flavor_dissection_P,eq:CC_F2L_ns_flavor_dissection_M} by making the replacement \mbox{$q^+_k \leftrightarrow q^-_k$} and write immediately
\begin{subequations}
	\begin{align}
		\F[][+]{3} &= \sum_{k=0}^{6}\sum_{j=1}^{6}\sum_{l=1}^{6} \F[k][+,ns,j,l]{3} \\
		\F[][-]{3} &= \sum_{k=0}^{6}\sum_{j=1}^{6}\sum_{l=1}^{6} \F[k][-,ns,j,l]{3}\,.
	\end{align}
\end{subequations}

\goodbreak
\section{Results}
\label{sec:results}

In this section, we present the numerical implementation of the \aSACOTchi{} scheme in
\apfelxx{}~\cite{Bertone:2013vaa,Bertone:2017gds,APFELppGithub} and show the impact on physical predictions using the scheme. First, we focus on the effect of the $\chi$-scaling variable on the structure functions for neutral and charged currents at NNLO. We then examine charged-current cross sections for kinematics that have been measured at HERA, and also consider pseudo-data sets for the EIC. The final subsection highlights the NNLO effects for neutrino DIS data sets from NuTeV, CCFR and Chorus, which are measured at lower $Q^2$ where the impact of heavy quarks can be significant. 

\subsection{Numerical Implementation}

\begin{table*}[tb]
	\centering
	\caption{The available VFN-schemes for the individual structure functions in \apfelxx{}~\cite{Bertone:2013vaa,Bertone:2017gds,APFELppGithub} at NLO and NNLO. The black check marks indicate the structure functions that were included in the public release of the code, while the red check marks indicate the newly added structure functions.}
	\renewcommand{\arraystretch}{1.2}
	\begin{tabular}{|>{\raggedright}m{0.15\textwidth} 
			|>{\centering}m{0.075\textwidth}>{\centering}m{0.075\textwidth} >{\centering}m{0.075\textwidth} 
			|>{\centering}m{0.075\textwidth} >{\centering}m{0.075\textwidth}>{\centering}m{0.075\textwidth} 
			|}
		\hline
		\multicolumn{1}{|c|}{}& \multicolumn{3}{c|}{\textbf{neutral current}} & \multicolumn{3}{c|}{\textbf{charged current}} \tabularnewline
		VFNS & ${\F{2}}$ & ${\F{L}}$& ${\F{3}}$ & ${\F{2}}$& ${\F{L}}$ & ${\F{3}}$ \tabularnewline
		\hline
		\multicolumn{1}{c}{\rule{0pt}{12pt}\textbf{NLO}} & \multicolumn{6}{c}{}\tabularnewline
		\hline
		
		full ACOT &\newimplemented &-- &-- &-- &-- &-- \tabularnewline
		SACOT-$\chi$    &\newimplemented &\newimplemented &\newimplemented &\newimplemented &\newimplemented &\newimplemented \tabularnewline
		\hline
		\multicolumn{1}{c}{\rule{0pt}{12pt}\textbf{NNLO}}& \multicolumn{6}{c}{}\tabularnewline
		\hline
		ZM & \alreadyimplemented & \alreadyimplemented & \alreadyimplemented & \alreadyimplemented & \alreadyimplemented & \alreadyimplemented \tabularnewline
		FONLL-C & \alreadyimplemented & \alreadyimplemented & -- & -- & -- & --\tabularnewline 
		aSACOT-$\chi$ &\newimplemented &\newimplemented &\newimplemented &\newimplemented &\newimplemented &\newimplemented \tabularnewline
		
		\hline
	\end{tabular}
	\renewcommand{\arraystretch}{1.}
	\label{tab:HQ_schemes_new_in_apfelxx}
\end{table*}

The numerical implementation was performed in the open-source code \apfelxx{}~\cite{Bertone:2013vaa,Bertone:2017gds,APFELppGithub}. The code base allows for a numerically efficient evaluation of the structure functions by means of precomputed interpolation tables. The computational advantage lies in interpolating the PDFs with
\begin{equation}
	f(x) = \sum_{\alpha=0}^{N_x}w_{\alpha}(x) f(x_\alpha)\,,
	\label{eq:PDF_interpolation}
\end{equation}
on a fixed $x$-grid $\{x_\alpha\}$ and storing the time-expensive convolution of the Wilson coefficients with the interpolating functions $w_{\alpha}(x)$. At evaluation time, only the interpolation step has to be performed, which reduces to a simple matrix-vector multiplication -- an extremely fast operation on modern CPUs. This setup allows one to recompute the structure functions with different PDF sets very quickly. The details of the implementation have been deferred from the main text and can be found in \cref{sec:Details_of_the_numerical_implementation}. An overview of current and newly implemented schemes is given in \cref{tab:HQ_schemes_new_in_apfelxx}.

The \apfelxx{} framework currently provides structure functions in the \ZMVFNS{} for both neutral- and charged-current interactions up to NNLO.
Additionally, the massive neutral-current $F_2$ and $F_L$ structure functions in the NNLO scheme \FONLL[C]~\cite{Forte:2010ta} are also available in the code. 

We implemented the full ACOT scheme for neutral-current $\F{2}$ and the \SACOTchi{} scheme for the complete set $\{\F{2},\F{L},\F{3}\}$ for neutral and charged current at NLO. The fully massive coefficients have been taken from Refs.~\cite{Aivazis:1993kh,Aivazis:1993pi,Kretzer:1998ju}. The implementation has been compared to the \ncteq{} code-base (which was in turn benchmarked in, e.g., Ref.~\cite{SM:2010nsa}) and agrees to better than 1\textperthousand{} with an evaluation speed that is $\mathcal{O}(100)$ faster. The remaining difference can be attributed to the interpolation effects, and can thus be made arbitrarily small.

At NNLO we implemented the $\{\F{2},\F{L},\F{3}\}$ structure functions for both neutral- and charged-current interactions in the \aSACOTchi{} scheme.
It is to be noted that for the numerical implementation we do not use the exact formulas for the NNLO massless coefficients presented in Refs.~\cite{vanNeerven:1991nn,Zijlstra:1991qc,Zijlstra:1992qd}, but an $x$-space parametrization, which was provided in Refs.~\cite{vanNeerven:1999ca,vanNeerven:2000uj}. These are accurate enough for numerical applications, but significantly faster to evaluate. The neutral current structure functions have been found to agree to better than 1\textperthousand{} with an evaluation speed that is of $\mathcal{O}(500)$ faster compared to the benchmark code used in the original formulation in Ref.~\cite{Stavreva:2012bs}.

\subsection{Mass effects on structure functions at NNLO}
\label{sec:mass_effects_on_structure_functions_at_NNLO}

The strength of the impact of the dynamic mass at NNLO is controlled by the $n$-scaling variable $\chi(n)$. To compare this effect for the structure functions in an isolated setup, we form a ratio of the \aSACOTchi{} scheme and a combination made up of the standard SACOT-$\chi$ at NLO with the massless scheme at NNLO
\begin{equation}
	\frac{\fstrut\F[][\textbf{LO+NLO}]{\lambda}(SACOT_\chi) + \F[][\textbf{NNLO}]{\lambda}(aSACOT_\chi(n))}{\fstrut\F[][\textbf{LO+NLO}]{\lambda}(SACOT_\chi) + \F[][\textbf{NNLO}]{\lambda}(ZM)\hfill}
	\label{eq:ratio_structure_fcn}
\end{equation}
for all structure functions. Thereby, the ratio is not sensitive to the impact of using NNLO over NLO predictions, but to the mass effects introduced in the \aSACOTchi{} scheme as this is the only difference between numerator and denominator. 

\begin{figure*}
	\centering
	\includegraphics[width=0.33\textwidth]{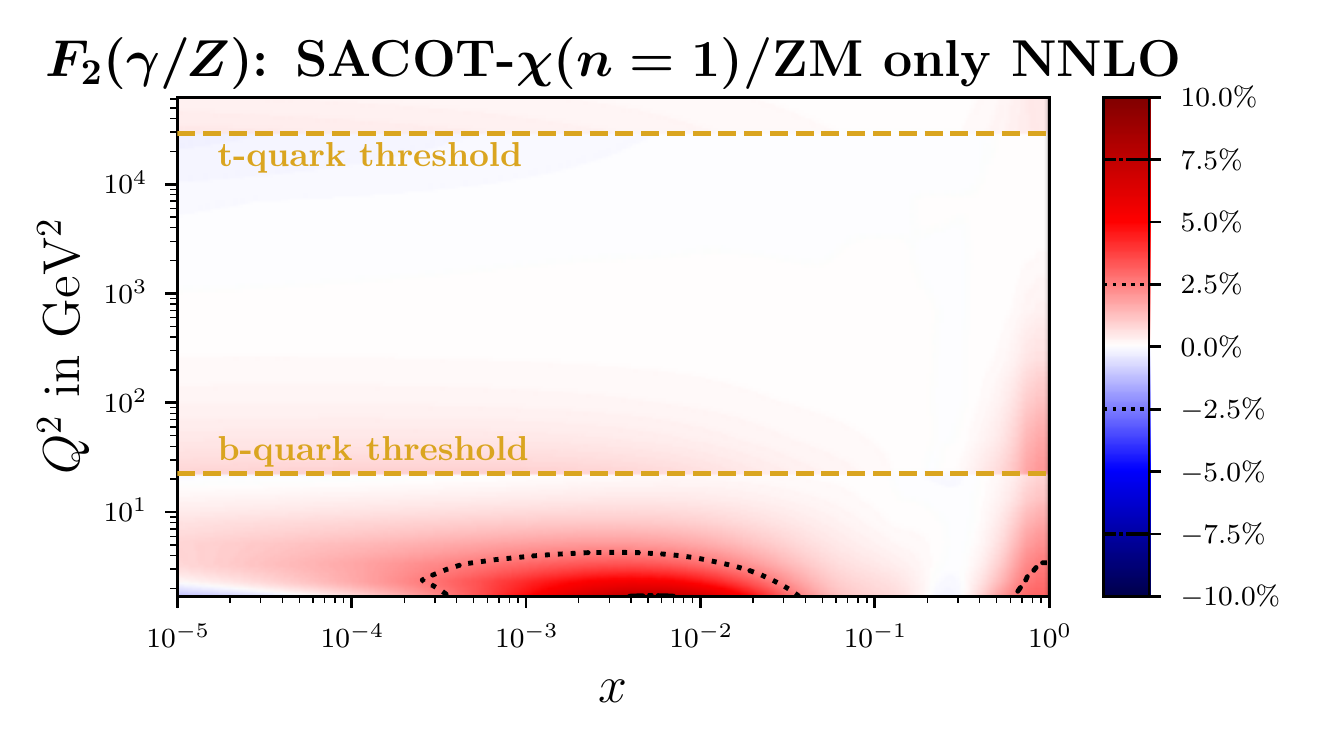}%
	\includegraphics[width=0.33\textwidth]{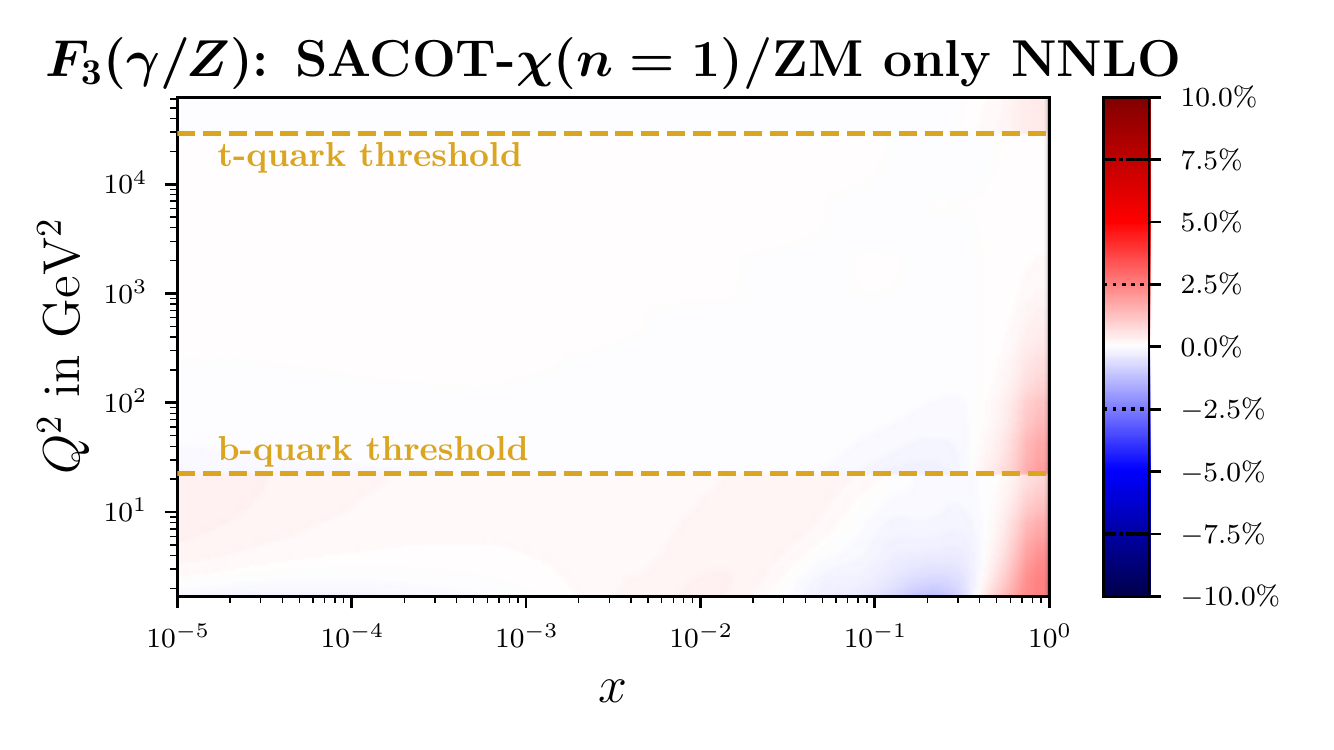}%
	\includegraphics[width=0.33\textwidth]{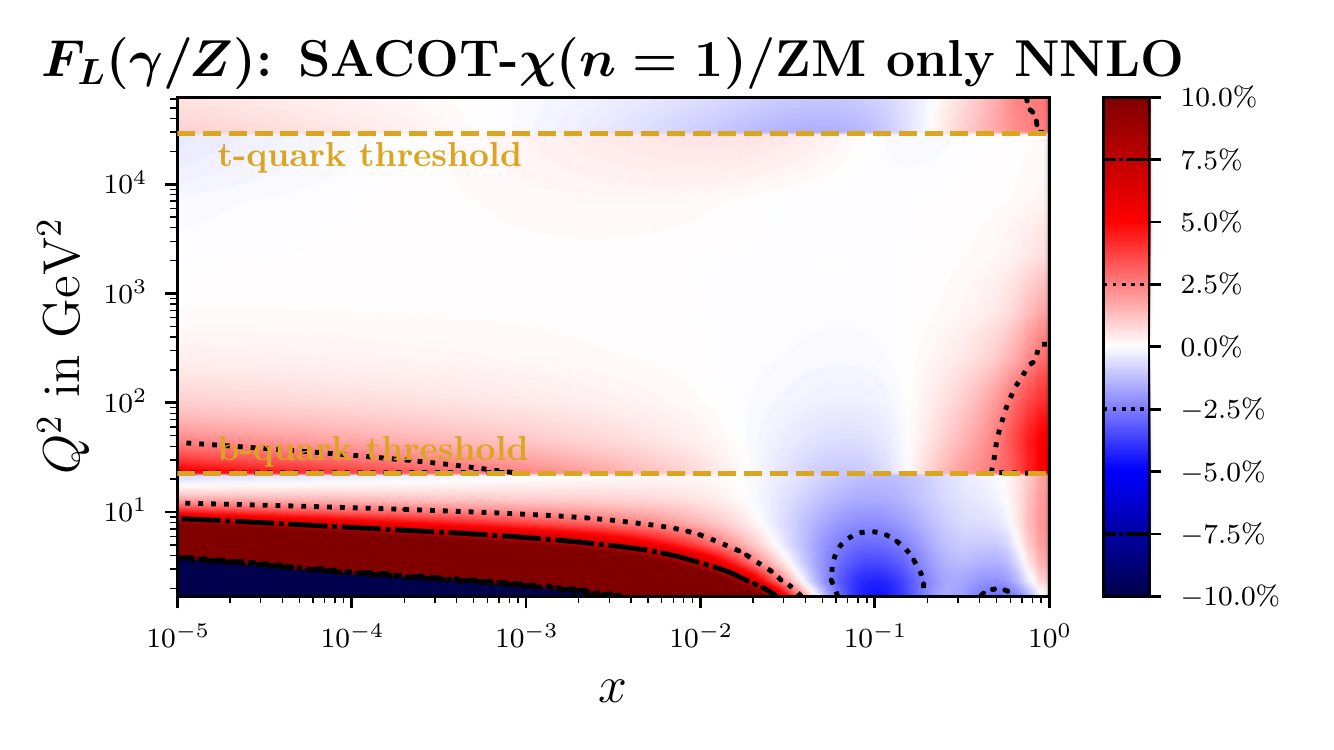}
	\includegraphics[width=0.33\textwidth]{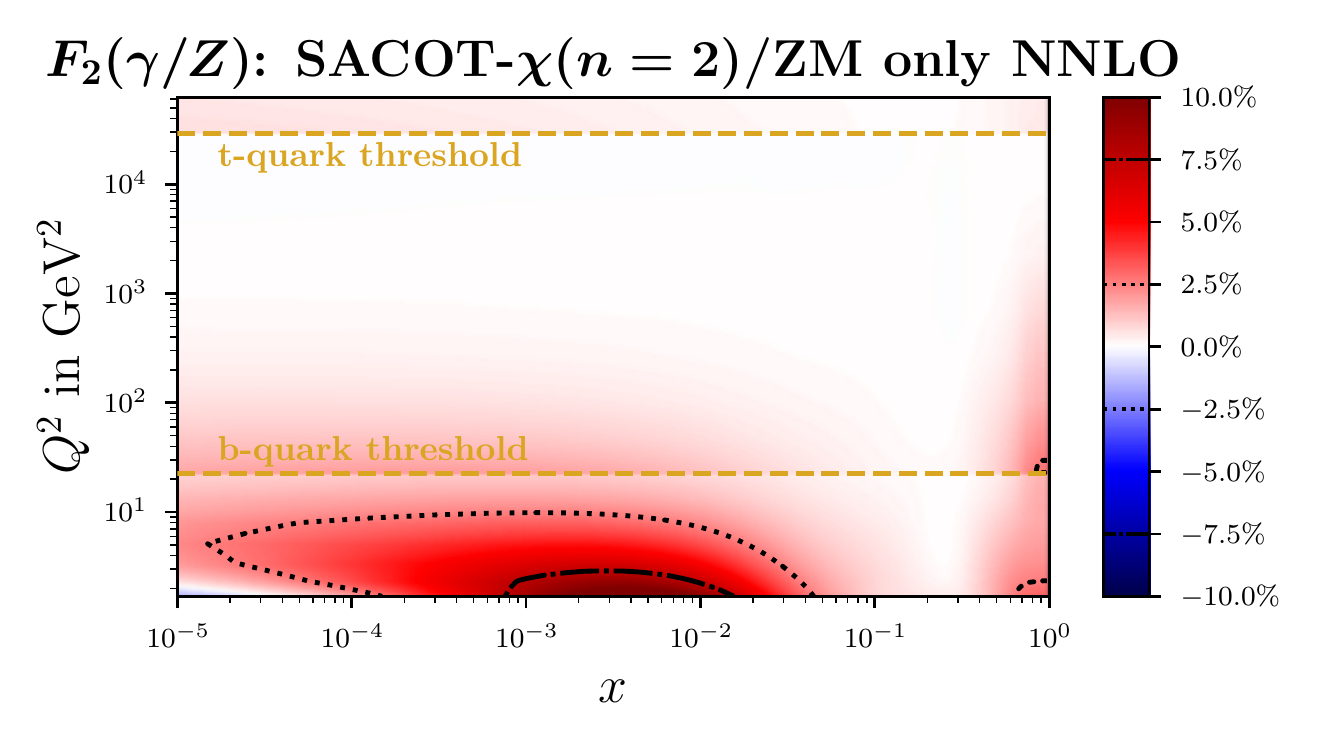}%
	\includegraphics[width=0.33\textwidth]{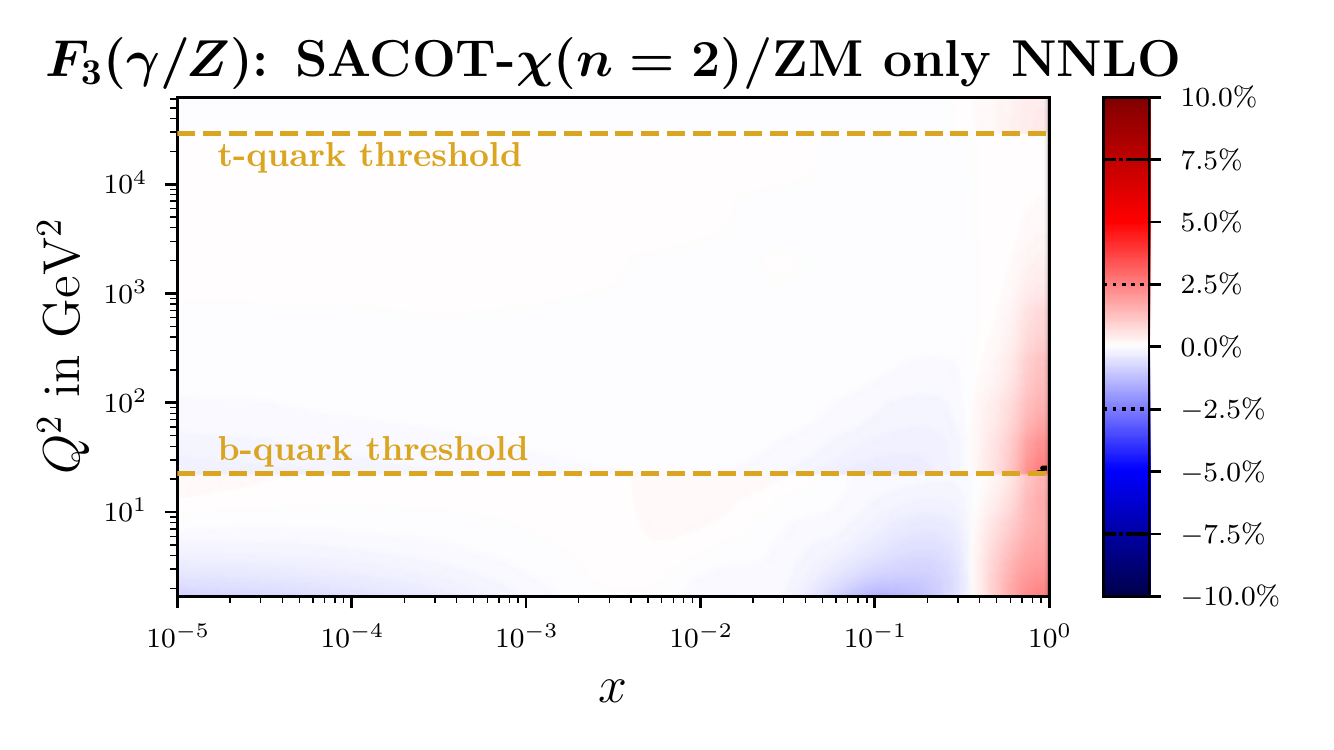}%
	\includegraphics[width=0.33\textwidth]{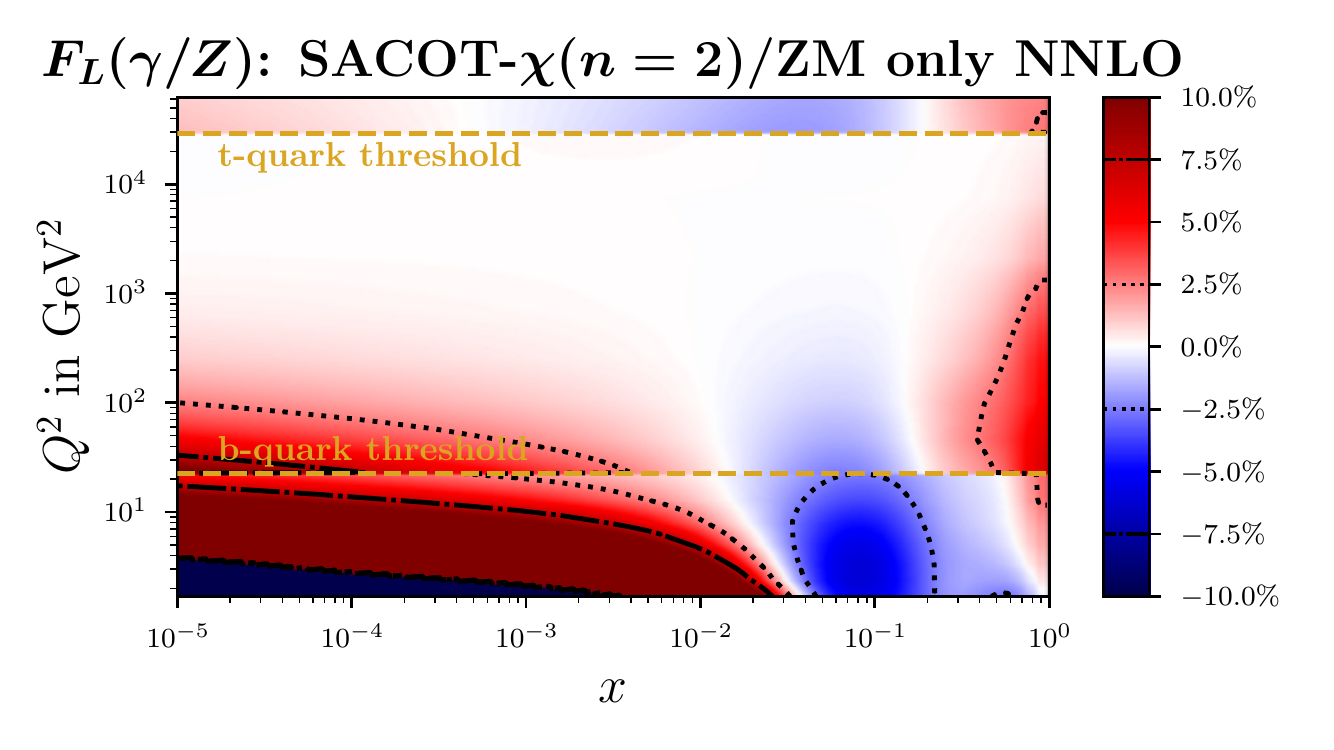}
	\includegraphics[width=0.33\textwidth]{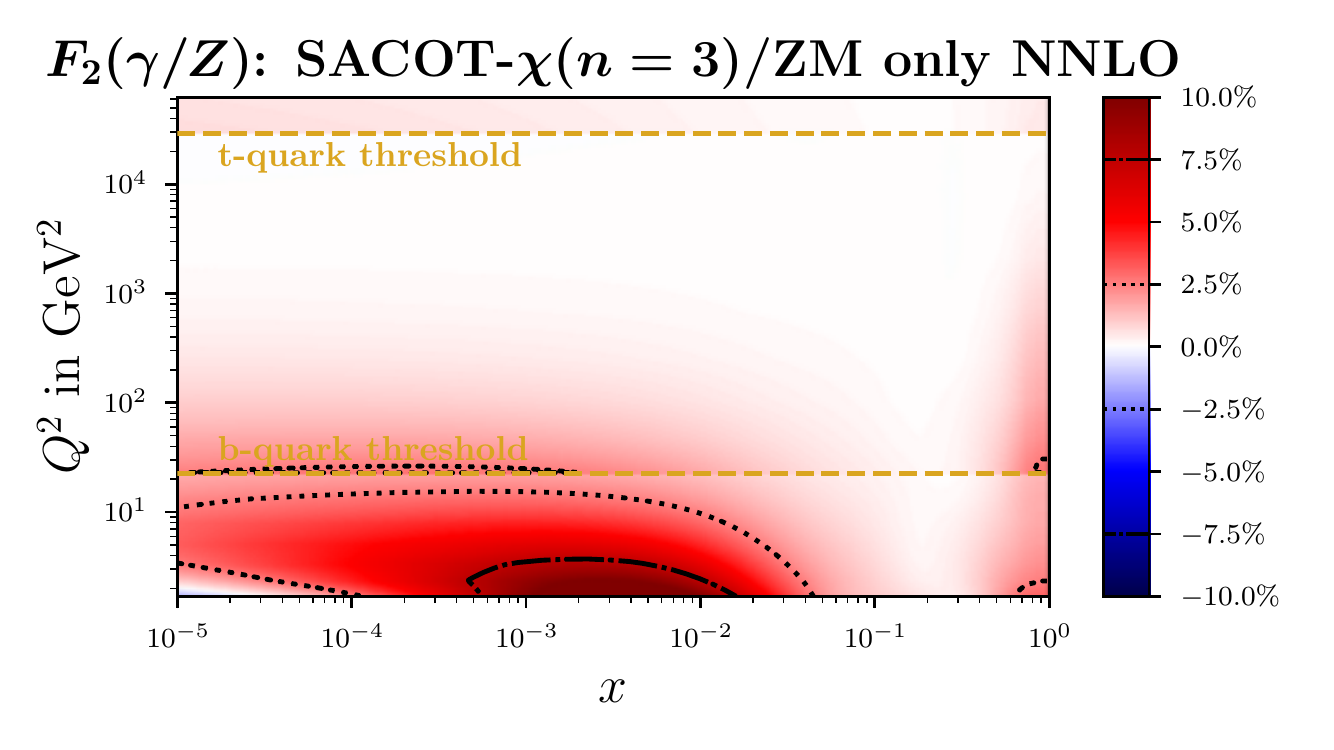}%
	\includegraphics[width=0.33\textwidth]{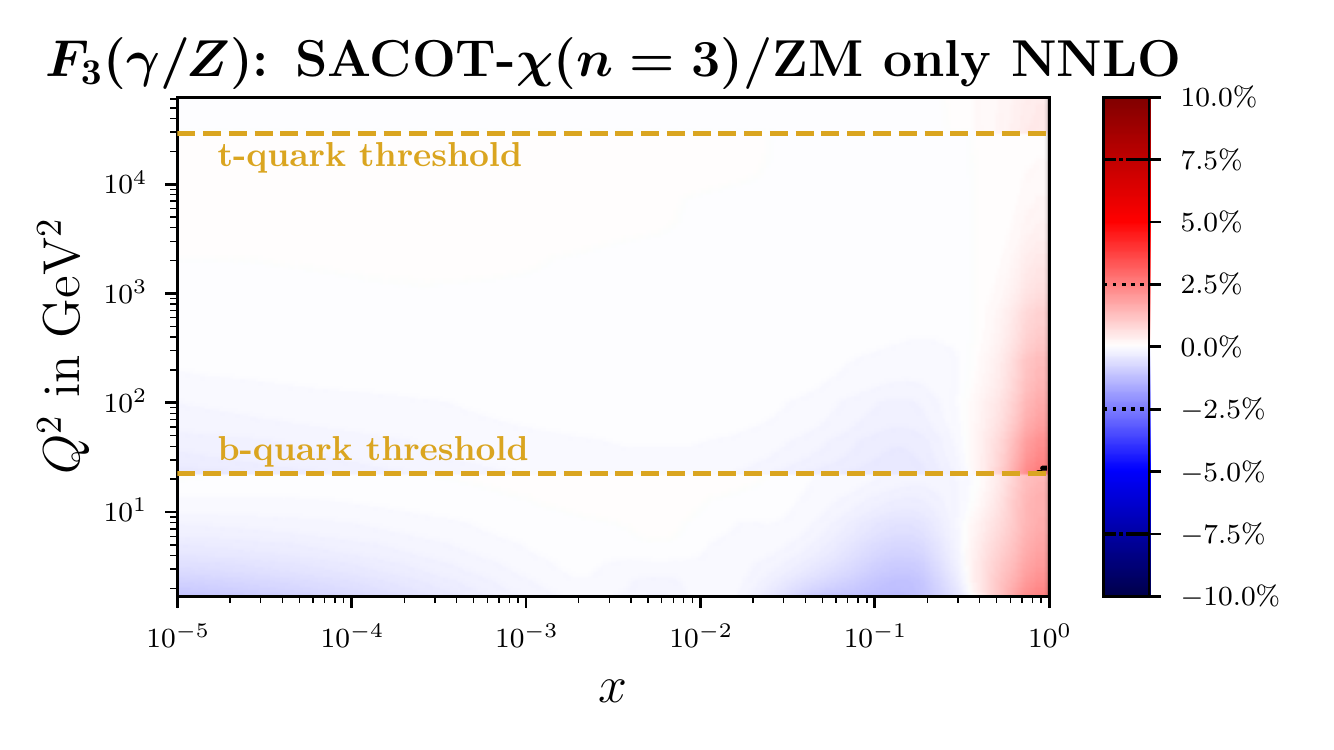}%
	\includegraphics[width=0.33\textwidth]{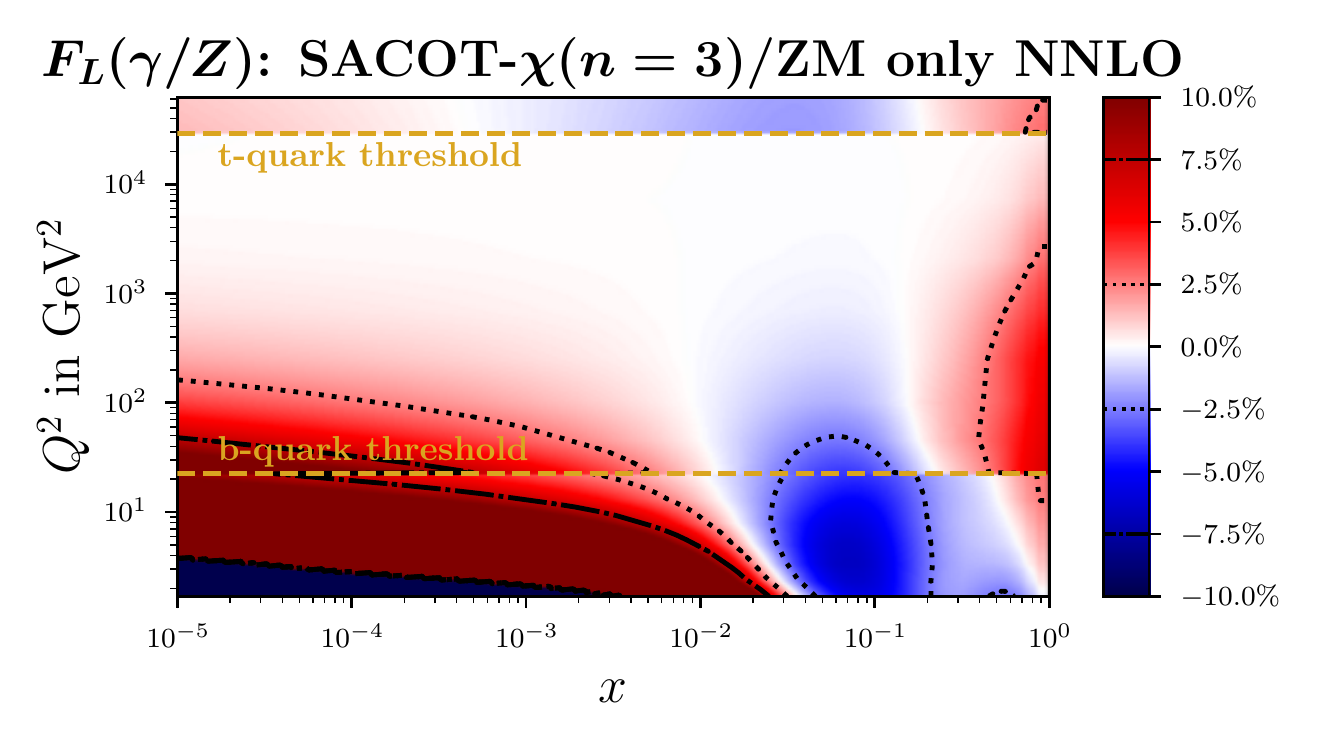}
	\caption{The ratio of neutral current structure functions $F_2,F_3$ and $F_L$ (from left to right) to the ZM NNLO-coefficient for $n=\{1,2,3\}$ (top to bottom). The ratio is defined in \cref{eq:ratio_structure_fcn}. The results are obtained with the \texttt{CT18}~\cite{Hou:2019efy} NNLO proton PDFs. We also indicate the mass thresholds for the bottom- and top-quark, as the ZM-coefficients are discontinuous at these values.}
	\label{fig:ratio_F23L_NC}
\end{figure*}

\begin{figure*}
	\centering
	\includegraphics[width=0.33\textwidth]{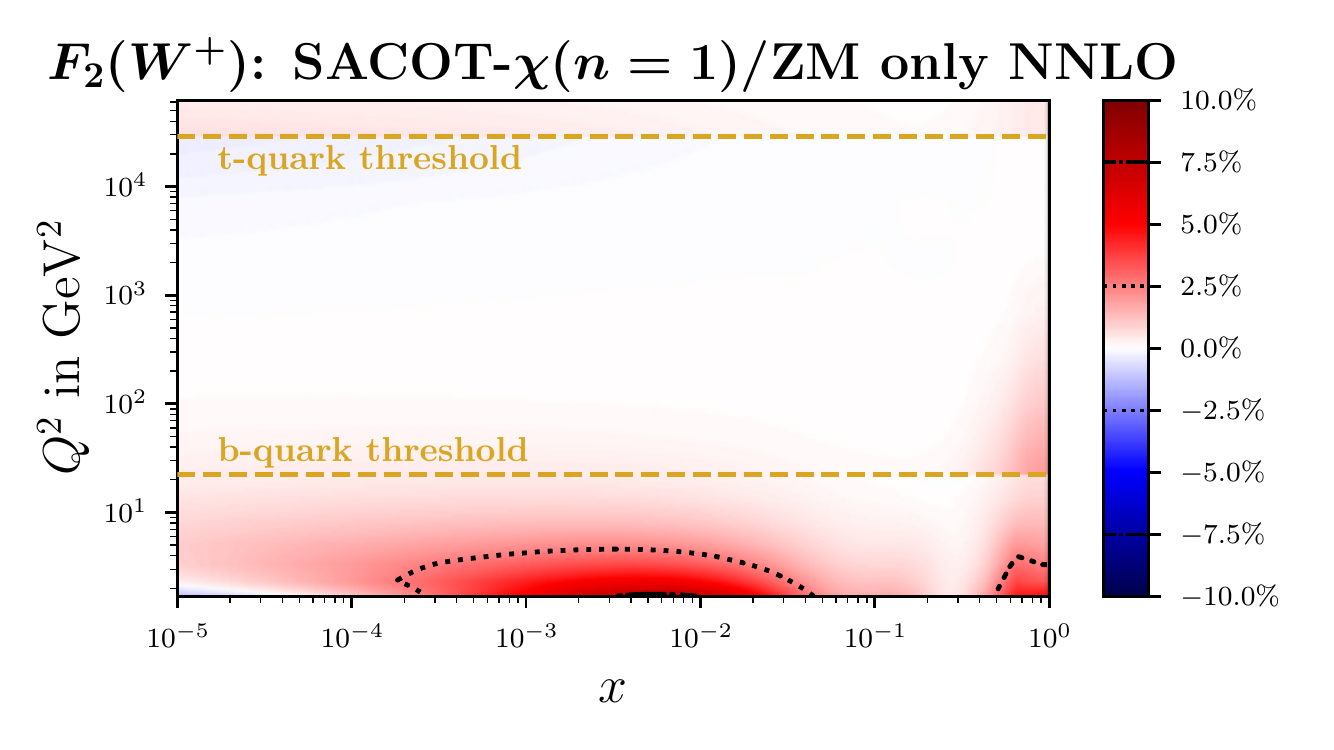}%
	\includegraphics[width=0.33\textwidth]{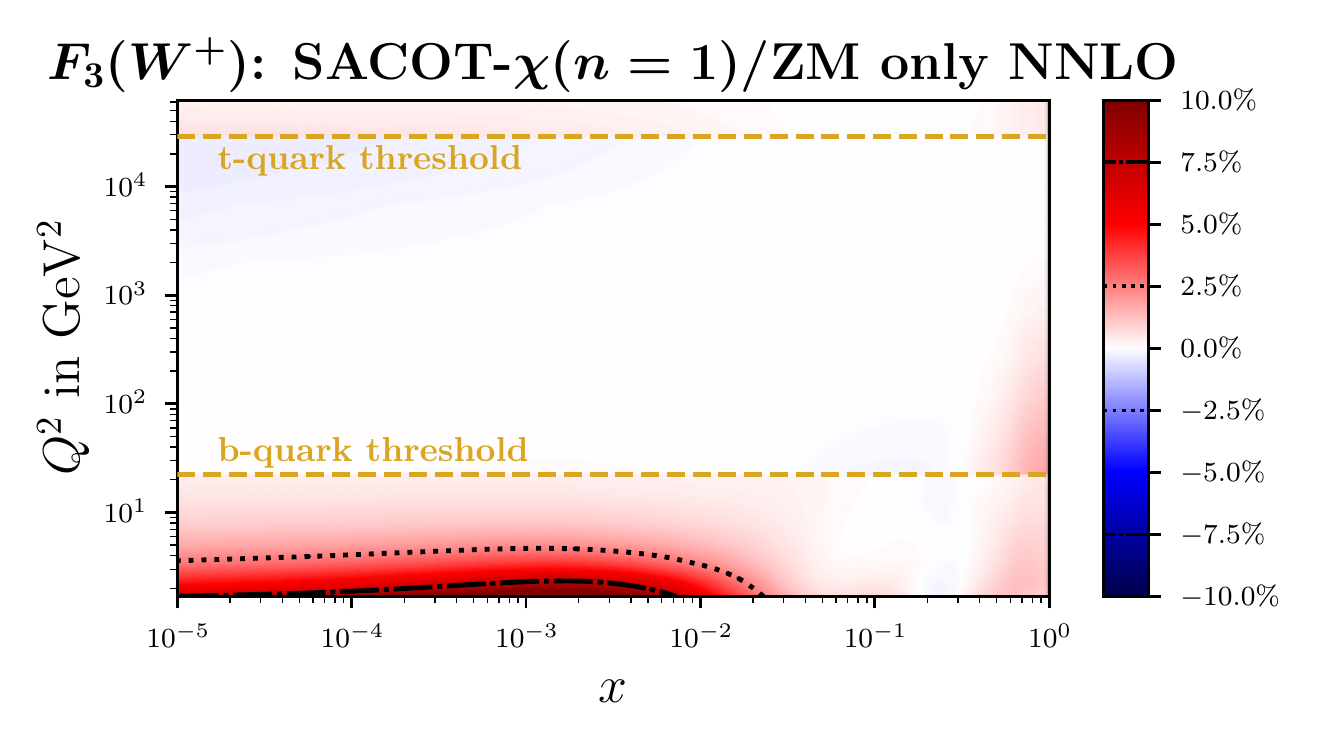}%
	\includegraphics[width=0.33\textwidth]{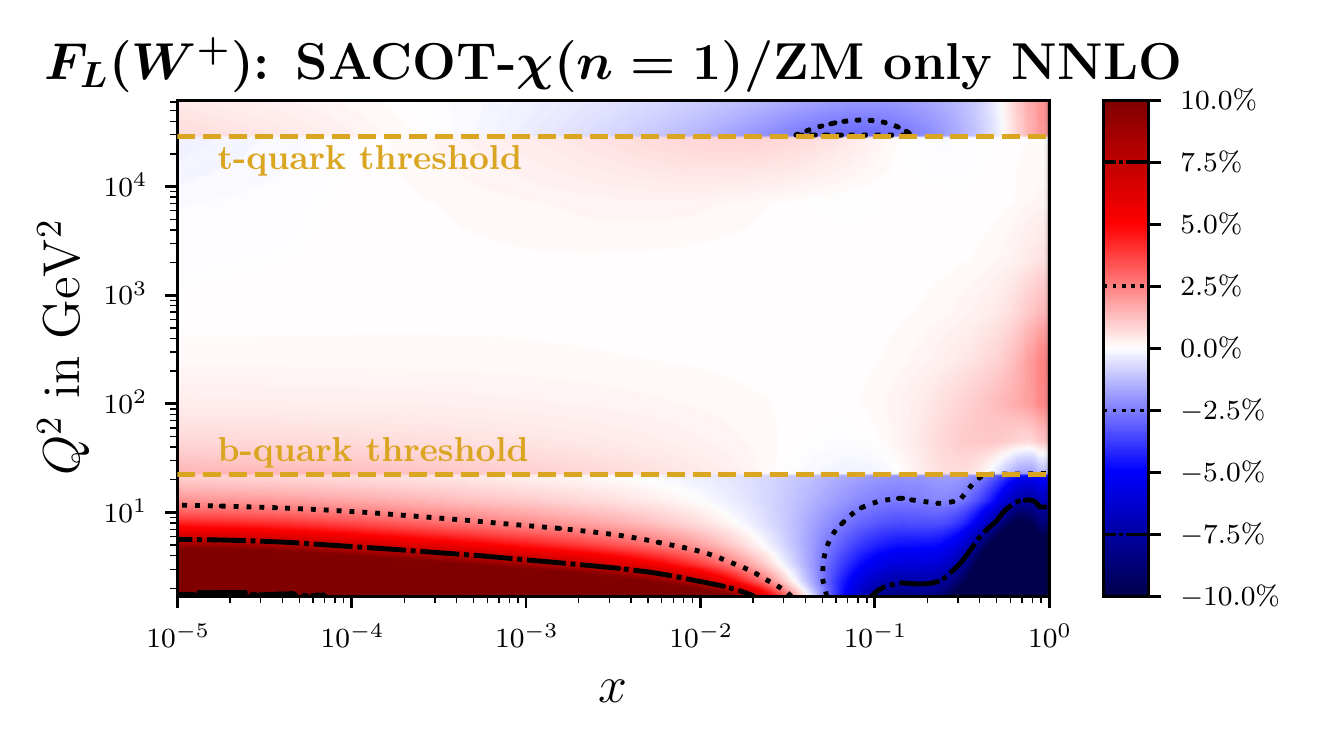}
	\includegraphics[width=0.33\textwidth]{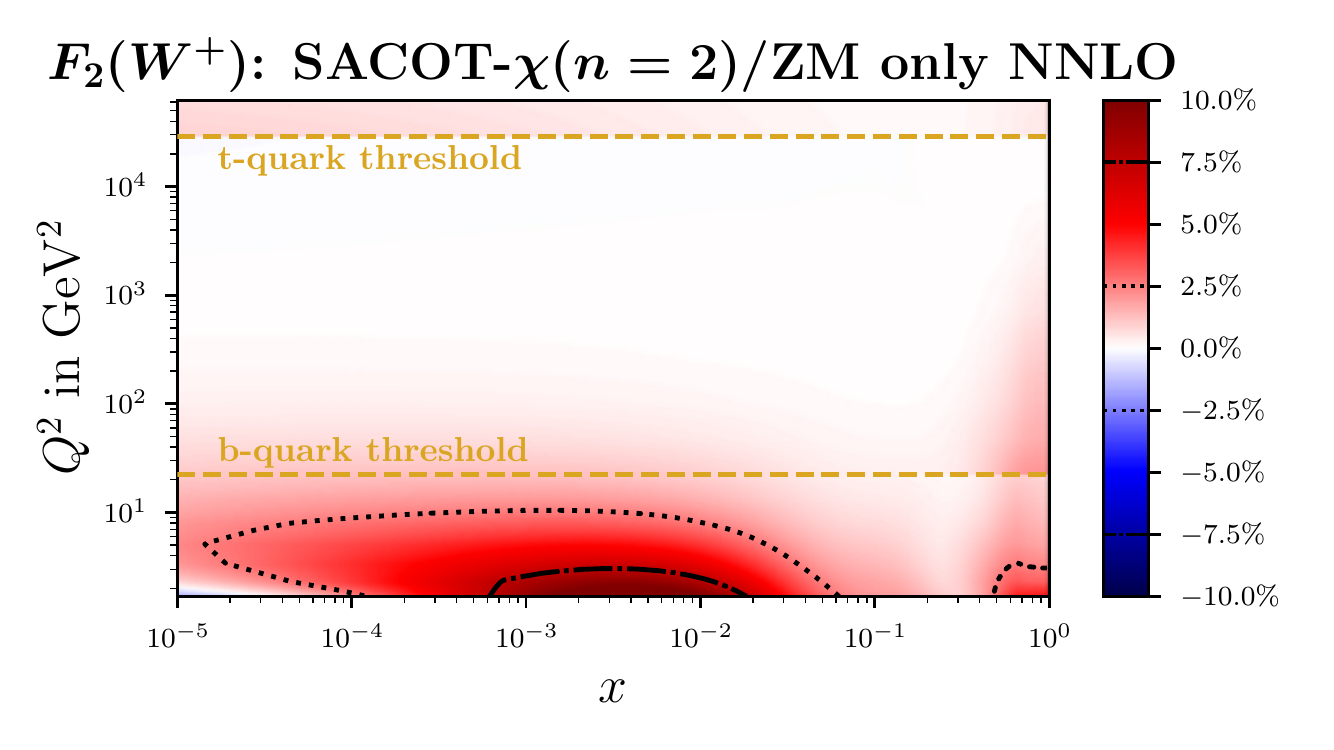}%
	\includegraphics[width=0.33\textwidth]{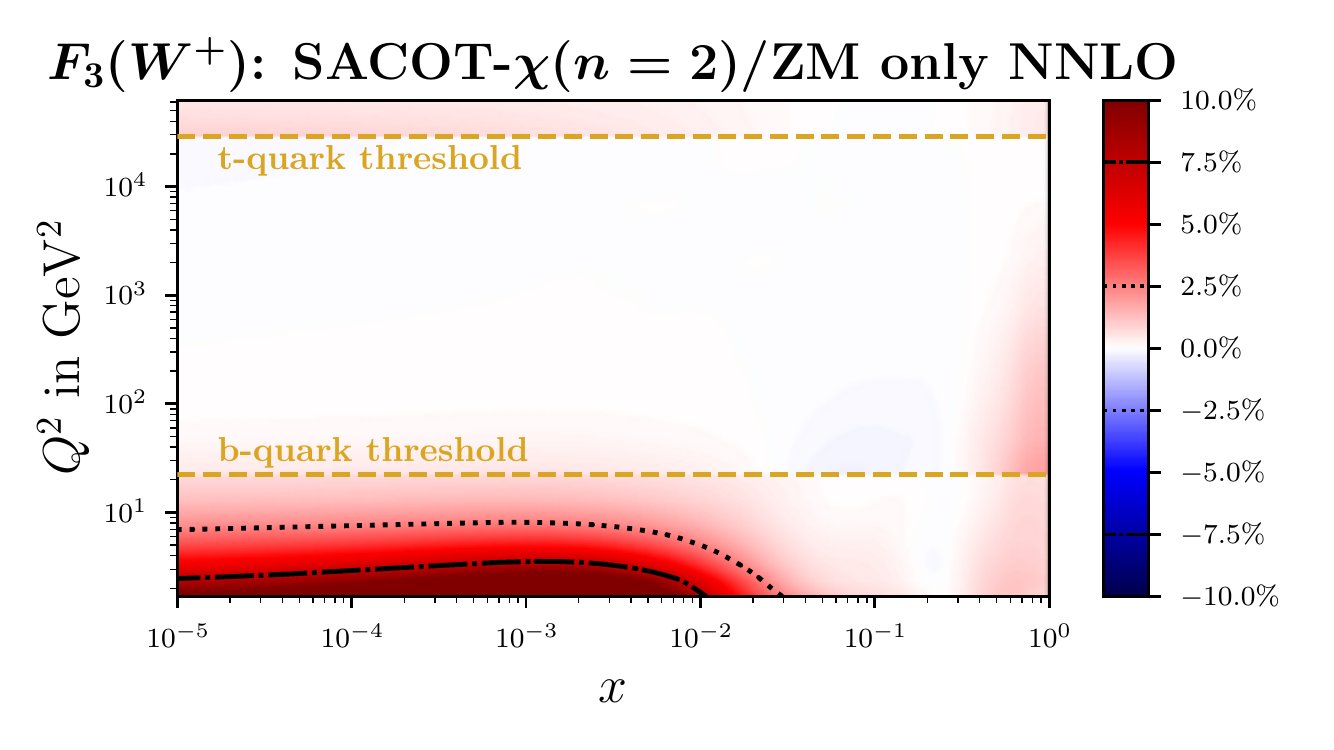}%
	\includegraphics[width=0.33\textwidth]{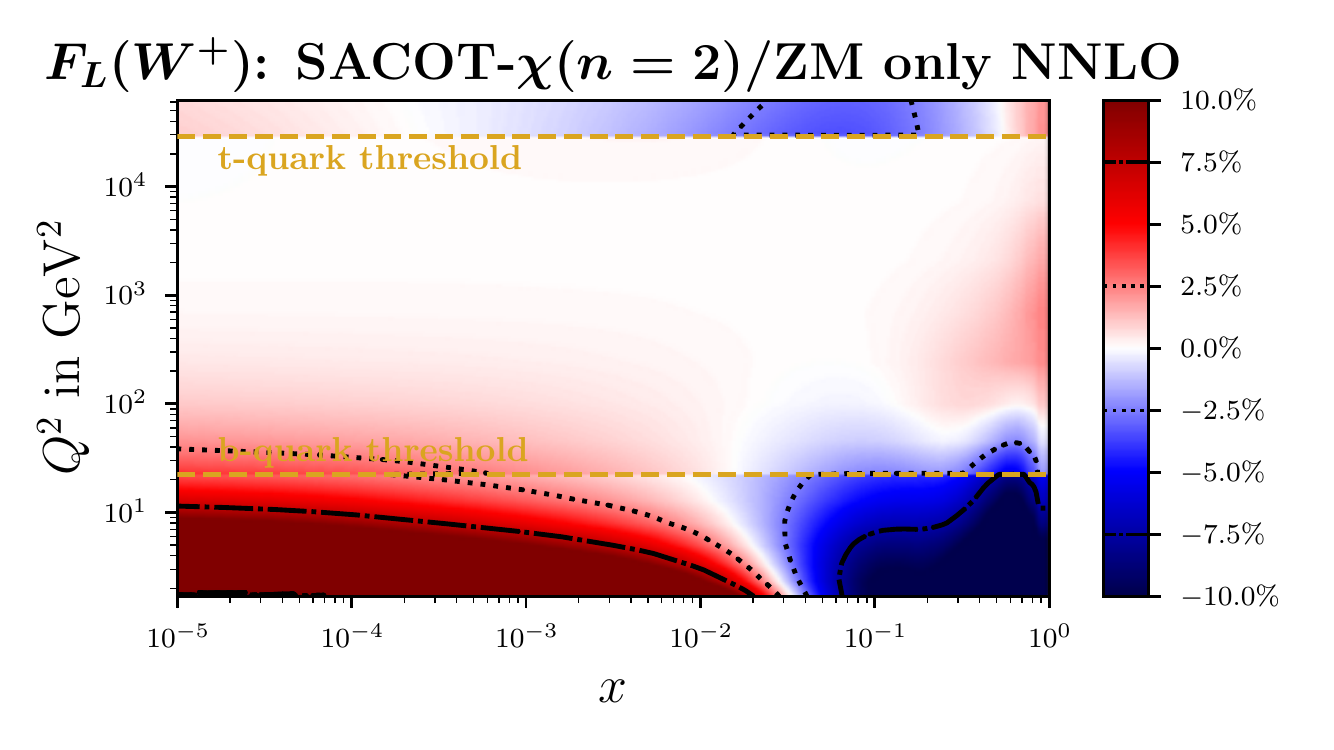}
	\includegraphics[width=0.33\textwidth]{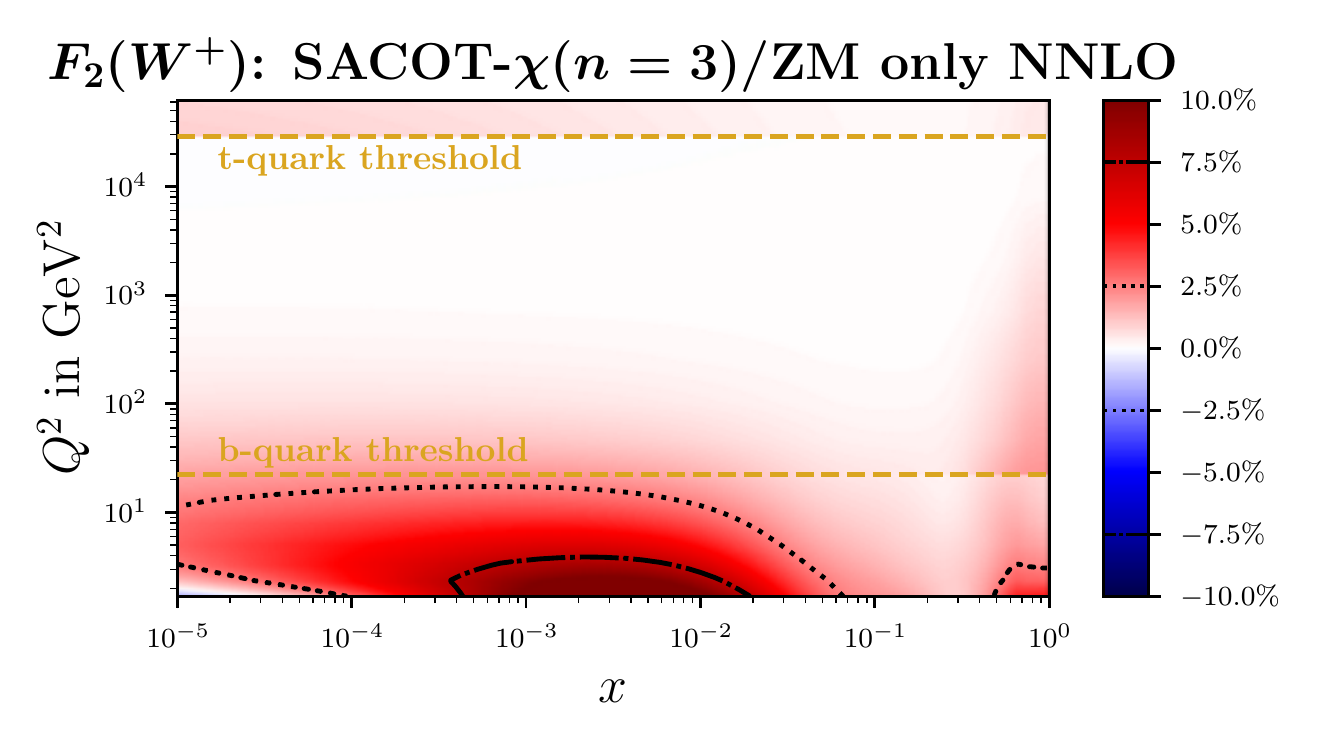}%
	\includegraphics[width=0.33\textwidth]{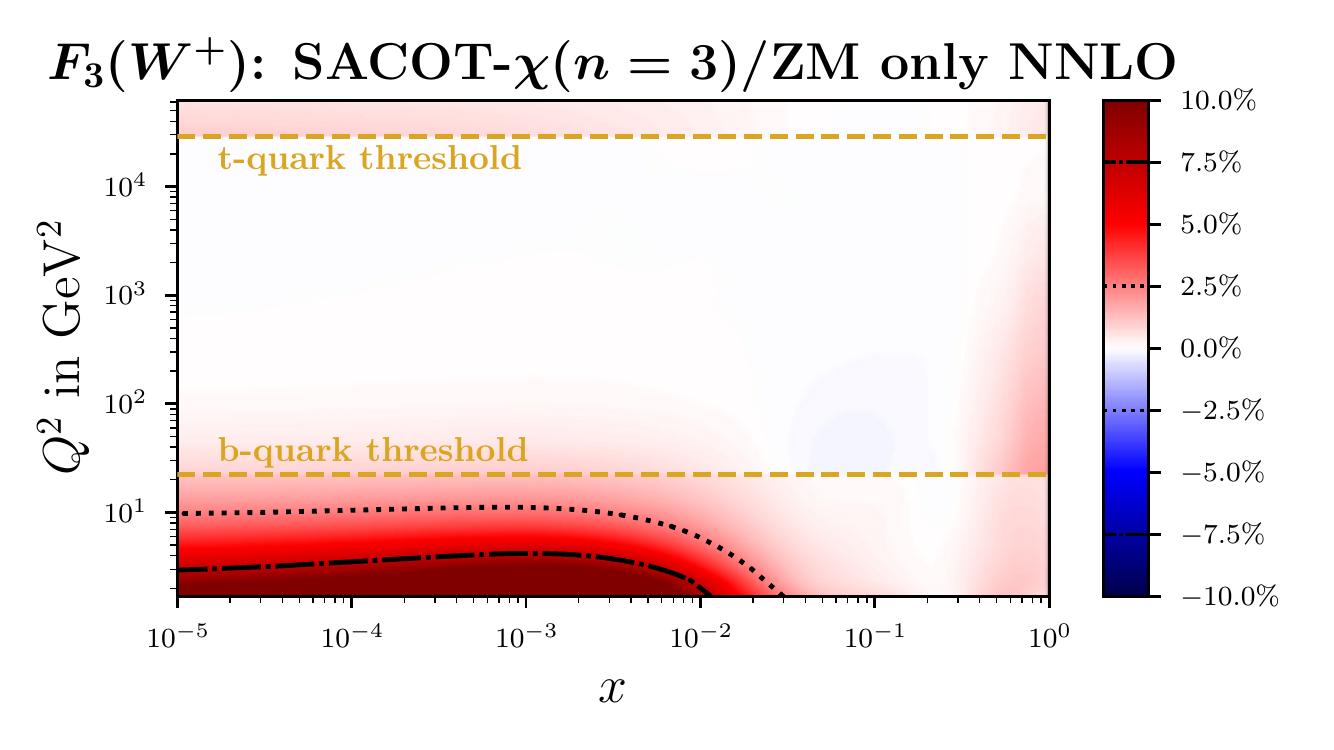}%
	\includegraphics[width=0.33\textwidth]{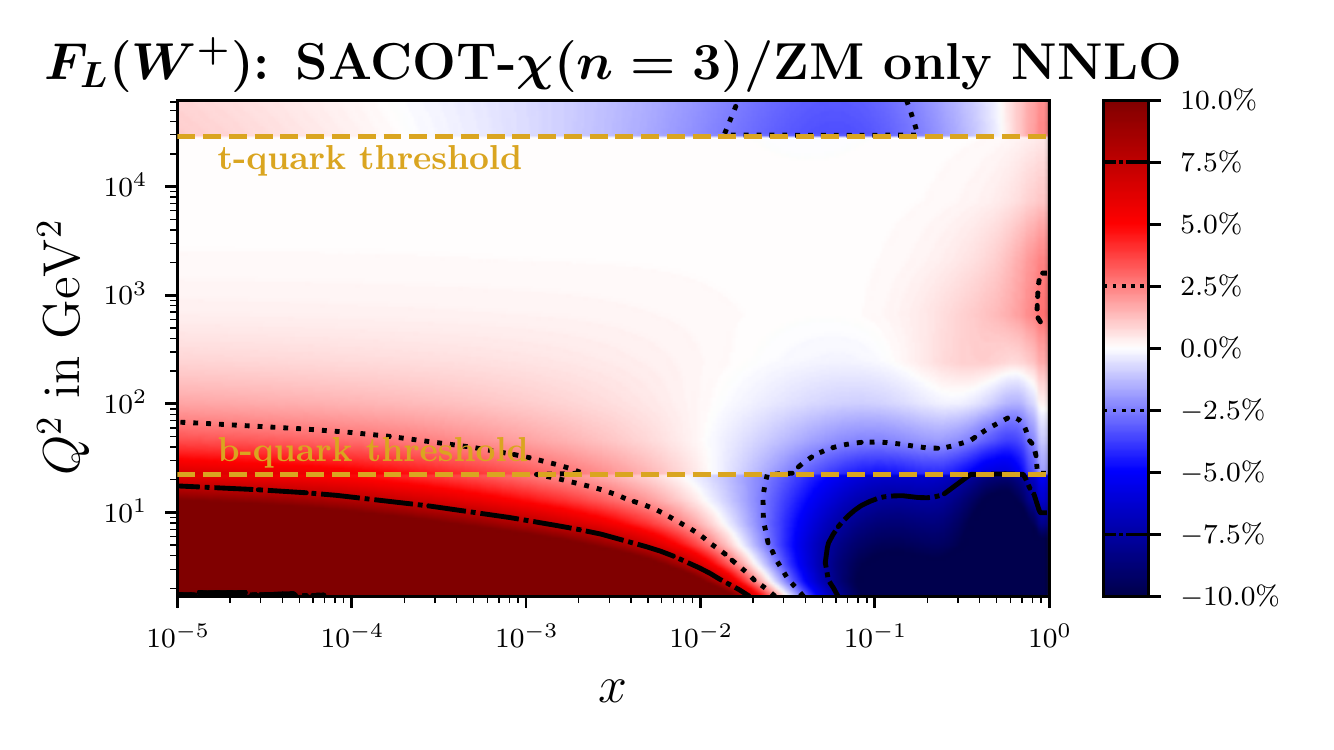}
	\caption{The same as \cref{fig:ratio_F23L_NC} but for charged current with a $W^+$ exchange.}
	\label{fig:ratio_F23L_WP}
\end{figure*}

\begin{figure*}
	\centering
	\includegraphics[width=0.33\textwidth]{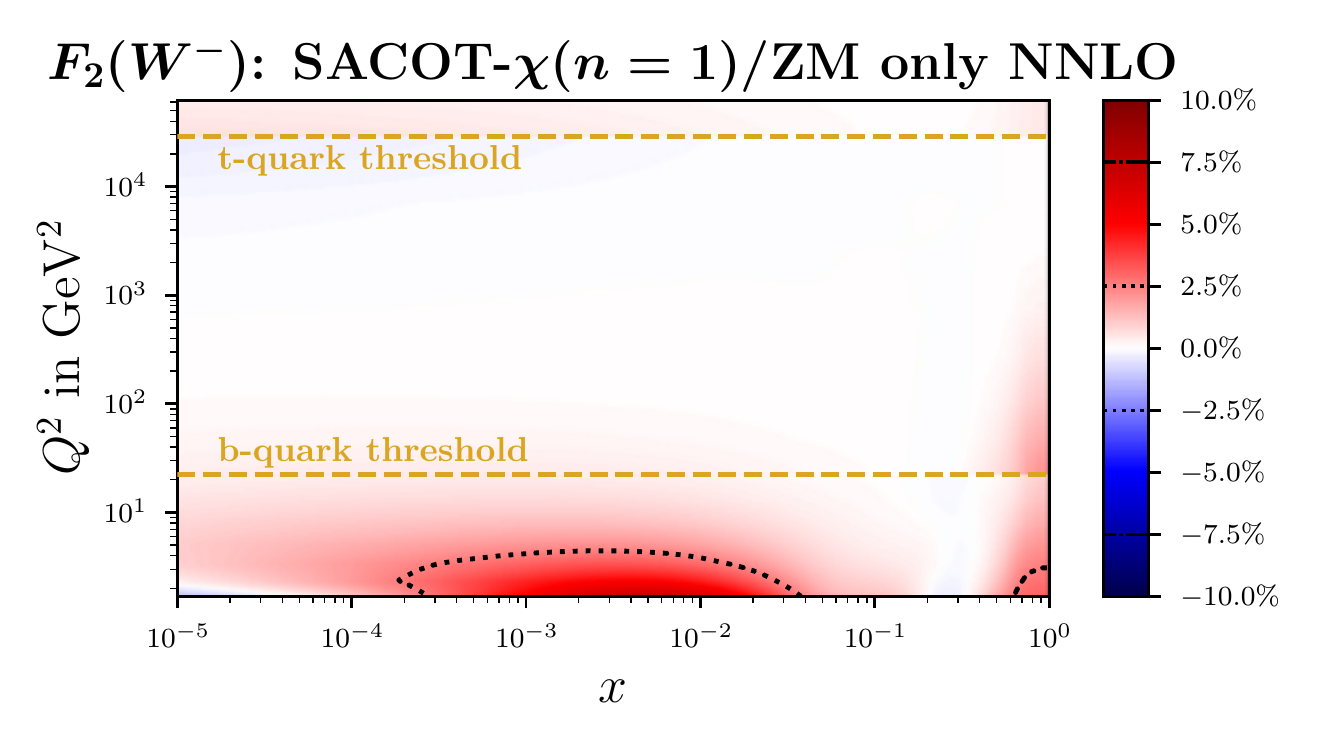}%
	\includegraphics[width=0.33\textwidth]{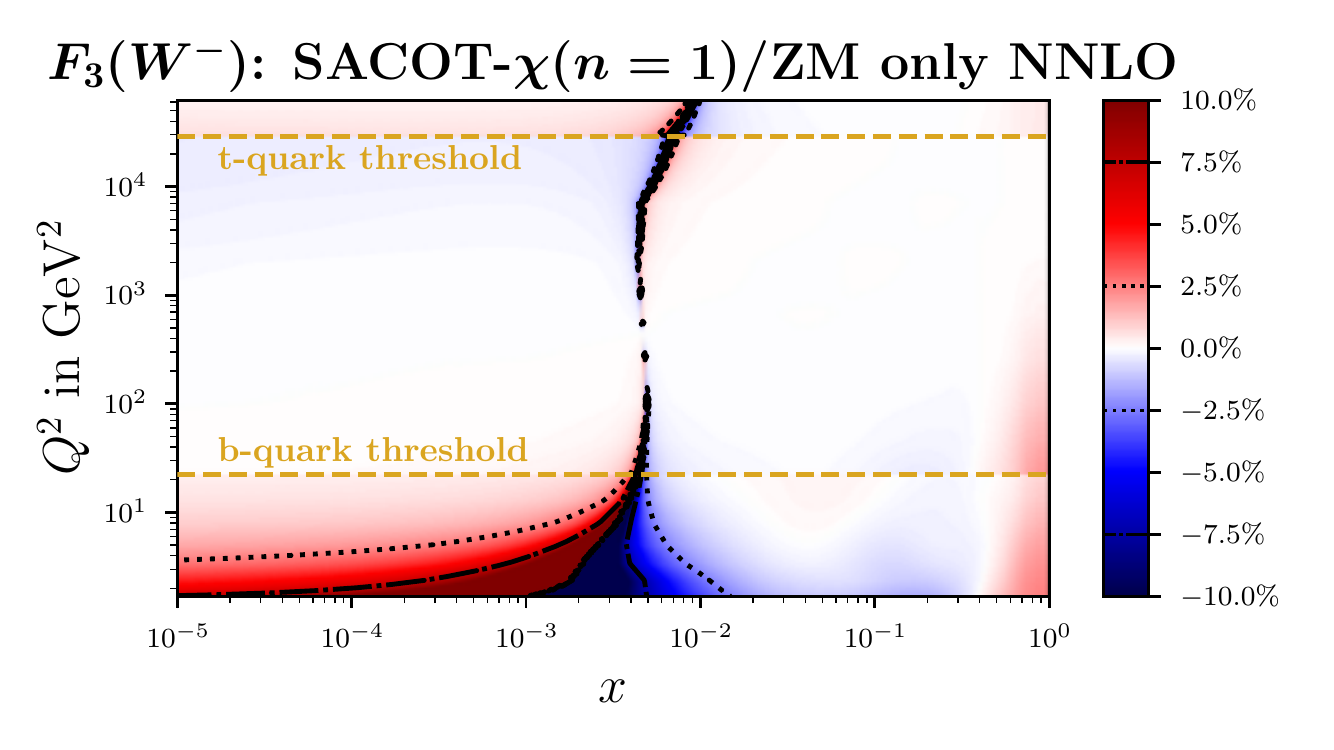}%
	\includegraphics[width=0.33\textwidth]{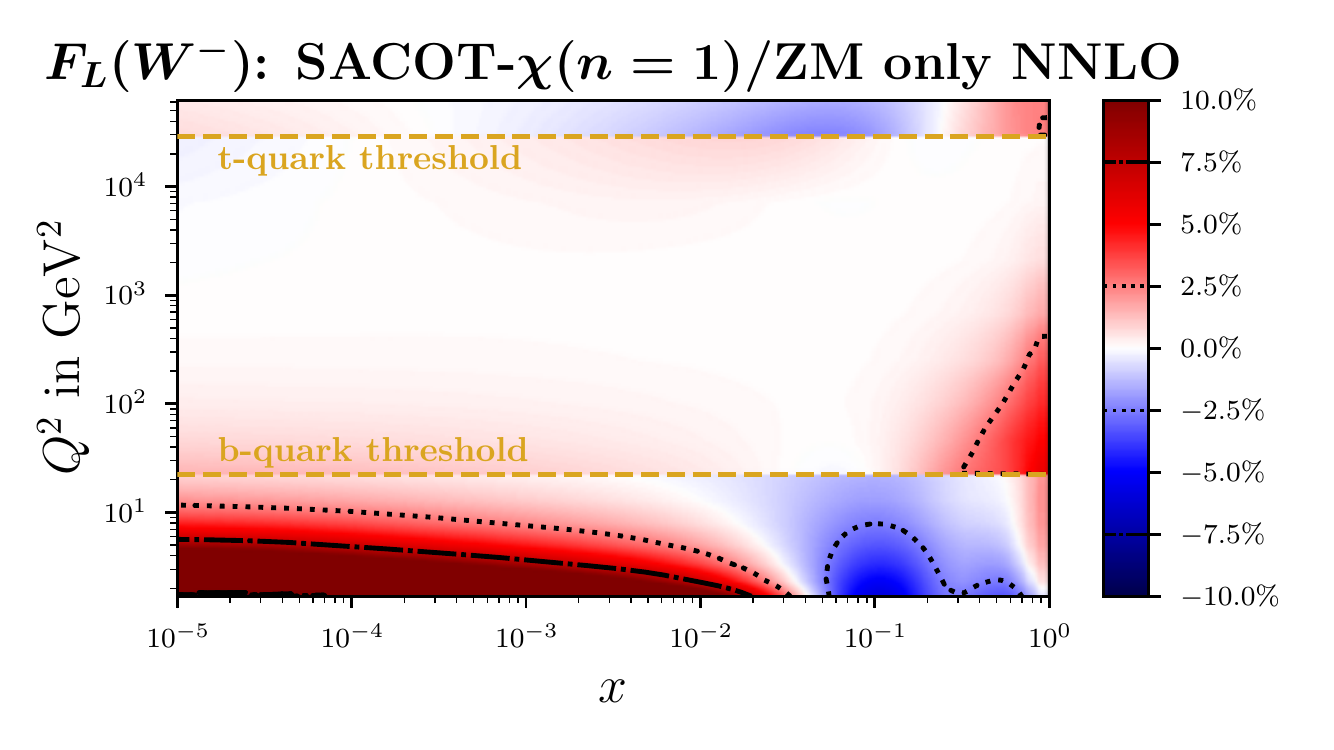}
	\includegraphics[width=0.33\textwidth]{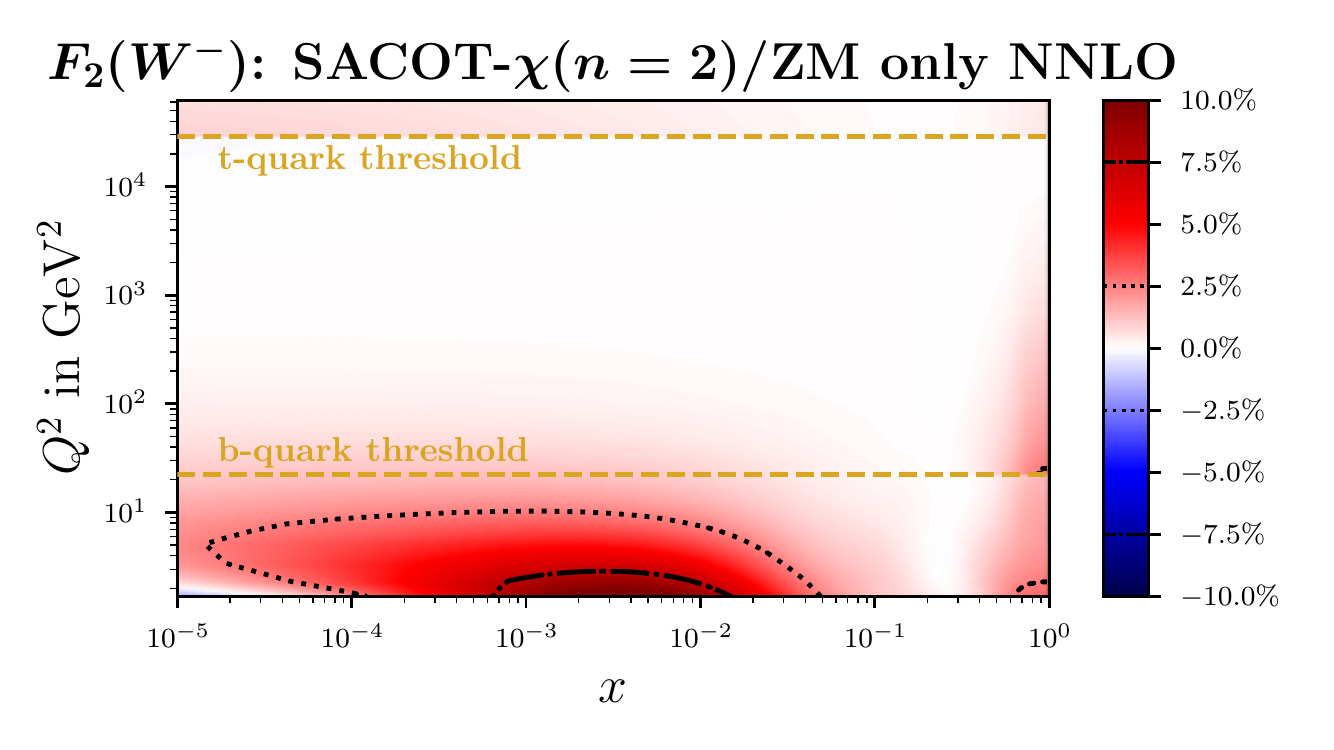}%
	\includegraphics[width=0.33\textwidth]{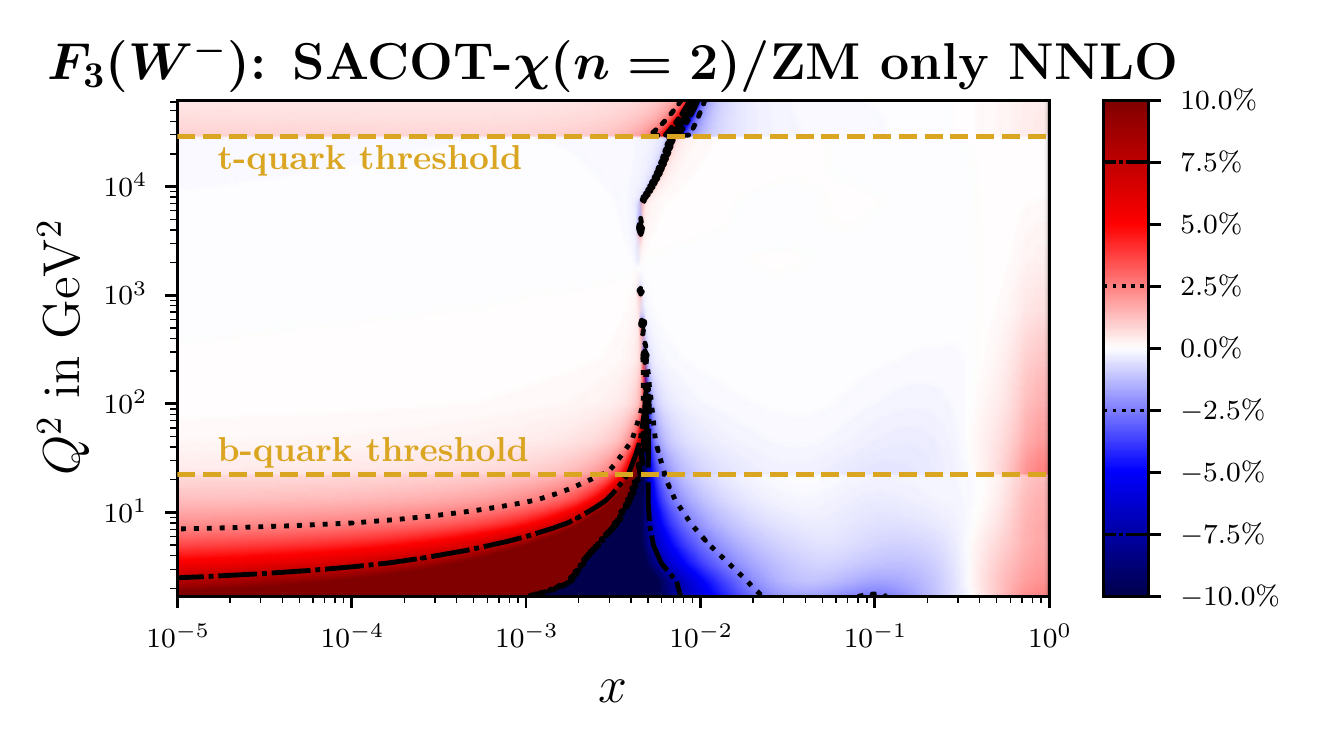}%
	\includegraphics[width=0.33\textwidth]{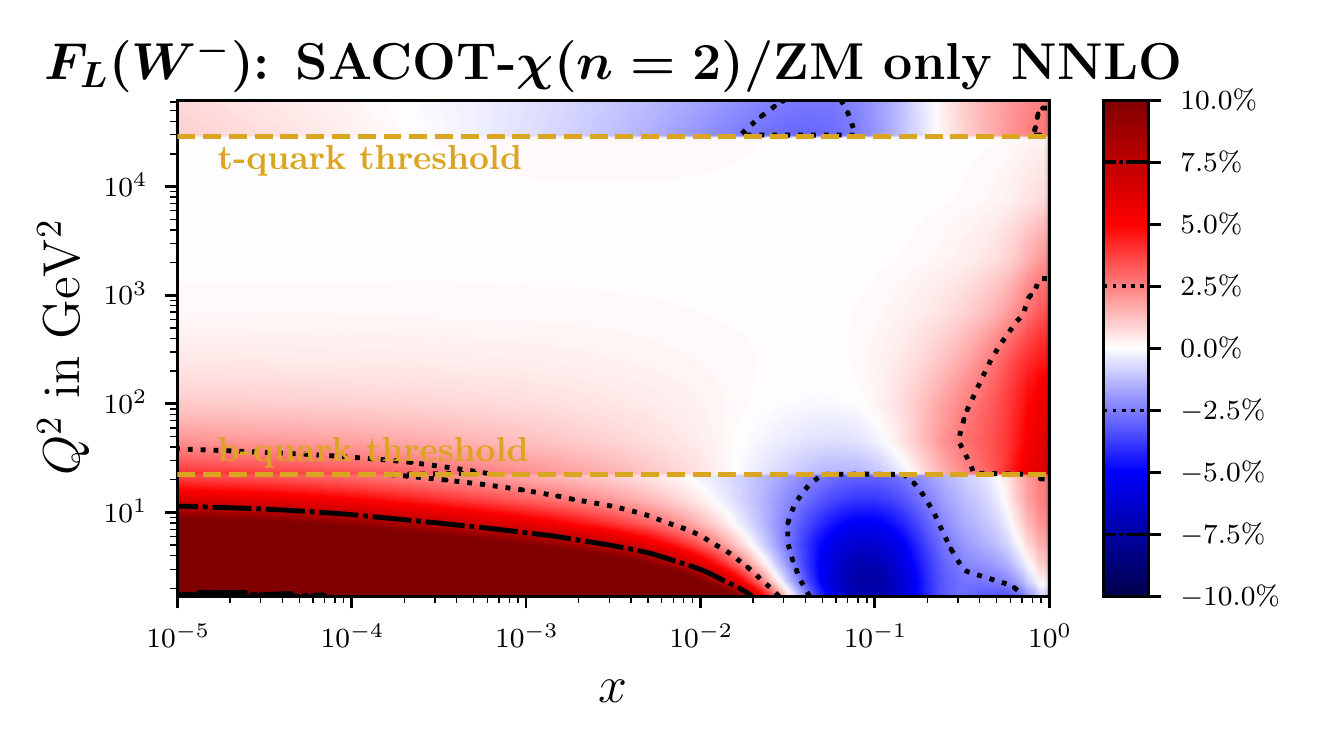}
	\includegraphics[width=0.33\textwidth]{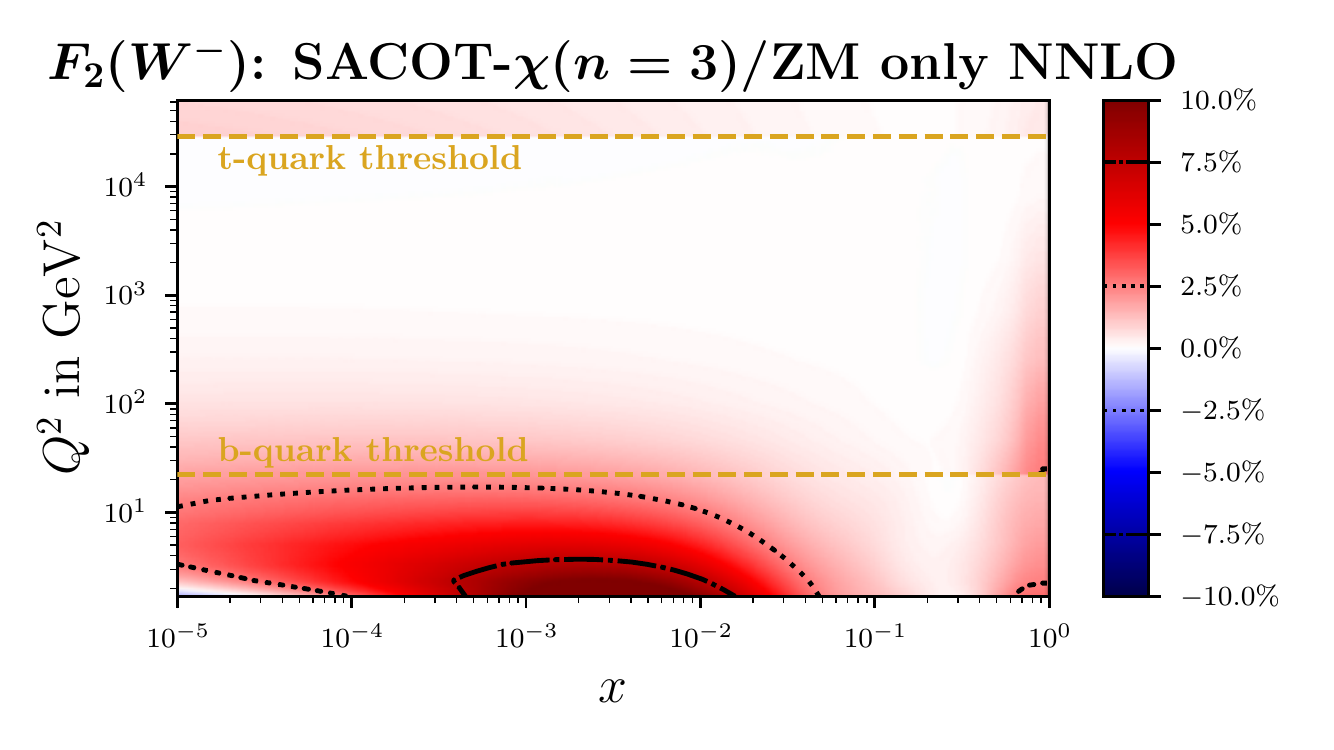}%
	\includegraphics[width=0.33\textwidth]{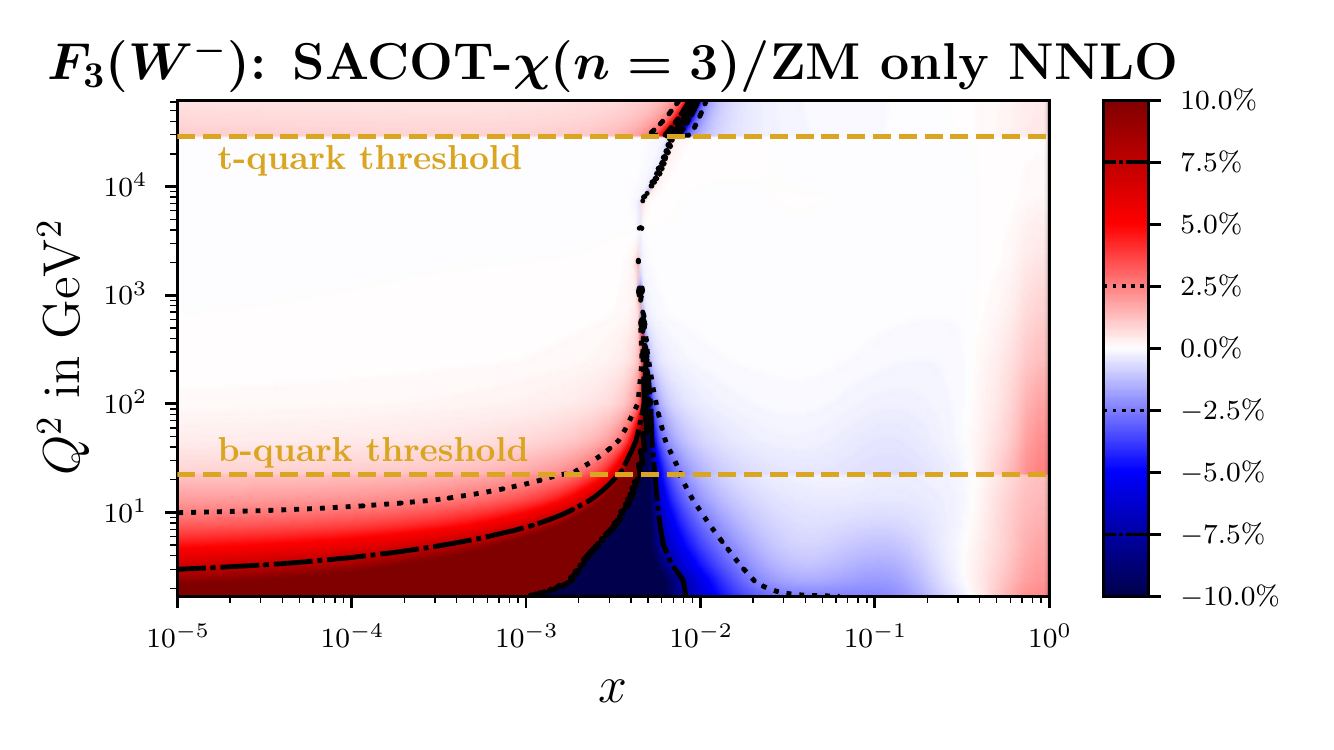}%
	\includegraphics[width=0.33\textwidth]{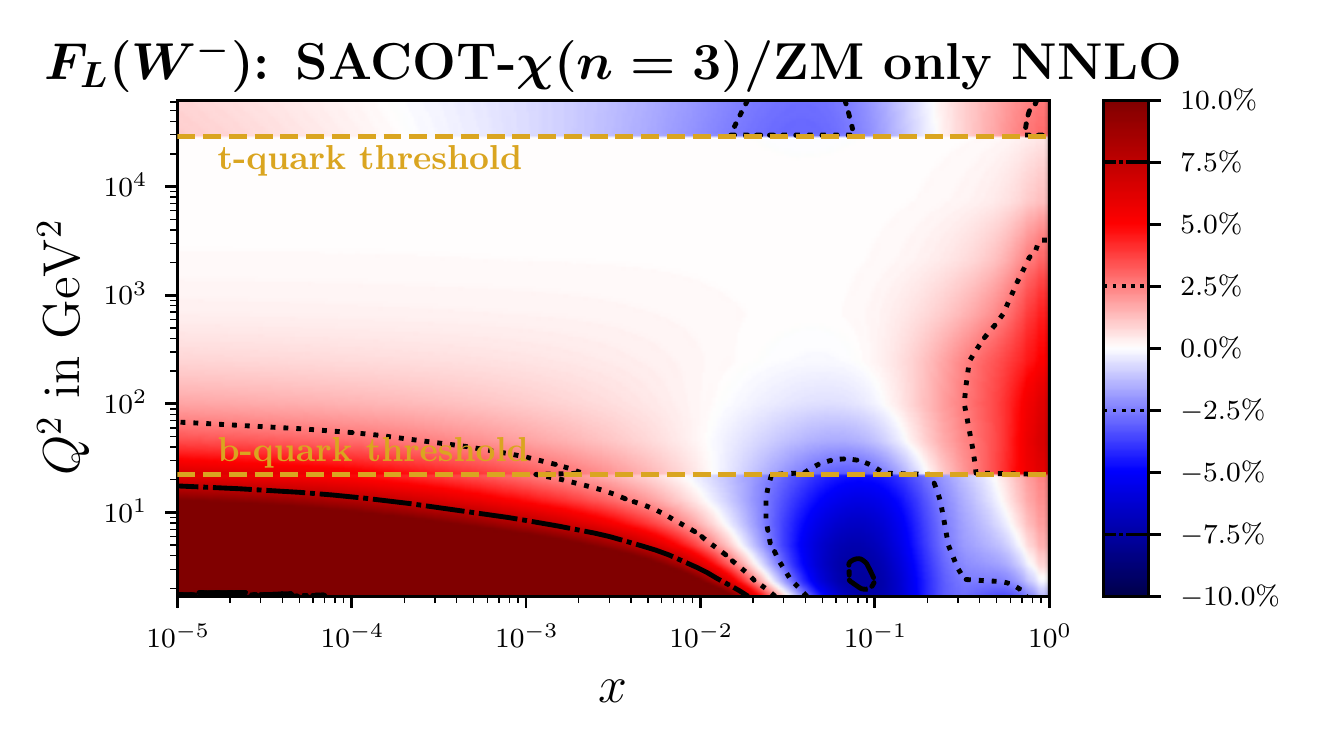}
	\caption{The same as \cref{fig:ratio_F23L_NC} but for charged current with a $W^-$ exchange. The ratio for $F_3(W^-)$ exhibits a distinct feature at $x\sim0.005$. This is a numerical artifact, as the structure function turns negative below this $x$-value and the ratio is not well-defined in this point (see \cref{subsec:comments_on_the_sign_of_F3}).}
	\label{fig:ratio_F23L_WM}
\end{figure*}

Structure functions depend solely on the (negative) virtuality of the vector boson $Q^2$ and the partonic longitudinal momentum fraction $x$. Therefore, we consider the ratio as a heat map on a double-logarithmic grid of $(400\times400)$-nodes corresponding to the ranges $Q^2\in [1.3^2,250^2]\,\text{GeV}^2$ and $x\in[10^{-5},1]$. Note that experimental determinations are additionally constrained by the center-of-mass collision energy $\sqrt{s}$ (or the energy of the incoming lepton $E_l$). This restricts the measurable region to a wedge in the bottom-right corner of the $(Q^2,x)$-plane, as the kinematics fulfill the relation
\begin{equation}
	Q^2 = xy(s-M^2)\quad\text{with}\quad 0\leq y\leq1\,,
\end{equation}
where $M$ is the mass of the struck hadron. Therefore, any $(Q^2,x)$-pair that lies above this linear relation cannot be measured. In the following, we display the ratio for the complete set of structure functions $\{F_2,F_3,F_L\}$ for $n=\{1,2,3\}$ in a $(3\times3)$-matrix. Neutral current interactions are considered in \cref{fig:ratio_F23L_NC} and charged current in \cref{fig:ratio_F23L_WP} for a $W^+$ and in \cref{fig:ratio_F23L_WM} for a $W^-$ exchange. The heat map is colored in red if the ratio is above unity, white if the ratio is equal to unity, and blue if the ratio is below unity. Dotted (dash-dotted) lines indicate contours of 2.5\% (7.5\%) difference. We use the \texttt{CT18}~\cite{Hou:2019efy} NNLO PDF set to evaluate the structure functions. We use the heavy quark masses as specified in the PDF determination.

Going from top to bottom in \cref{fig:ratio_F23L_NC} (neutral current), we can confirm the expectation that the strength of the mass effects increases with increasing $n$. More precisely, when the ratio is enhanced/suppressed in the first row, then the enhancement/suppression increases in the second and third row. Further, we find that these effects are strongest for low $Q^2$ values, as expected since the mass is divided by $Q$ in the scaling variable. Interestingly, we observe almost no difference between the massless NNLO corrections and \aSACOTchi{} for the structure function $\F{3}$. The strongest impact on the ratio is found for $\F{L}$, which vanishes in the \SACOTchi{} scheme at LO and is therefore effectively one order lower in the perturbative expansion.\footnote{Note that the LO contribution to $\F{L}$ does not vanish in the \ACOT{}-scheme or if target mass corrections are included~\cite{Ruiz:2023ozv}.} Thus the effects are enhanced relative to $\F{2}$ and $\F{3}$.

Moving to the charged current structure functions in \cref{fig:ratio_F23L_WP,fig:ratio_F23L_WM}, we note that mass effects are of similar size compared to neutral current structure functions, except for $\F{3}$. For $\F{3}(W^+)$, the extent in the $(Q^2,x)$-plane and the relative size of mass effects are similar compared to $\F{2}$. In the case of a $W^-$ exchange, we observe a numerical artifact at $x\sim 0.005$. This is a result of the ratio being ill-defined in these kinematics since $\F{3}(W^-)$ crosses zero at this value of $x$ (see \cref{subsec:comments_on_the_sign_of_F3}). 

\subsection{Physical cross sections}

Having implemented all charged current structure functions, the effect of the $n$-scaling can be investigated for the full cross section. In the following, we compute theoretical predictions for measurements taken at HERA and the upcoming EIC. 

\begin{figure*}
	\centering
	\includegraphics[width=0.5\textwidth]{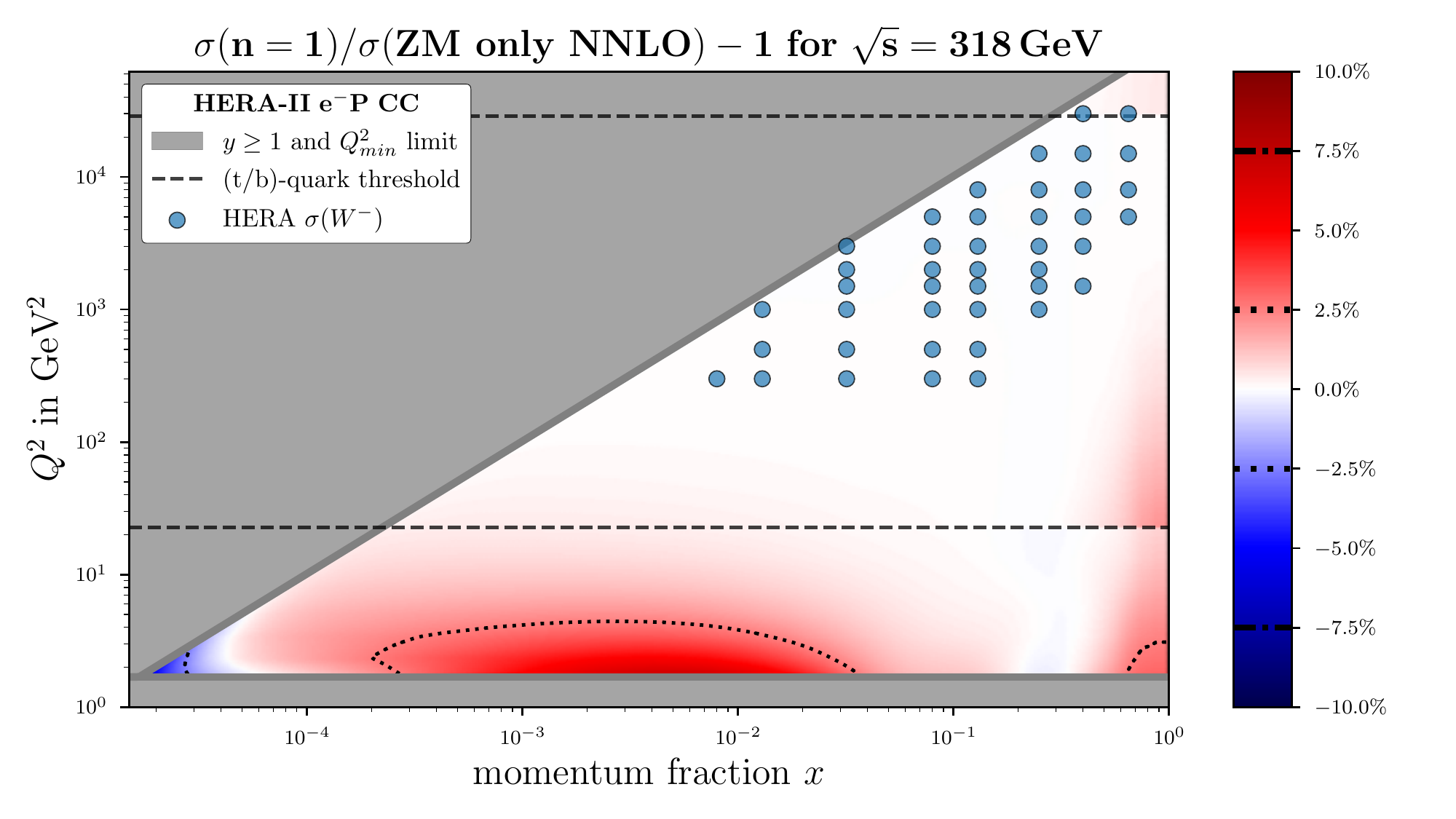}%
	\includegraphics[width=0.5\textwidth]{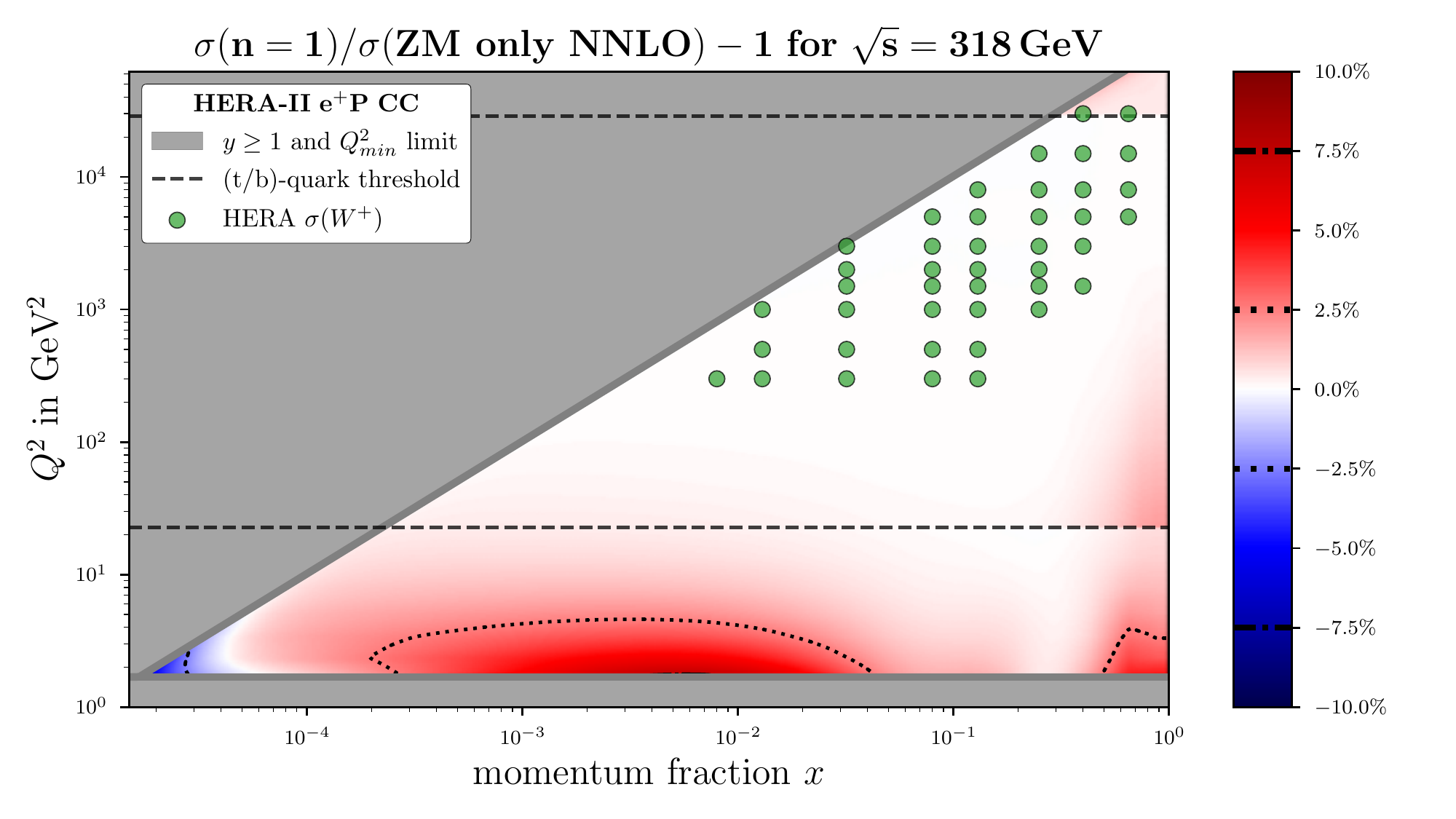}
	\includegraphics[width=0.5\textwidth]{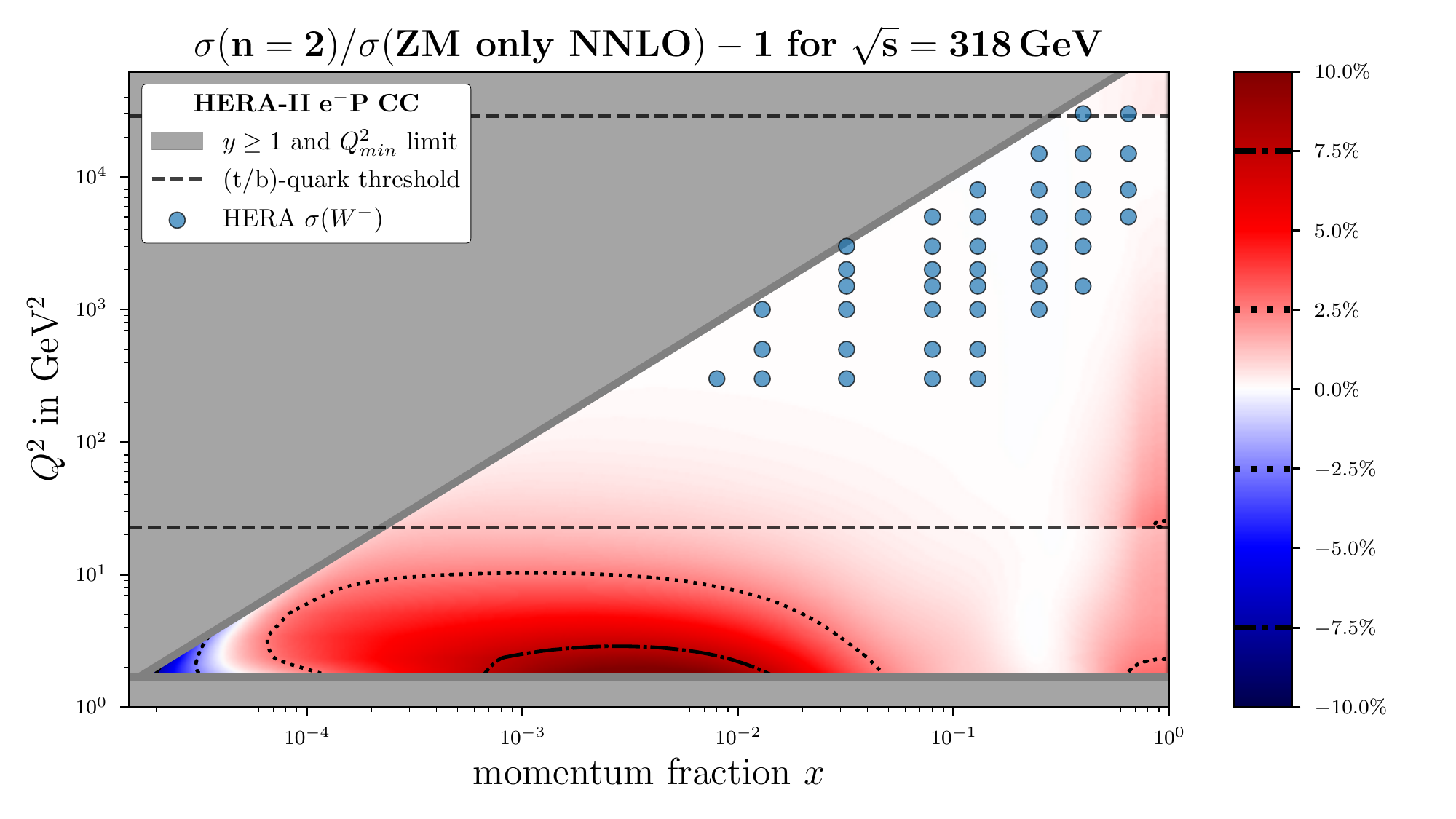}%
	\includegraphics[width=0.5\textwidth]{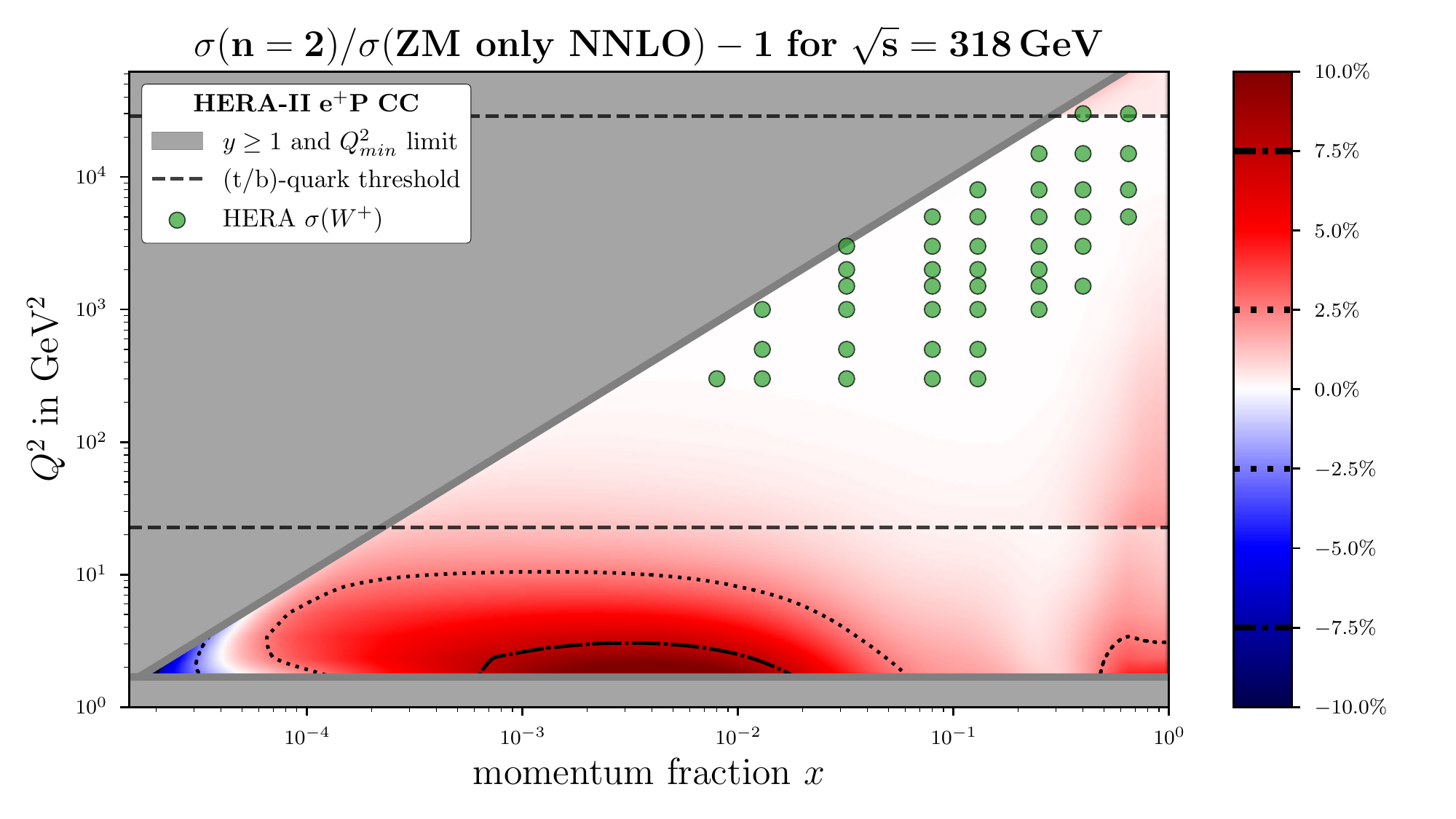}
	\includegraphics[width=0.5\textwidth]{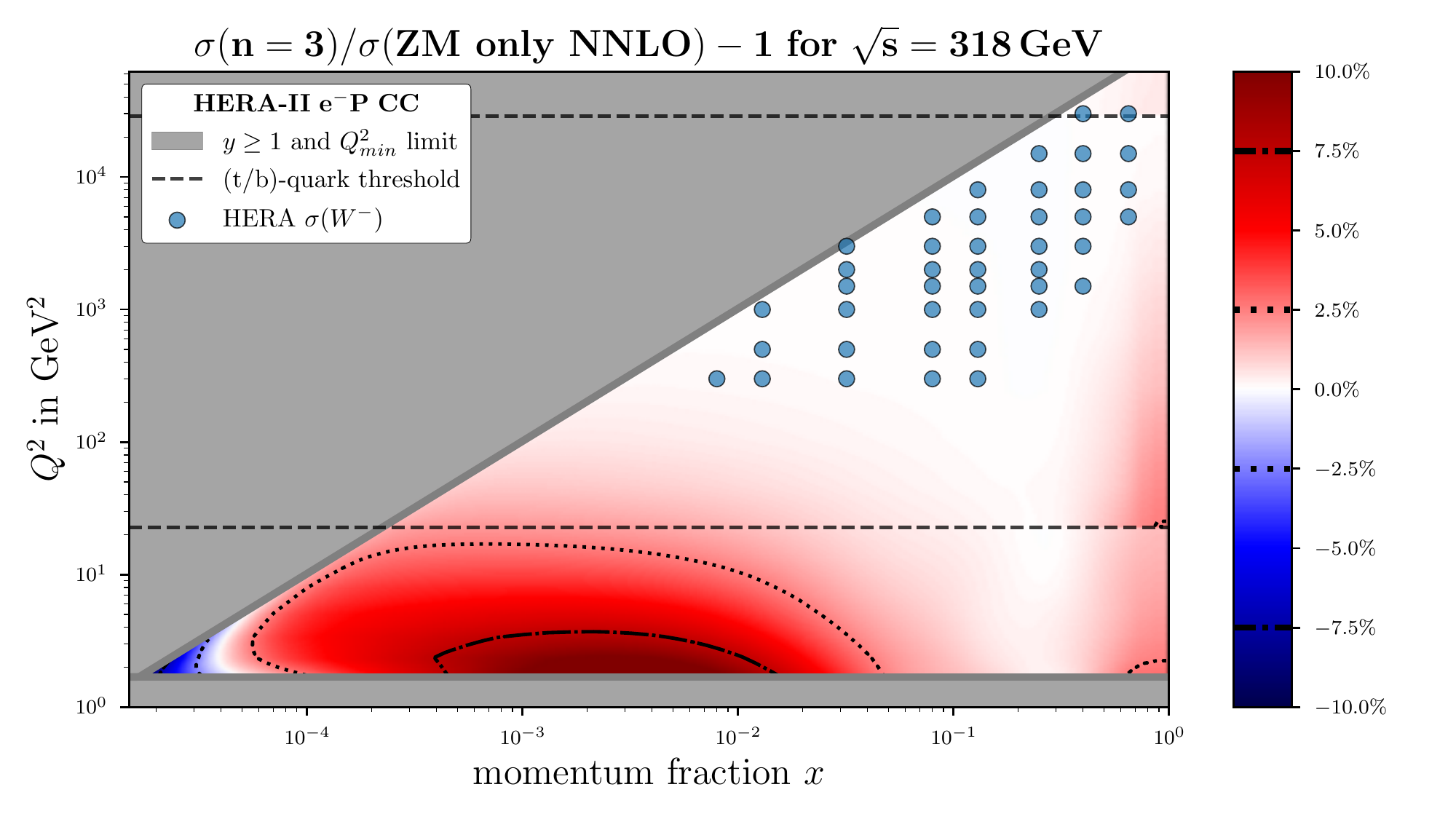}%
	\includegraphics[width=0.5\textwidth]{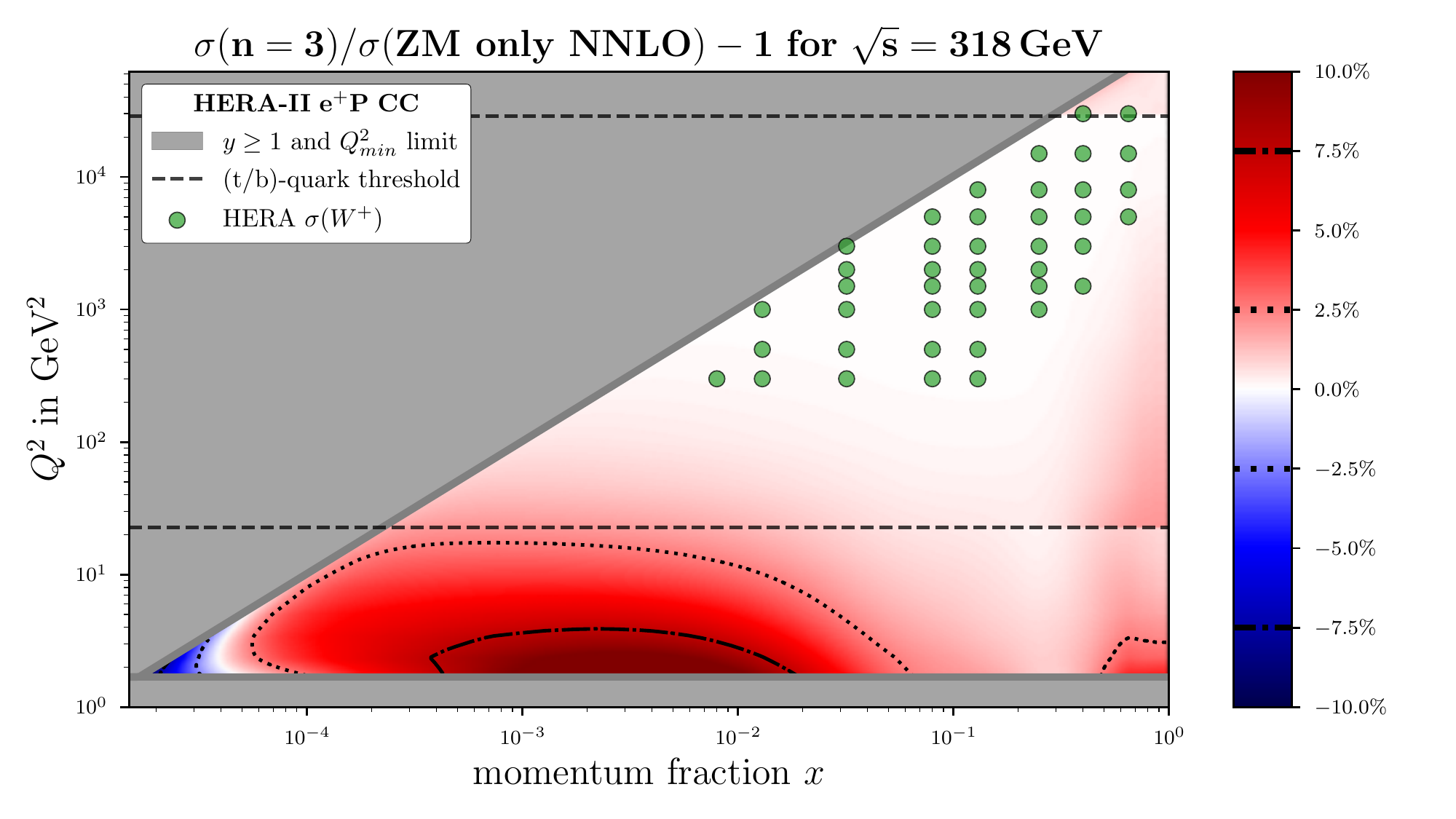}
	
	\caption{The reduced CC DIS cross-section measurements from HERA for an incoming $e^-$ (left\review{, blue circles}) and $e^+$ (right\review{, green circles}). The CMS energy is set to $s=318^2$ GeV${}^2$. From top to bottom we set the scaling variable $n=\{1,2,3\}$.}
	\label{fig:HQ_heat_maps_HERA_CC}
\end{figure*}

We first study the mass effects only in the NNLO correction. 
We define a ratio similar to the structure-function ratio of the previous subsection 
\begin{equation}
	\frac{\fstrut\sigma^{\textbf{LO + NLO}}(SACOT_\chi) + \sigma^{\textbf{NNLO}}(aSACOT_\chi(n))}{\fstrut \sigma^{\textbf{LO + NLO}}(SACOT_\chi) + \sigma^{\textbf{NNLO}}(ZM)\hfill} \,.
	\label{eq:ratio_sigma_sacot-chi_n_ZM_only_NNLO}
\end{equation}
In \cref{fig:HQ_heat_maps_HERA_CC} we show the ratio in the $(Q^2,x)$-plane as a heat map.
In order to make a connection with a specific experiment, we consider the measurements at HERA~\cite{H1:2015ubc}, with $\sqrt{s}=318$ GeV.

Unphysical regions are indicated by gray patches. The predictions were again obtained from the \texttt{CT18} NNLO PDF set, which has an initial scale of $Q_{min}=1.3$ GeV. The data set consists of measurements of an incoming $e^-$, shown in the left column \review{with the location of the measurements as blue circles}, and an incoming $e^+$, shown in the right column \review{as green circles}. We find that the data points are taken at such a high $Q^2$ that the mass effect is negligible. In fact, performing a $\chi^2$-evaluation of the data sets with the ZM scheme at all orders compared to using \aSACOTchi{} (i.e.~with mass effects at every order) results in a difference of 0.2 $\chi^2$-points for the 82 charged-current data points. This indicates that mass effects are not resolved by the measurements.

\begin{figure*}
	\centering
	\includegraphics[width=0.5\textwidth]{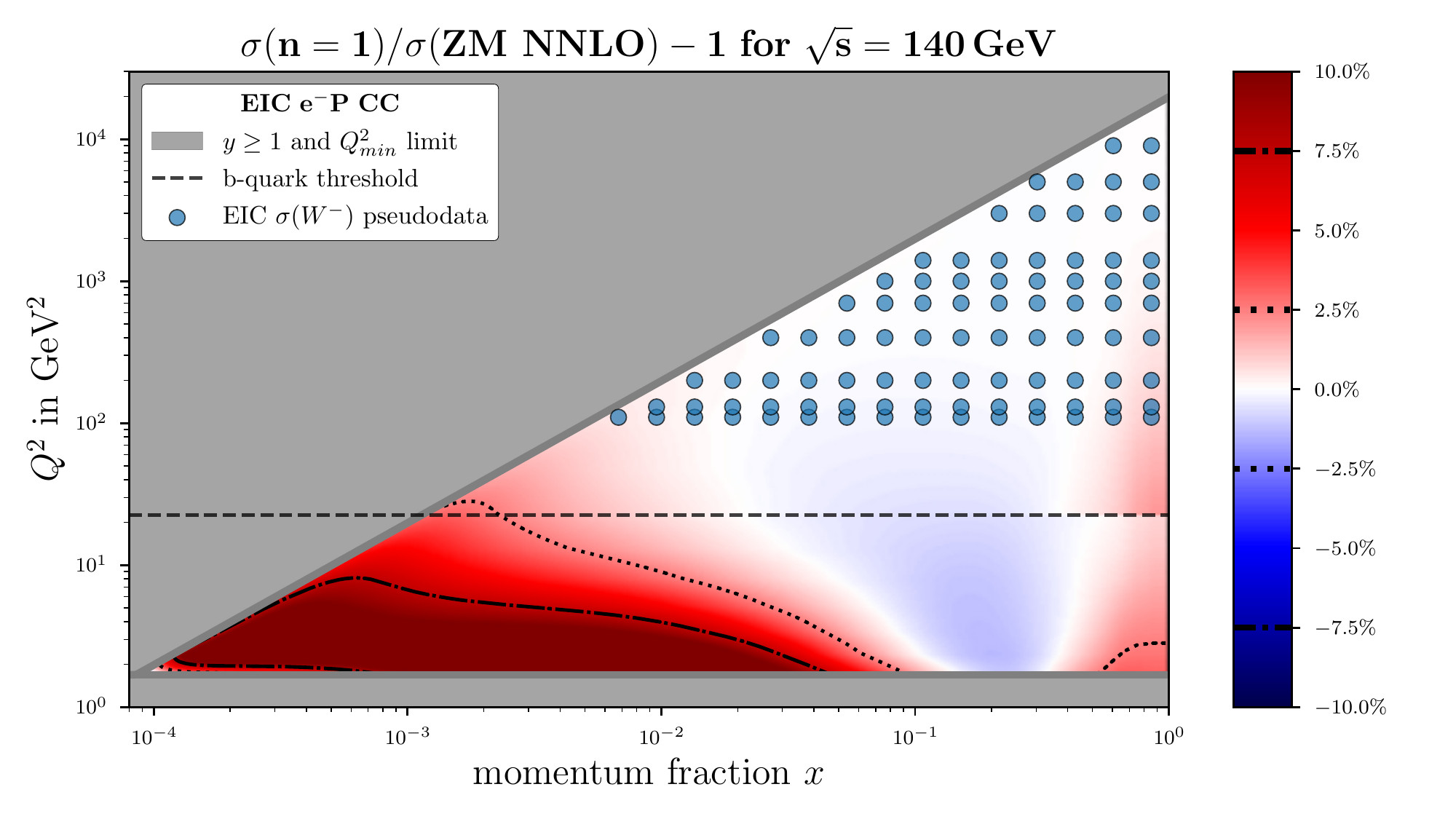}%
	\includegraphics[width=0.5\textwidth]{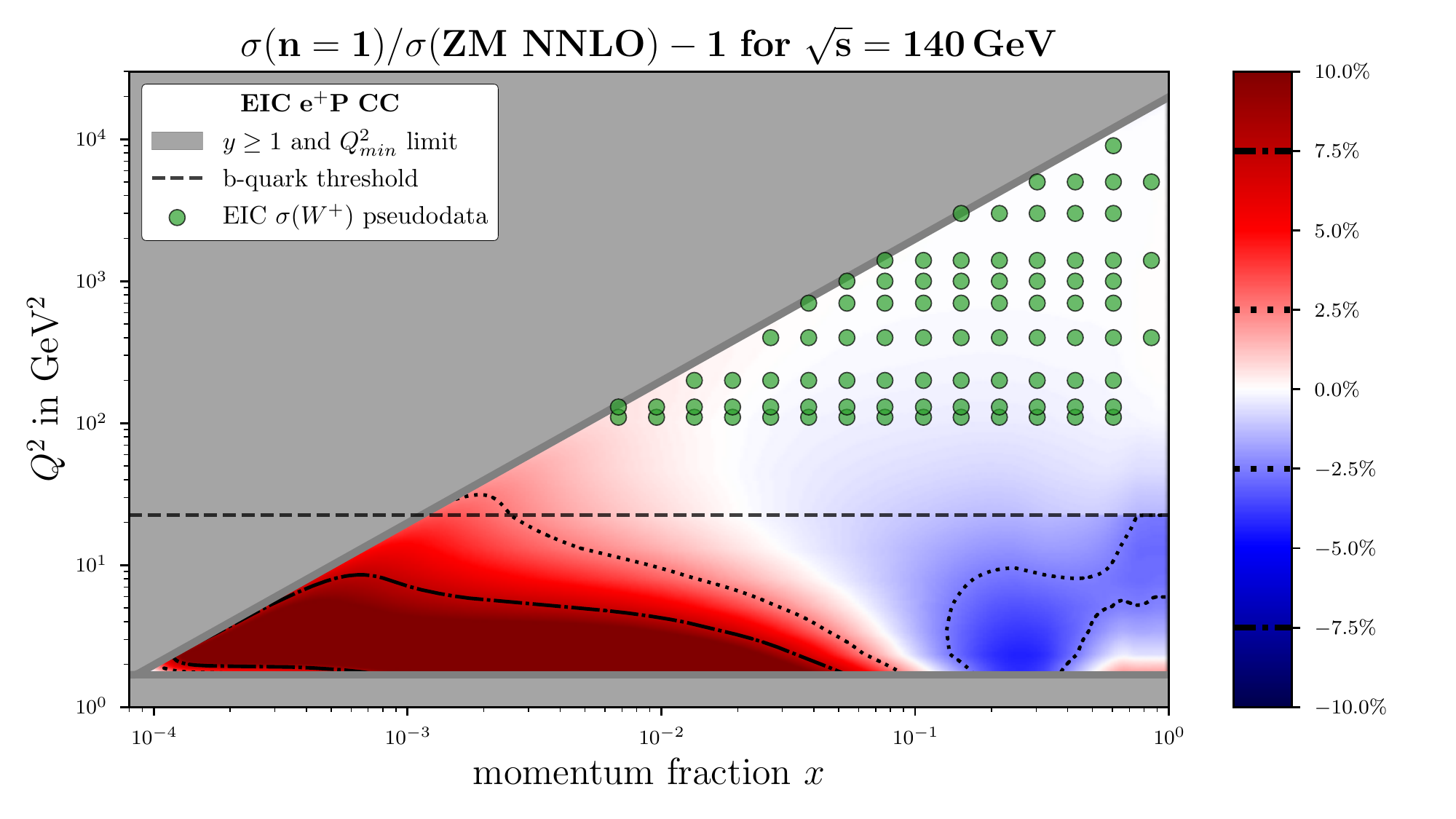}
	\includegraphics[width=0.5\textwidth]{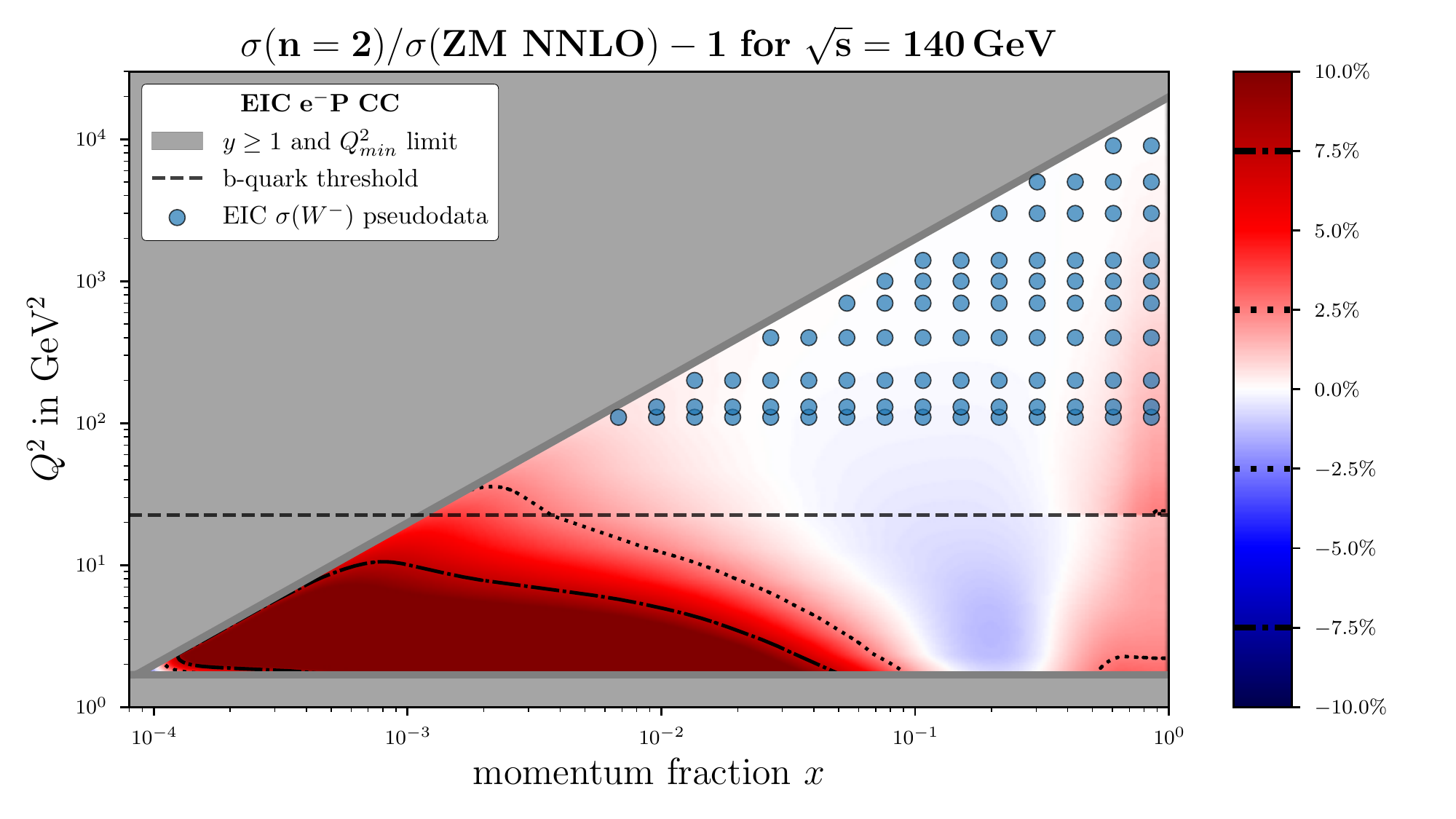}%
	\includegraphics[width=0.5\textwidth]{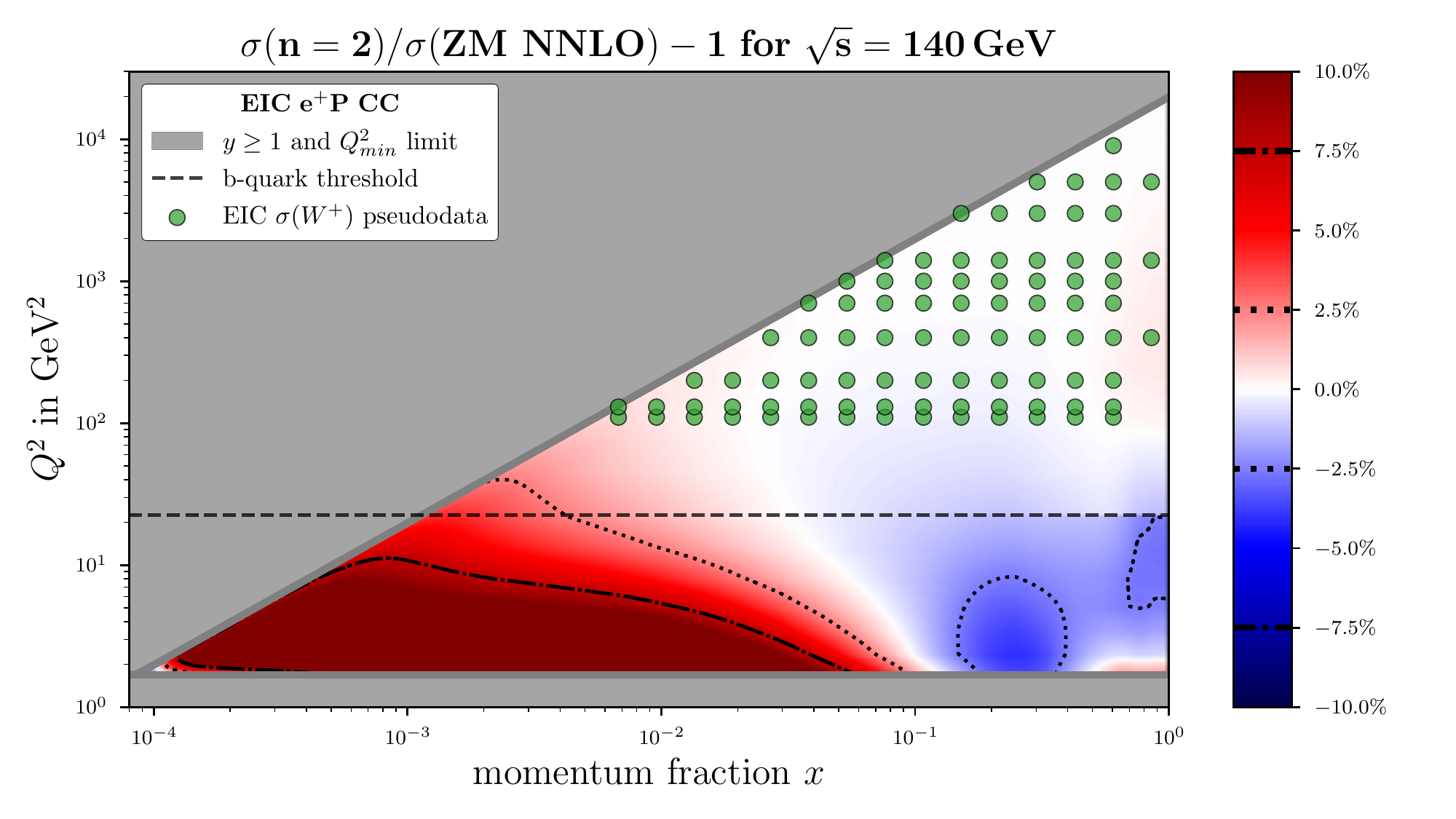}
	\includegraphics[width=0.5\textwidth]{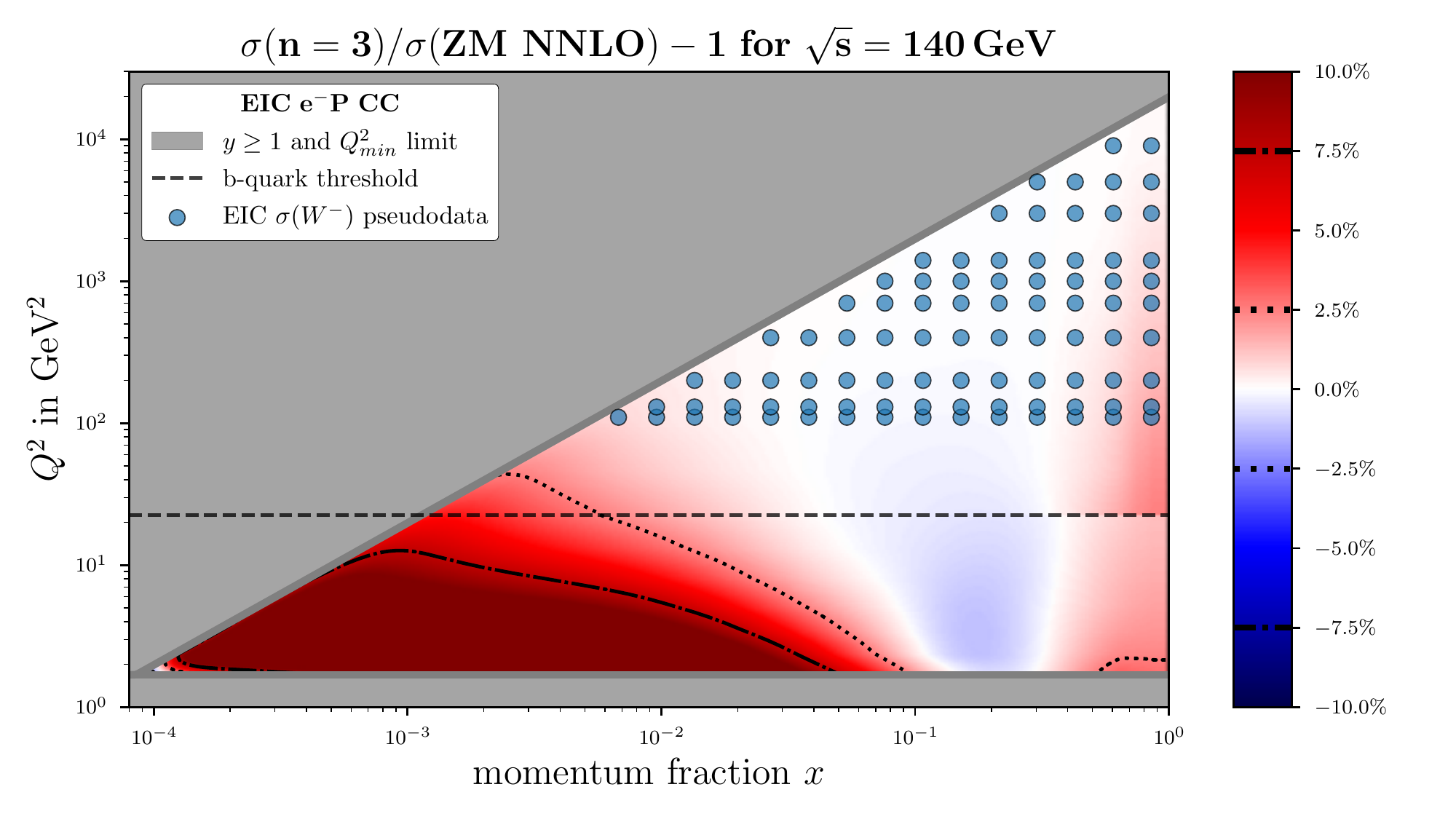}%
	\includegraphics[width=0.5\textwidth]{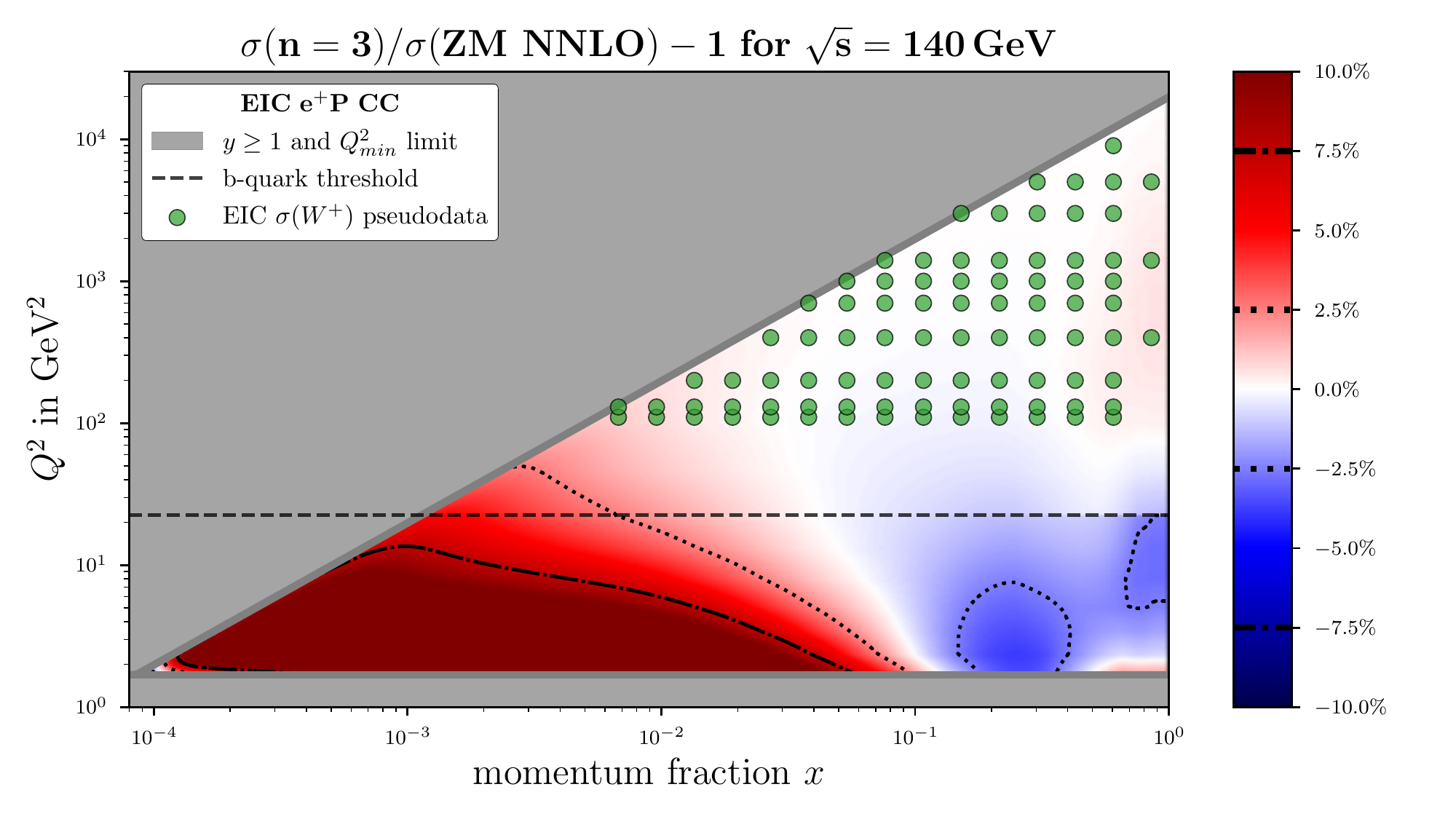}
	
	\caption{We display pseudo-data for charged current interactions at the EIC~\cite{AbdulKhalek:2021gbh,Khalek:2021ulf}. The left column assumes an incoming $e^-$ \review{(location of pseudo-data as blue circles)} and the right column an $e^+$ \review{(green circles)} with the CMS energy set to $s=140^2$ GeV${}^2$. The heat maps show \cref{eq:ratio_sigma_sacot-chi_n_ZM_NNLO} of \aSACOTchi{} against the ZM scheme (at all orders up to NNLO). From top to bottom we set the scaling variable to $n=\{1,2,3\}$.}
	\label{fig:heat_maps_EIC_CC}
\end{figure*}

Now we investigate the impact of including mass effects in the whole tower of contributions. 
For that we use the example of EIC instead.
We modify the heat map to display the ratio
\begin{equation}
	\frac{\fstrut\sigma^{\textbf{LO + NLO}}(SACOT_\chi) + \sigma^{\textbf{NNLO}}(aSACOT_\chi(n))}{\fstrut\sigma^{\textbf{LO + NLO + NNLO}}(ZM)}\,,
	\label{eq:ratio_sigma_sacot-chi_n_ZM_NNLO}
\end{equation}
i.e.~we compare the \aSACOTchi{} against the massless predictions at all orders. 
For the EIC predictions, we use the pseudo-data sets presented in Refs.~\cite{AbdulKhalek:2021gbh,Khalek:2021ulf}.
The results for EIC kinematics are shown in \cref{fig:heat_maps_EIC_CC}. The layout is the same as in the former HERA plots. Comparing the two we find that mass effects are more pronounced but, in spite of the lower $Q^2$ compared to the HERA data, the impact is less than $2.5\%$. In the case of an incoming $e^-$, data might be sensitive to mass effects in the lowest $Q^2$ bins and high-$x$ region, but only for an experimental accuracy of $\sim1\%$. 

\subsection{Neutrino DIS cross sections}

\begin{table*}[tb]
	\centering
	\caption{$\chi^2$-goodness of fit criterion per number of data points for the NuTeV~\cite{NuTeV:2005wsg}, CCFR~\cite{CCFRNuTeV:2000qwc,Yang:2001rm} and Chorus~\cite{CHORUS:2008vjb} neutrino and anti-neutrino data sets for various mass schemes. The upper entry in each row gives the $\chi^2$ per number of points and the lower entry, given in slanted font, the difference to the \SACOTnlo{} scheme per number of data points. The values are obtained with the Iron (NuTeV and CCFR) and Lead (Chorus) \texttt{nCTEQ15HQ}~\cite{Duwentaster:2022kpv} sets. We note that the absolute $\chi^2$ values depend on the PDF set used, but the qualitative trend especially the difference w.r.t.~the \SACOTnlo{} scheme does not.
		Also note that ``\ZMonlyNNLO{}'' is equivalent to ``\SACOT{0}''.
	}
	\renewcommand{\arraystretch}{1.2}
	\begin{tabular}{|>{\raggedright}m{0.25\textwidth}
			|>{\centering}m{0.075\textwidth}>{\centering}m{0.075\textwidth} 
			|>{\centering}m{0.075\textwidth}>{\centering}m{0.075\textwidth} 
			|>{\centering}m{0.075\textwidth}>{\centering}m{0.075\textwidth} 
			|}
		\hline
		\multicolumn{1}{|c|}{}& \multicolumn{2}{c|}{\textbf{NuTeV}} & \multicolumn{2}{c|}{\textbf{CCFR}} & \multicolumn{2}{c|}{\textbf{Chorus}} \tabularnewline
		& $\bm\nu$ & $\bm{\anti{\nu}}$& $\bm\nu$ & $\bm{\anti{\nu}}$& $\bm\nu$ & $\bm{\anti{\nu}}$ \tabularnewline
		Data Points       & 1371 & 1146 & 1282 & 1273 & 534 & 534 \tabularnewline
		\hline
		\multicolumn{1}{c}{\rule{0pt}{12pt}\textbf{NLO}} & \multicolumn{6}{c}{}\tabularnewline
		\hline
		
		\multirow{2}*{\ZMNLO{}}     &\,3.68 &\,1.56 &\,1.59 &\,1.04 &\,1.74 &\,1.33 \tabularnewline
		&\,\numhighl{0.74} &\,\numhighl{0.20} & \numhighl{-0.17} & \numhighl{-0.09} & \numhighl{-0.16} & \numhighl{-0.05} \tabularnewline
		\hline
		\multirow{2}*{\SACOTnlo}    &\,2.95 &\,1.36 &\,1.75 &\,1.13 &\,1.90 &\,1.38 \tabularnewline
		&  --   &  --   &  --   &  --   &  --   &  --   \tabularnewline
		\hline
		\multicolumn{1}{c}{\rule{0pt}{12pt}\textbf{NNLO}}& \multicolumn{6}{c}{}\tabularnewline
		\hline
		\multirow{2}*{\ZMNNLO{}}    &\,3.10 &\,1.51 &\,2.54 &\,1.24 &\,1.89 &\,1.35 \tabularnewline
		&\,\numhighl{0.15} &\,\numhighl{0.15} &\,\numhighl{0.79} &\,\numhighl{0.12} & \numhighl{-0.01} & \numhighl{-0.03} \tabularnewline
		\hline
		\multirow{2}*{\ZMonlyNNLO}  &\,2.81 &\,1.65 &\,2.77 &\,1.38 &\,2.10 &\,1.48 \tabularnewline
		& \numhighl{-0.13} &\,\numhighl{0.29} &\,\numhighl{1.02} &\,\numhighl{0.26} &\,\numhighl{0.20} &\,\numhighl{0.09} \tabularnewline
		\hline
		\multirow{2}*{\SACOT{1}}    &\,2.60 &\,1.50 &\,2.52 &\,1.32 &\,1.92 &\,1.40 \tabularnewline
		& \numhighl{-0.35} &\,\numhighl{0.14} &\,\numhighl{0.77} &\,\numhighl{0.19} &\,\numhighl{0.01} &\,\numhighl{0.02} \tabularnewline
		\hline
		\multirow{2}*{\SACOT{2}}    &\,2.50 &\,1.41 &\,2.37 &\,1.27 &\,1.84 &\,1.37 \tabularnewline
		& \numhighl{-0.44} &\,\numhighl{0.05} &\,\numhighl{0.62} &\,\numhighl{0.14} & \numhighl{-0.06} & \numhighl{-0.01} \tabularnewline
		\hline
		\multirow{2}*{\SACOT{3}}    &\,2.48 &\,1.38 &\,2.32 &\,1.25 &\,1.84 &\,1.37 \tabularnewline
		& \numhighl{-0.46} &\,\numhighl{0.02} &\,\numhighl{0.56} &\,\numhighl{0.12} & \numhighl{-0.06} & \numhighl{-0.01} \tabularnewline
		\hline
	\end{tabular}
	\renewcommand{\arraystretch}{1.}
	\label{tab:DISNEU_chi2_table}
\end{table*}
As discussed in the previous section, mass effects are dominant in the low-$Q^2$ regime. Neutrino DIS measurements have been performed in this region to a precision that is comparable to the relative size of mass effects. 
To show this, we consider measurements from the NuTeV~\cite{NuTeV:2005wsg} and CCFR~\cite{CCFRNuTeV:2000qwc,Yang:2001rm} Collaborations taken on an iron target. Additionally, we include measurements from the Chorus Collaboration~\cite{CHORUS:2008vjb} taken on a lead target.
In \cref{tab:DISNEU_chi2_table}, we evaluate the $\chi^2$-function to quantify the quality of the description of different data sets by theory calculations in different schemes. We additionally split the data sets into incoming $\nu$ (left) and $\anti{\nu}$ (right).
For the $\chi^2$ evaluation we use the \texttt{nCTEQ15HQ} NLO PDFs~\cite{Duwentaster:2022kpv} for the appropriate nuclei and cut data points measured below the initial scale of the PDF set, i.e.~$Q_\textrm{min} = 1.3$~GeV. As before, we use the heavy quark masses as specified in the PDF determination. The available correlated uncertainties for NuTeV and Chorus have been consistently accounted for.
We do not apply target-mass corrections~\cite{Ruiz:2023ozv} or higher-twist effects of any kind.

The table lists the number of data points after cuts and the $\chi^2$-value per number of data points (upper value) alongside with the difference w.r.t.~the \SACOTnlo{} scheme (lower value) for the schemes:
\begin{enumerate}  
	\setlength{\itemindent}{0.5cm}
	\setlength\itemsep{0em}
	\item ZM at LO and NLO
	\item \SACOTnlo{}
	\item ZM at LO, NLO and NNLO
	\item 
	\mbox{\SACOTnlo{} + ZM at NNLO} 
	\mbox{\qquad (called ``ZM only NNLO'')} 
	\mbox{\qquad  equivalent to  \aSACOTchi{} with $n=0$}  
	\item[5.-7.] \aSACOTchi{} with $n\in\{1,2,3\}$
\end{enumerate}
We choose to highlight the difference with respect to the \SACOTnlo{} scheme, since this is the current state-of-the-art choice to predict neutrino DIS cross-section measurements in nuclear PDF extractions, see e.g.~Ref.~\cite{Muzakka:2022wey,Klasen:2023uqj}.

We also note that PDF extractions depend on the heavy-quark mass scheme used in the fit because the numerical optimization cannot distinguish between PDF information and heavy-quark mass effects. Therefore, the minimizer tends to ``overfit'' the PDFs to the employed mass scheme (see Ref.~\cite{Thorne:2008xf}) and in order to make conclusive statements about which mass scheme describes the data best a new global fit with each of the mass schemes would be necessary. This is beyond the scope of this paper.  The \SACOTnlo{} scheme was used in the \texttt{nCTEQ15HQ} extraction for the DIS predictions. 

We repeated the same procedure with different PDF sets and found that, although the absolute $\chi^2$ values differ from one set to another, the qualitative trend, especially the difference w.r.t.~the \mbox{\SACOTnlo}, remains the same.

For the NuTeV and CCFR measurements, we observe a significant impact on the $\chi^2$-value when moving from a NLO scheme to a NNLO scheme. Indeed, $\chi^2$-values per number of data points may differ by an amount that ranges between $12\%$ and $102\%$ . Furthermore, we notice that the difference between using massless and massive schemes for the NNLO contribution (\SACOTnlo{} + ZM NNLO vs. \SACOT{1}) is important as well, since $\chi^2$-value shifts by up to $\sim 341$ units are observed. The impact on the Chorus data set is less pronounced, which can be explained by the fact that the data lies at higher $Q^2$ values and is less precise. We conclude that the experimental neutrino data is sensitive to the quark-mass effects. In order to fully understand the effect of heavy quarks at NNLO, in the following we investigate the NuTeV data in more detail.

\begin{figure*}
	\centering
	\includegraphics[width=0.5\textwidth]{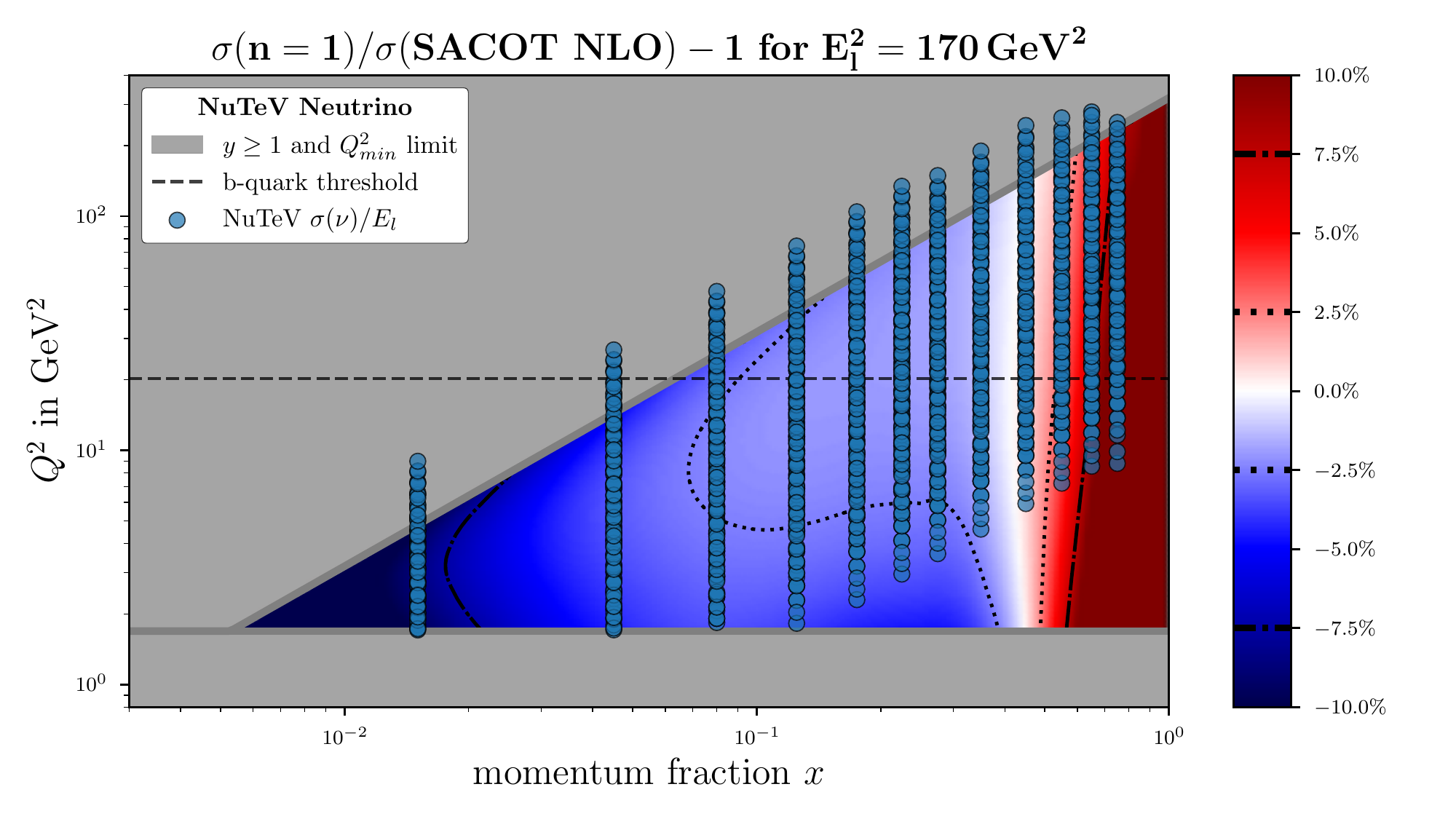}%
	\includegraphics[width=0.5\textwidth]{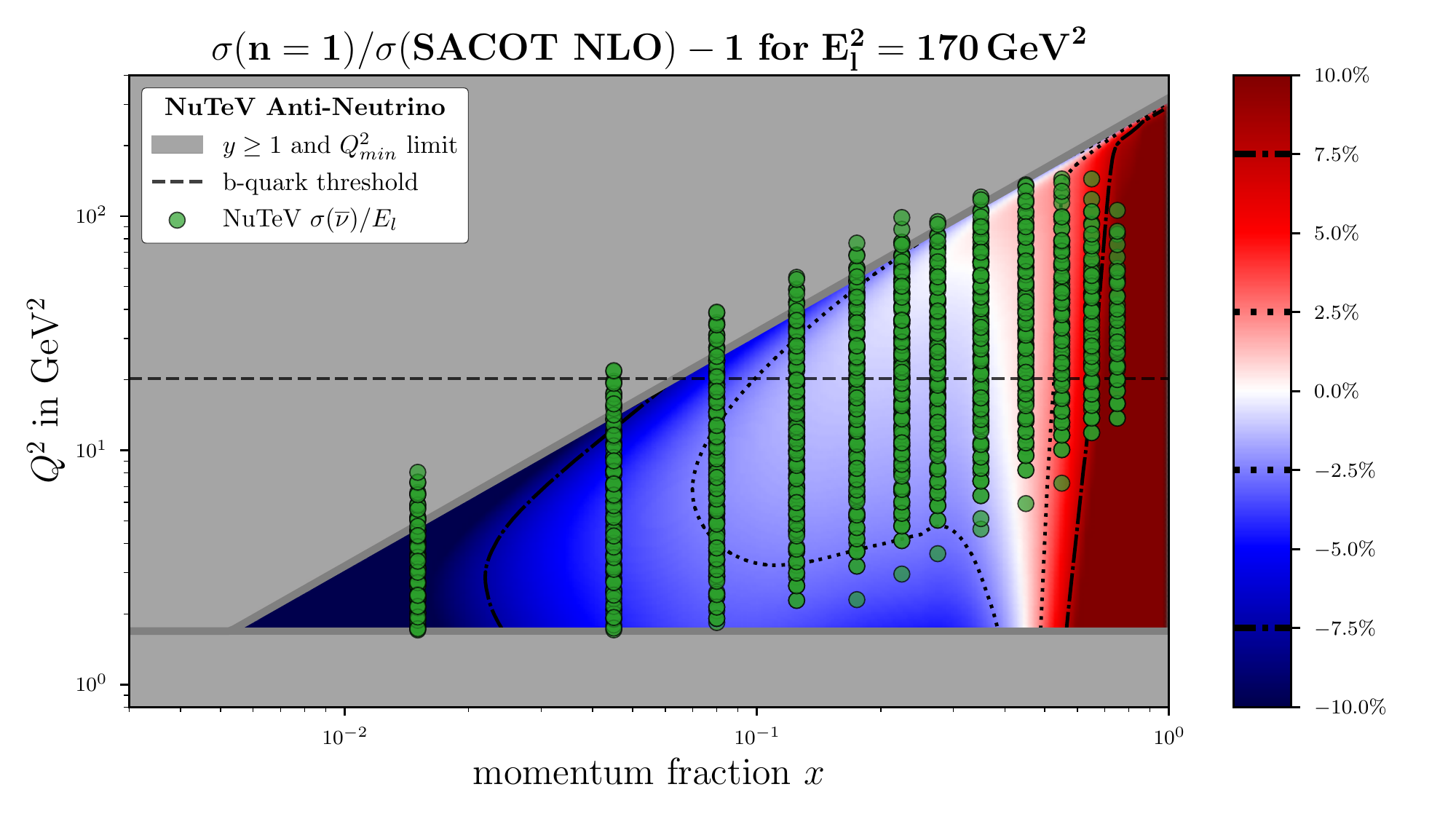}
	\includegraphics[width=0.5\textwidth]{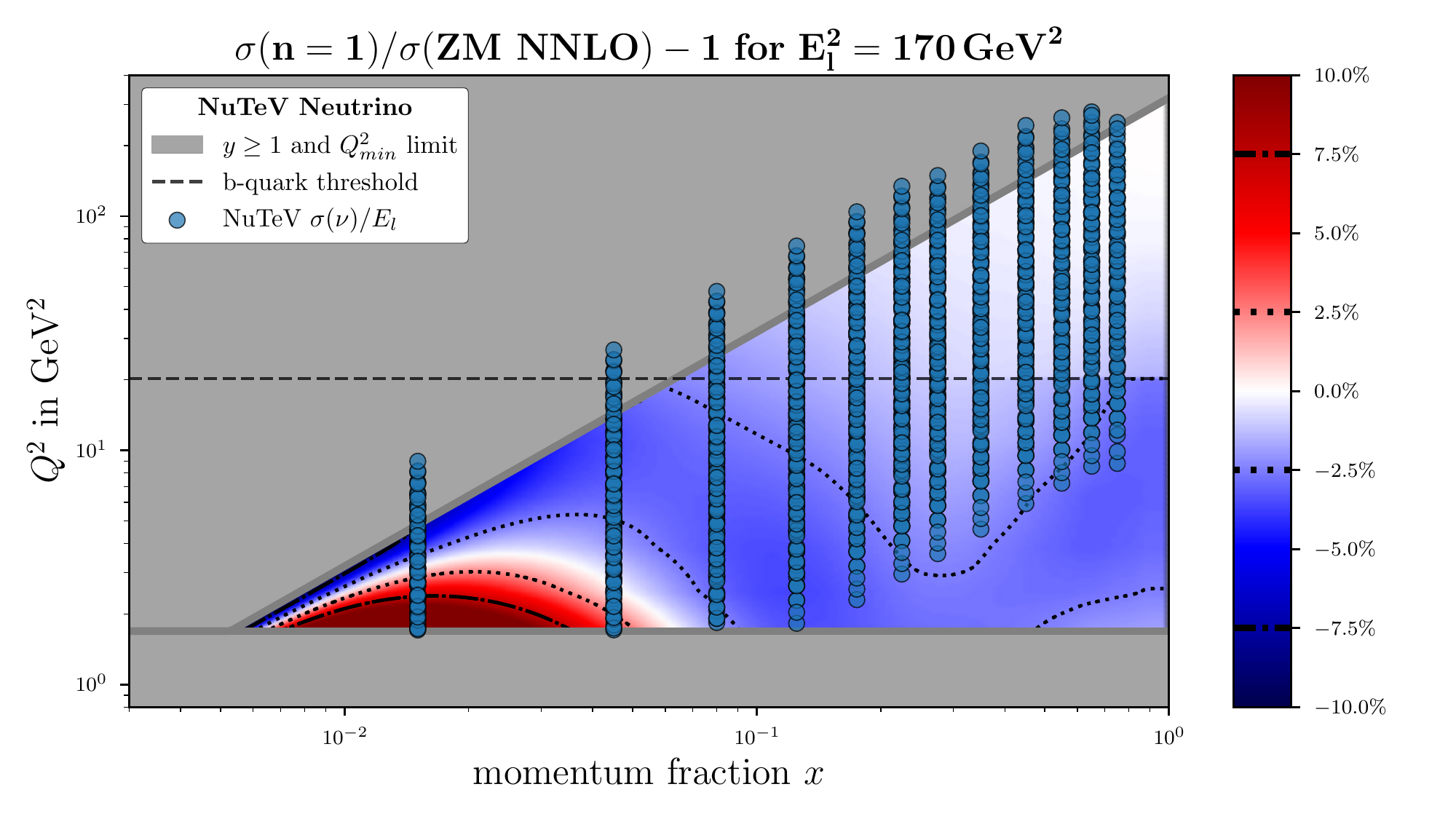}%
	\includegraphics[width=0.5\textwidth]{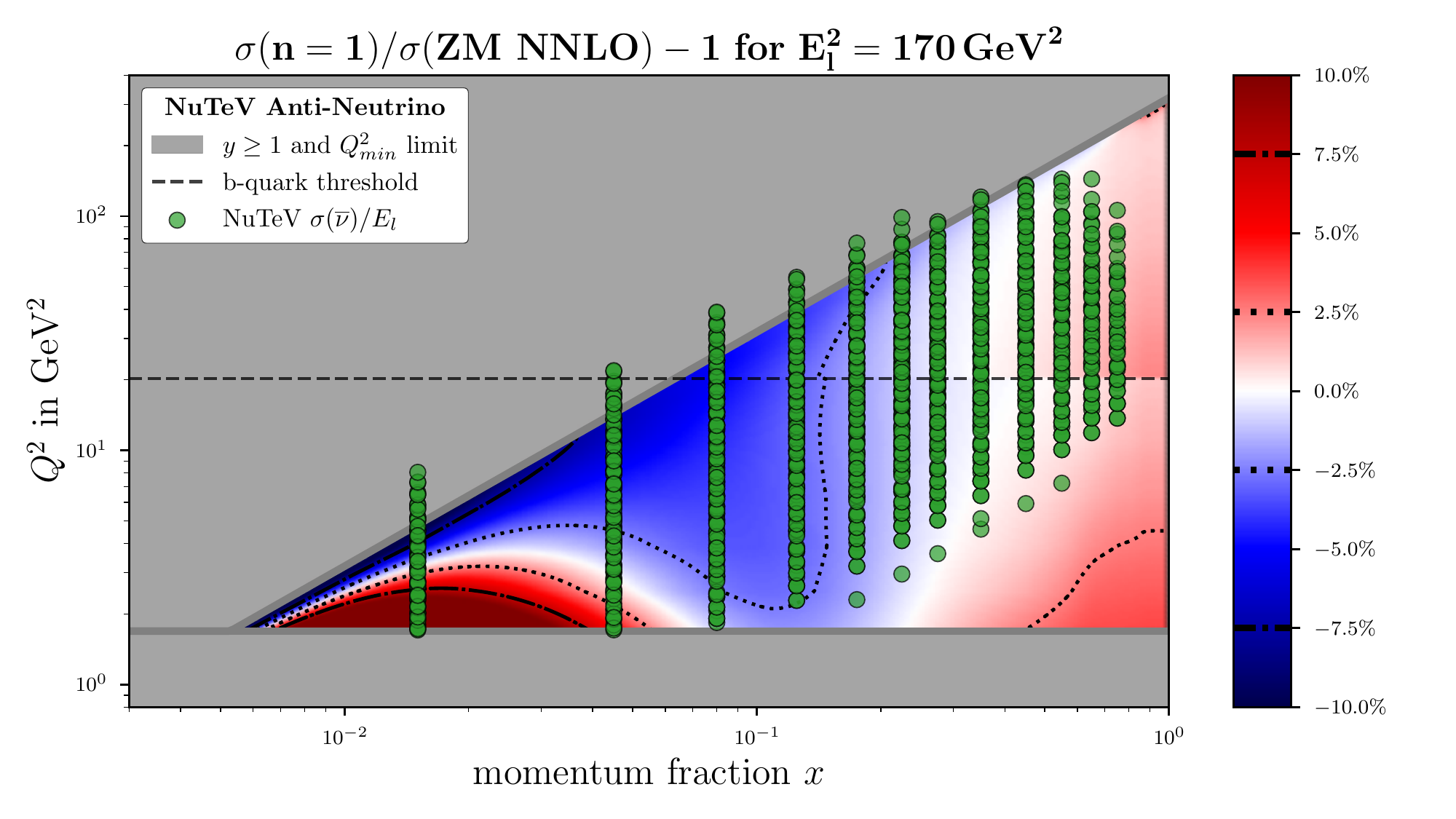}
	\includegraphics[width=0.5\textwidth]{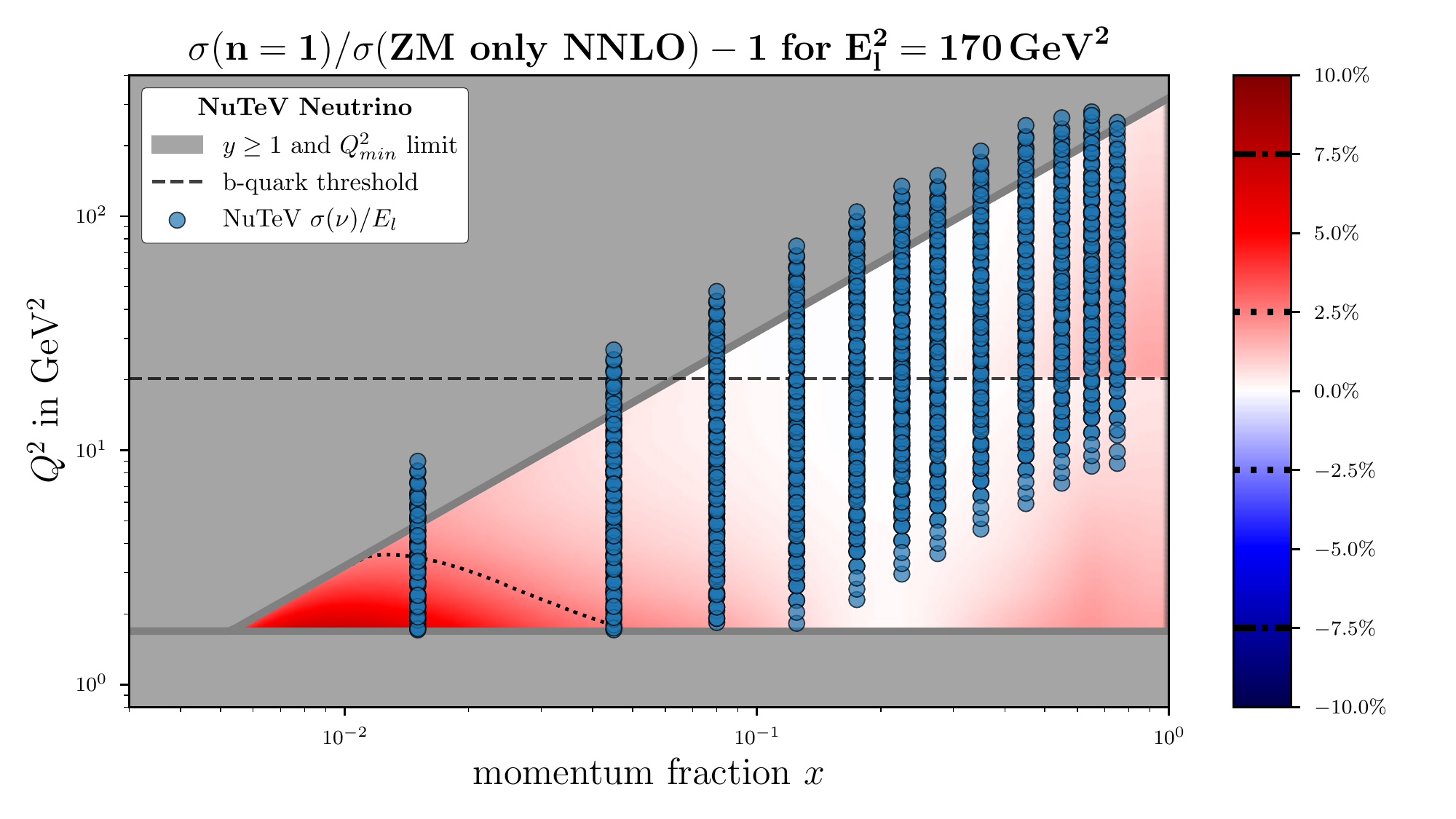}%
	\includegraphics[width=0.5\textwidth]{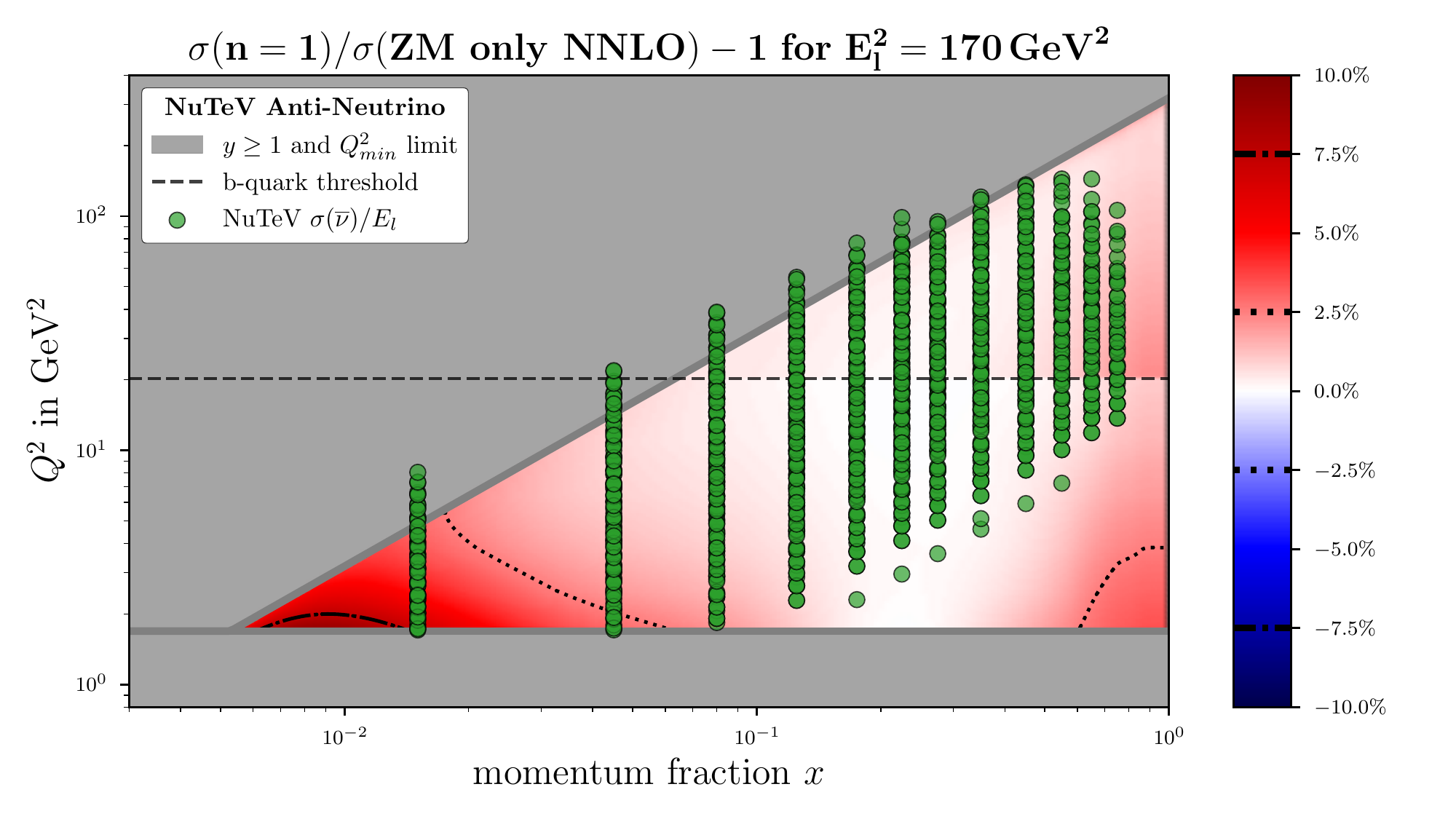}
	\caption{The ratio of \SACOT{1} to \SACOTnlo{} in the first row, \ZMNNLO{} in the second row and \ZMonlyNNLO{} in the last row with NuTeV inspired kinematics in the $(x,Q^2)$-plane as a heat map. The left column displays the ratio for an incoming $\nu$ and the right column an incoming $\overline{\nu}$ alongside with the corresponding measurements. We assume a lepton energy of $E^2_l=170$ GeV${}^2$, which is the median of the data set. The blue/green circles are the positions of every data available (not only those with $E^2_l=170$ GeV${}^2$). The gray patches indicate the kinematic limits for the prediction, with $Q^2_{min}$ originating from the parametrization scale of the \texttt{nCTEQ15HQ} Iron-PDF set used here.}
	\label{fig:CC_NuTeV_n1_over_all}
\end{figure*}

In \cref{fig:CC_NuTeV_n1_over_all}, the \SACOT{1} scheme is compared to \SACOTnlo{}, ZM (all orders) and ZM only in the NNLO contribution from top to bottom. These ratios are obtained by setting the incoming lepton energy $E_l$ to the median value of the NuTeV data set: $E^2_l = 170$~GeV${}^2$. The blue (incoming neutrino) and green (incoming anti-neutrino) circles display the position of the individual data points, including those that do not belong to the median lepton energy. The gray patches indicate regions that are excluded, either due to the initial scale of the PDF set ($Q_\textrm{min} = 1.3$~GeV) or due to kinematic constraints ($y\geq 1$). The dotted and dash-dotted lines indicate the 2.5\% and 7.5\% contours of the ratio. The ratio is evaluated on the same $(400\times 400)$-grid as the structure functions in \cref{sec:mass_effects_on_structure_functions_at_NNLO}.

Focusing on the upper row first, the heat map clearly indicates why the difference in $\chi^2$ is so significant:  1)~almost the full range of $(Q^2,x)$-pairs, that capture the scheme differences (i.e.~ratio $\neq 1$, colored patches) is covered by the data points; 2)~further, the experimental precision is comparable to the size of the effect. 

The second row indicates the total mass effects for all orders and, although the ratio is less than $5\%$ from unity for most of the kinematic region, mass effects are still significant for almost every data point.

Finally, the last row shows the mass effects arising from the NNLO corrections. The absolute size decreases even more compared to the first and second row, but can reach more than $2.5\%$ and affects a large amount of data points.

Overall, we conclude that mass effects are essential for describing the neutrino DIS measurements, and a more detailed investigation is required.
This is in particular important in view of the fact that all current nuclear PDFs~\cite{Muzakka:2022wey,AbdulKhalek:2022fyi,Eskola:2021nhw} use NLO calculations for predictions for the $\nu$-DIS data, and they have problems with describing these measurements~\cite{Schienbein:2009kk,Paukkunen:2010hb,Kovarik:2010uv,Muzakka:2022wey}.

\section{Conclusions}\label{sec:conclusions}

We extended the definition of the \aSACOTchi{} scheme for DIS structure functions to charged-current interactions at NNLO and performed the corresponding calculations. The basic principles of the neutral current definition can be transferred to the charged current case by accounting for the more involved flavor decomposition. Furthermore, we defined $\F{3}$ which was missing in the neutral current definition of the scheme. With this extension, the three most important structure functions are now available for all interactions and the corresponding cross section can be calculated for the first time in this scheme.

The numerical implementation has been performed in the open-source framework of \apfelxx{}, which allows for an efficient evaluation of structure functions by means of interpolation techniques. The implementation will be made publicly available with an update of the code accompanying this paper. 
Additionally, as \apfelxx{} is integrated into the \xfitter{} framework~\cite{Alekhin:2014irh,xFitter:2022zjb}, 
these results can also be studied with the \xfitter{} package.

The results of the \aSACOTchi{} calculation have been compared in detail to the available zero-mass calculations and in the context of physical cross-section predictions for HERA, the upcoming EIC and neutrino-DIS measurements by the NuTeV, CCFR and Chorus collaborations. We concluded that the charged-current measurements at HERA and the EIC are at momentum transfers too large for the heavy-quark mass effects to be resolved. However, in the case of neutrino-DIS the process energy is lower and mass effects play a significant role in the description of the data, which can have important phenomenological effects, e.g.~on nuclear PDFs. A more detailed investigation is required, \review{since the measurements have been proven to be sensitive to other kinematic effects like target mass corrections~\cite{Ruiz:2023ozv} or general higher twist effects. A systematic nuclear PDF fit at NNLO precision with the these effects included (and then individually turned off) could  also shed new light on the issue.}

\clearpage
\appendix

\section{Basic principles of the numerical implementation}
\label{sec:Details_of_the_numerical_implementation}

This appendix gives a brief introduction to the numerical methods used in the implementation of structure functions in \apfelxx{}. A more detailed description of the technology can be found in Refs.~\cite{Bertone:2013vaa,Bertone:2017gds} or alternatively~\cite{Carrazza:2020gss,christopher_schwan_2025_15635174}. 

The computation of structure functions reduces to Mellin convolutions between a Wilson coefficient $\coef{\lambda}$ and a parton distribution $f$. The integral to solve is structured as 
\begin{equation}
    \left[\coef{\lambda}\otimes f\right](x,Q^2) = \myIntFrac{x}{1}{z}{z} \coef{\lambda}\left(z,Q^2\right)f\left(\frac{x}{z},Q^2\right)\,.
    \label{eq:mellin_conv_appendix}
\end{equation}
Since the Wilson coefficient consists of non-trivial functions and distributions, these integrals are not straightforward to compute and the numerical evaluation is time intensive. In practical applications, parton distributions change constantly (e.g.~in a global PDF fit) and thus repeated evaluations of these integrals are required. Conversely, Wilson coefficients do not change. The technology employed in \apfelxx{} is a ``PDF independent'' formulation of the integrals, where the integration over the Wilson coefficients is precomputed and the relevant information is stored in look-up tables.

The definition of these look-up tables is first given at fixed $Q^2$, and in the following generalized to variable $Q^2$.

\subsection{Interpolation for fixed \texorpdfstring{$Q^2$}{Q²}}

In this subsection, we consider \cref{eq:mellin_conv_appendix} for fixed values of $Q^2$, which we temporarily drop from our notation. In order to remove the parton distributions from the integration, we interpolate $f(x)$ on a predefined $x$-grid 
\begin{equation}
    g_x = \{x_0,\dots,x_{N_x}\}
\end{equation}
with interpolating functions $w_\alpha(x)$ as given in \cref{eq:PDF_interpolation}. The interpolating functions are uniquely defined by the grid $g_x$ through Lagrange interpolation of arbitrary degree. By inserting the interpolated PDF into \cref{eq:mellin_conv_appendix}, we find
\begin{align}
    \left[\coef{\lambda}\otimes f\right](x) &= \sum^{N_x}_\alpha\myIntFrac{x}{1}{z}{z}\coef{\lambda}\left(z\right) w_\alpha\left(\frac{x}{z}\right)f(x_\alpha) \notag\\
    &=\sum^{N_x}_\alpha f(x_\alpha)\myIntFrac{x}{1}{z}{z}\coef{\lambda}\left(z\right) w_\alpha\left(\frac{x}{z}\right)\,,
    \label{eq:mellin_convolution_pdf_interpolated}
\end{align}
which removes the PDF dependence from the integrand. 

Finally, to obtain a prediction for the Mellin convolution at every $x$-value without the need of recalculating, we interpolate the convolution on the same grid $g_x$:
\begin{equation}
    \left[\coef{\lambda}\otimes f\right](x) = \sum_\beta^{N_x} w_\beta(x) \left[\coef{\lambda}\otimes f\right](x_\beta)\,.
\end{equation}
Inserting in \cref{eq:mellin_convolution_pdf_interpolated} yields the final interpolation table $\Gamma_{\alpha\beta}$
\begin{equation}
    \left[\coef{\lambda}\otimes f\right](x,Q^2) = \sum_\beta^{N_x}\sum_\alpha^{N_x} w_\beta(x) f(x_\alpha,Q^2) \Gamma_{\alpha\beta}(Q^2)
    \label{eq:mellin_convolution_full_interpolation}
\end{equation}
with
\begin{equation}
    \Gamma_{\alpha\beta}(Q^2) = \myIntFrac{x_\beta}{1}{z}{z} \coef{\lambda}(z,Q^2) w_\alpha\left(\frac{x_\beta}{z}\right)\,,
\end{equation}
where, for completeness, we have reintroduced the $Q^2$ dependence.

\subsection{Interpolation for any \texorpdfstring{$Q^2$}{Q²}}

In order to predict the structure functions at any $Q^2$ without needing to recalculate $\Gamma_{\alpha\beta}(Q^2)$, we interpolate our results from above on a \textit{separate} $Q^2$-grid $g_Q$:
\begin{equation}
    g_Q = \{Q^2_0,\dots,Q^2_{N_Q}\}\,,
\end{equation}
with associated interpolation functions given by $\tilde{w}(Q^2)$. Thus to calculate \cref{eq:basic_mellin_convolution} at any $Q^2$ we use
\begin{equation}
    \left[\coef{\lambda}\otimes f\right](x,Q^2) = \sum_\gamma^{N_Q}\tilde{w}_\gamma(Q^2)\left[\coef{\lambda}\otimes f\right](x,Q^2_\gamma)\,.
\end{equation}
And with the shorthand notation 
\begin{align}
    \Gamma_{\alpha\beta\gamma} &= \Gamma_{\alpha\beta}(Q_\gamma^2)\notag\\
     &=\myIntFrac{x_\beta}{1}{z}{z} \coef{\lambda}(z,Q^2_\gamma) w_\alpha\left(\frac{x_\beta}{z}\right)
\end{align}
we find the final formula 
\begin{align}
    &\left[\coef{\lambda}\otimes f\right](x,Q^2) = \notag\\
    &\qquad\sum_\gamma^{N_Q}\sum_\beta^{N_x}\sum_\alpha^{N_x}\tilde{w}_\gamma(Q^2)w_\beta(x)f(x_\alpha,Q^2_\gamma)\Gamma_{\alpha\beta\gamma}\,.
        \label{eq:mellin_convolution_full_interpolation_2}
\end{align}
The entries of $\Gamma_{\alpha\beta\gamma}$ are numbers that can be computed once and for all and stored. Thus, to arrive at a final evaluation of the integral it is sufficient to evaluate the simple vector-matrix operations of \cref{eq:mellin_convolution_full_interpolation_2}.

\twocolumngrid
\goodbreak
\section{Feynman diagrams for CC DIS at NNLO}
\label{sec:feynman}

\begin{figure}[tbh]
	\centering
	\subfloat[Graphs independent of $n_f$.\label{fig:NNLO_ns_CC_no_nf}]{
		\shortstack{
			\includegraphics[width=0.32\linewidth]{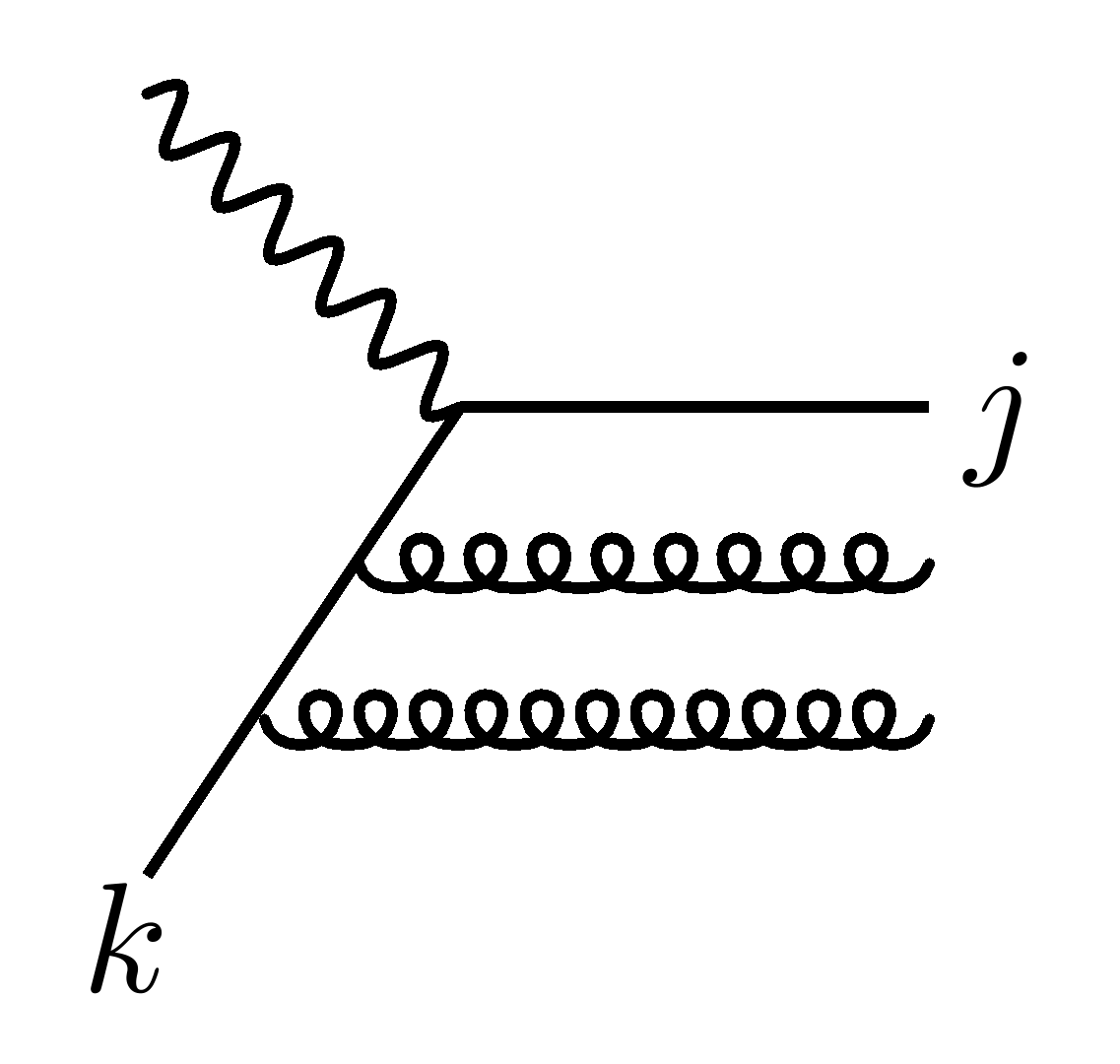}%
			\includegraphics[width=0.32\linewidth]{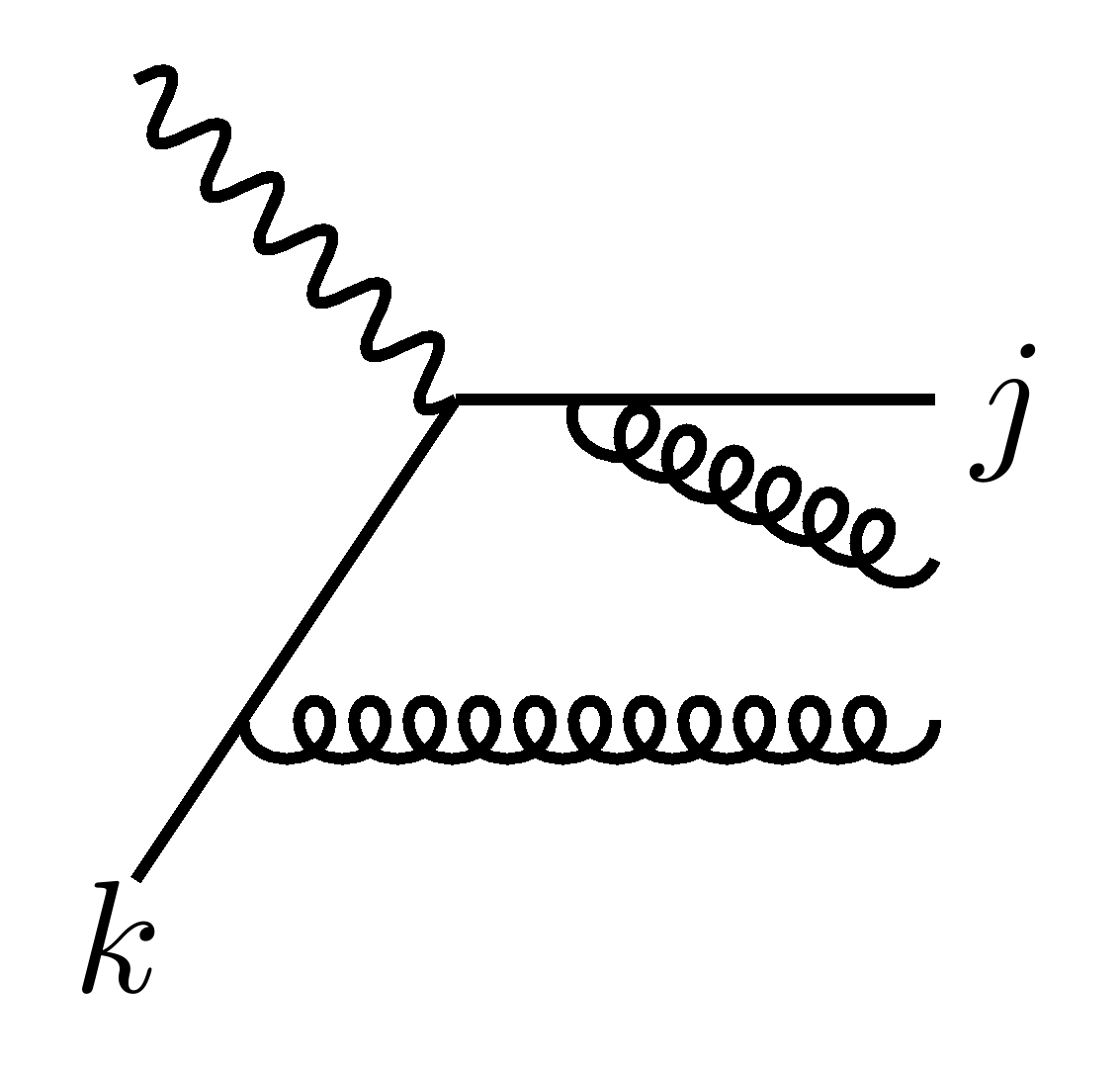}%
			\includegraphics[width=0.32\linewidth]{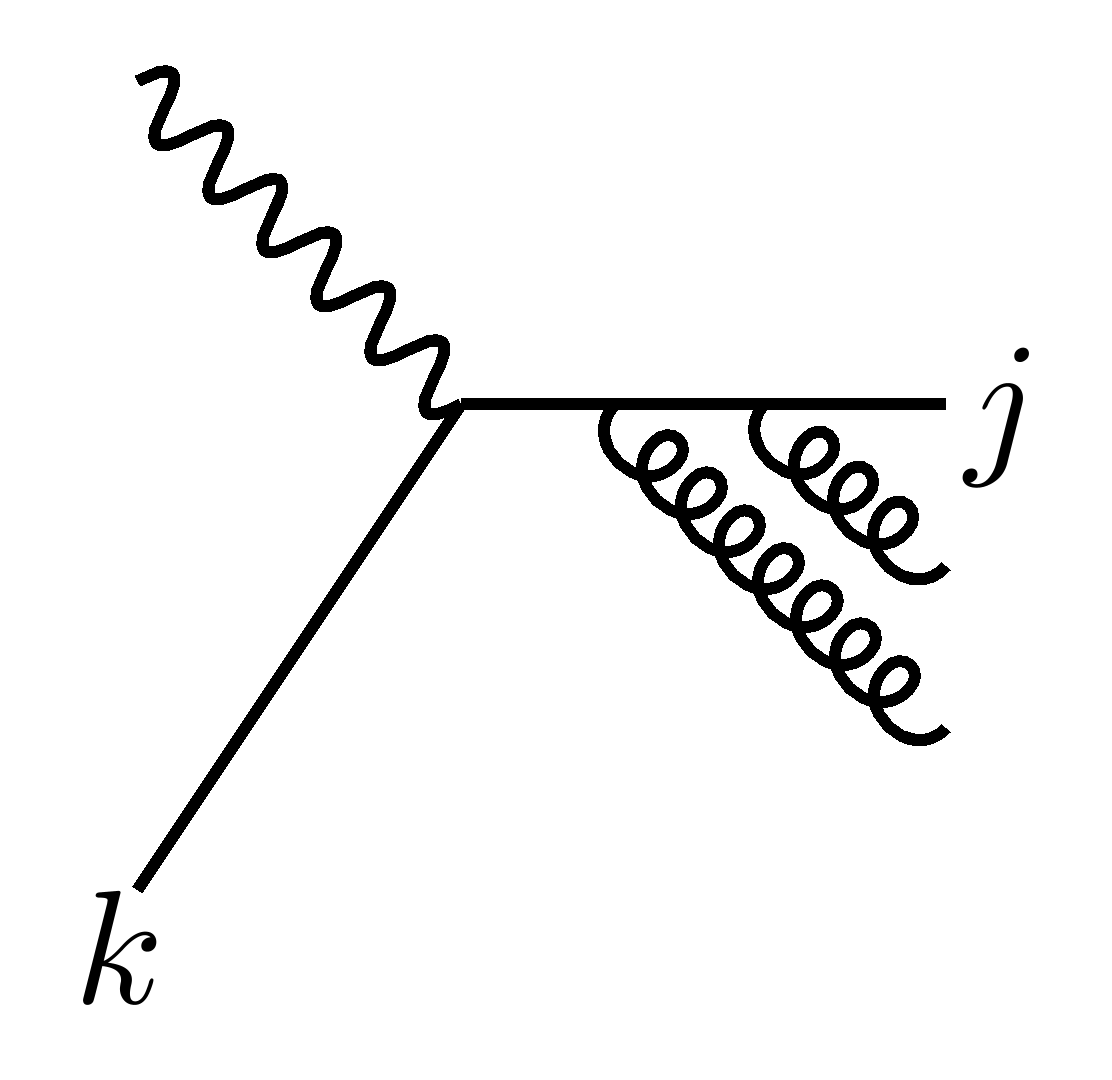}\\
			\includegraphics[width=0.32\linewidth]{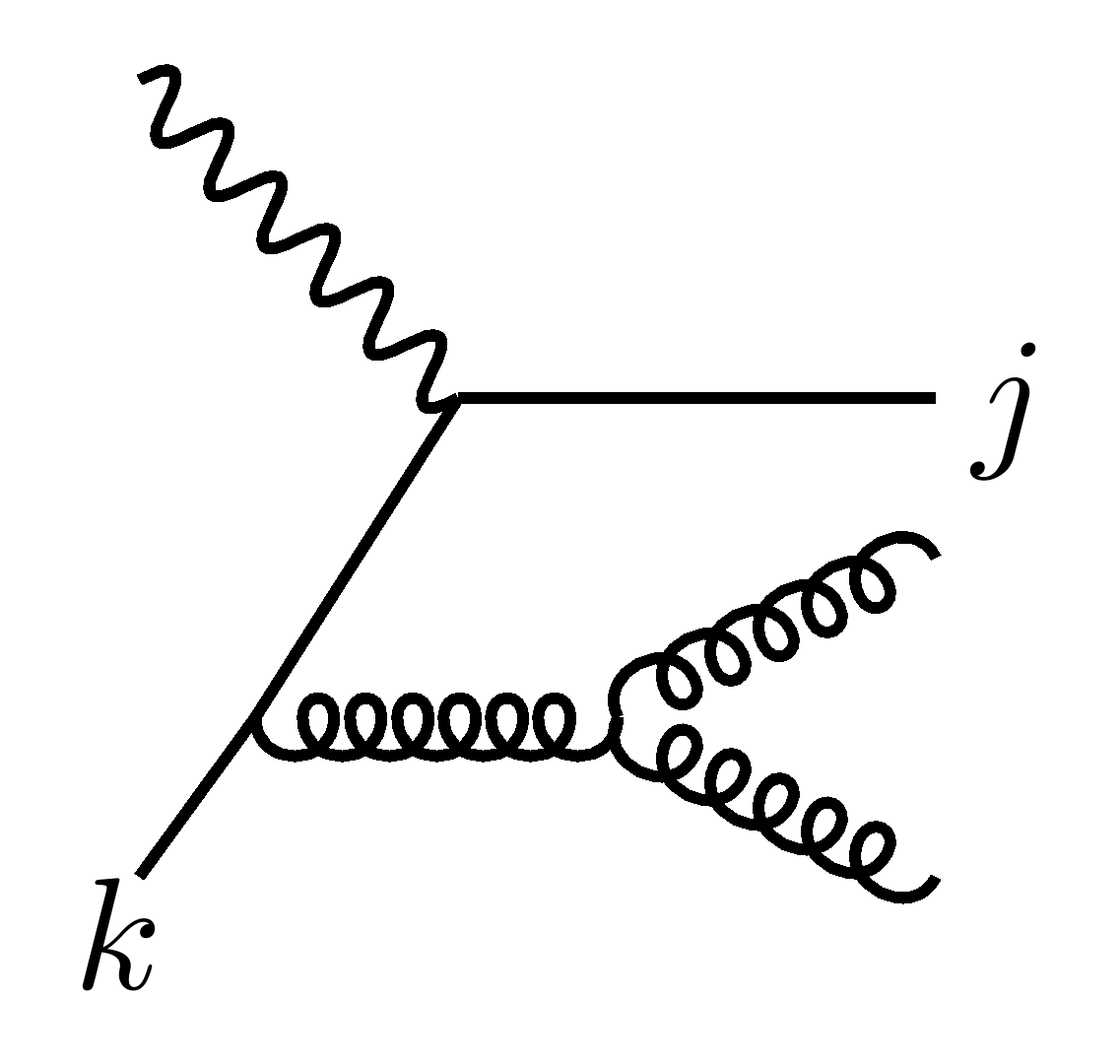}%
			\includegraphics[width=0.32\linewidth]{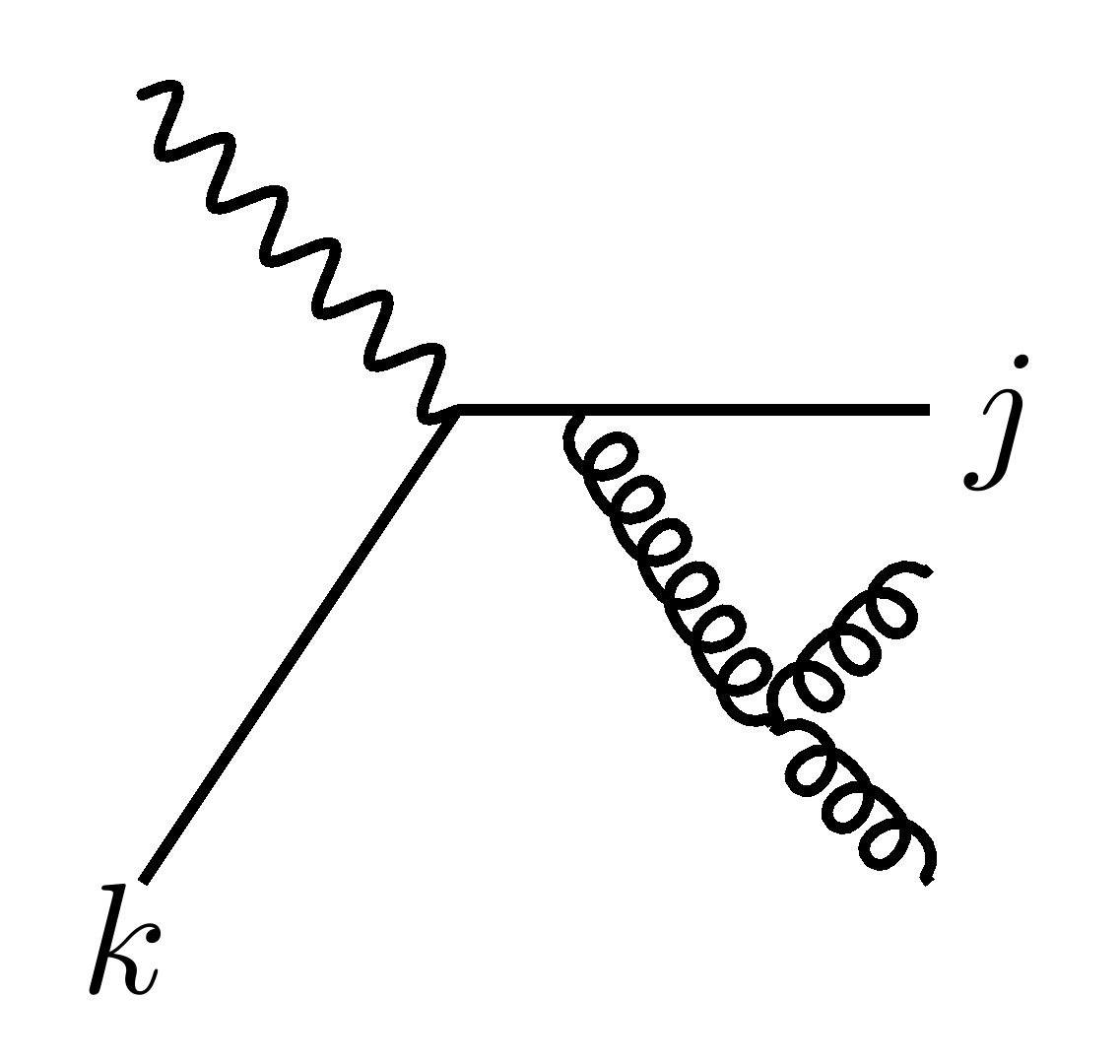}%
	}}
	
	\subfloat[Graphs proportional to $n_f$.\label{fig:NNLO_ns_CC_nf}]{
		\includegraphics[width=0.32\linewidth]{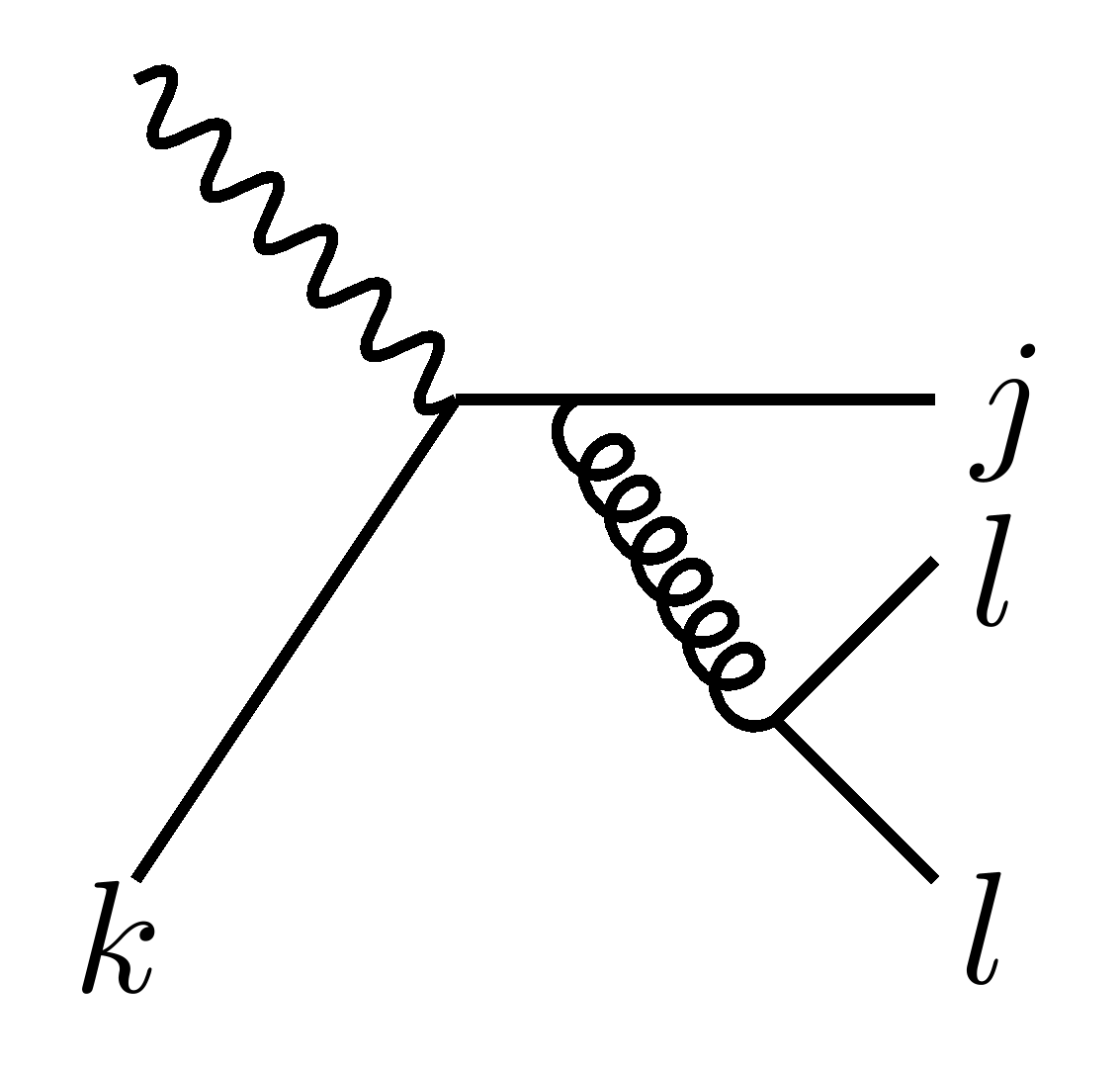}
		\includegraphics[width=0.32\linewidth]{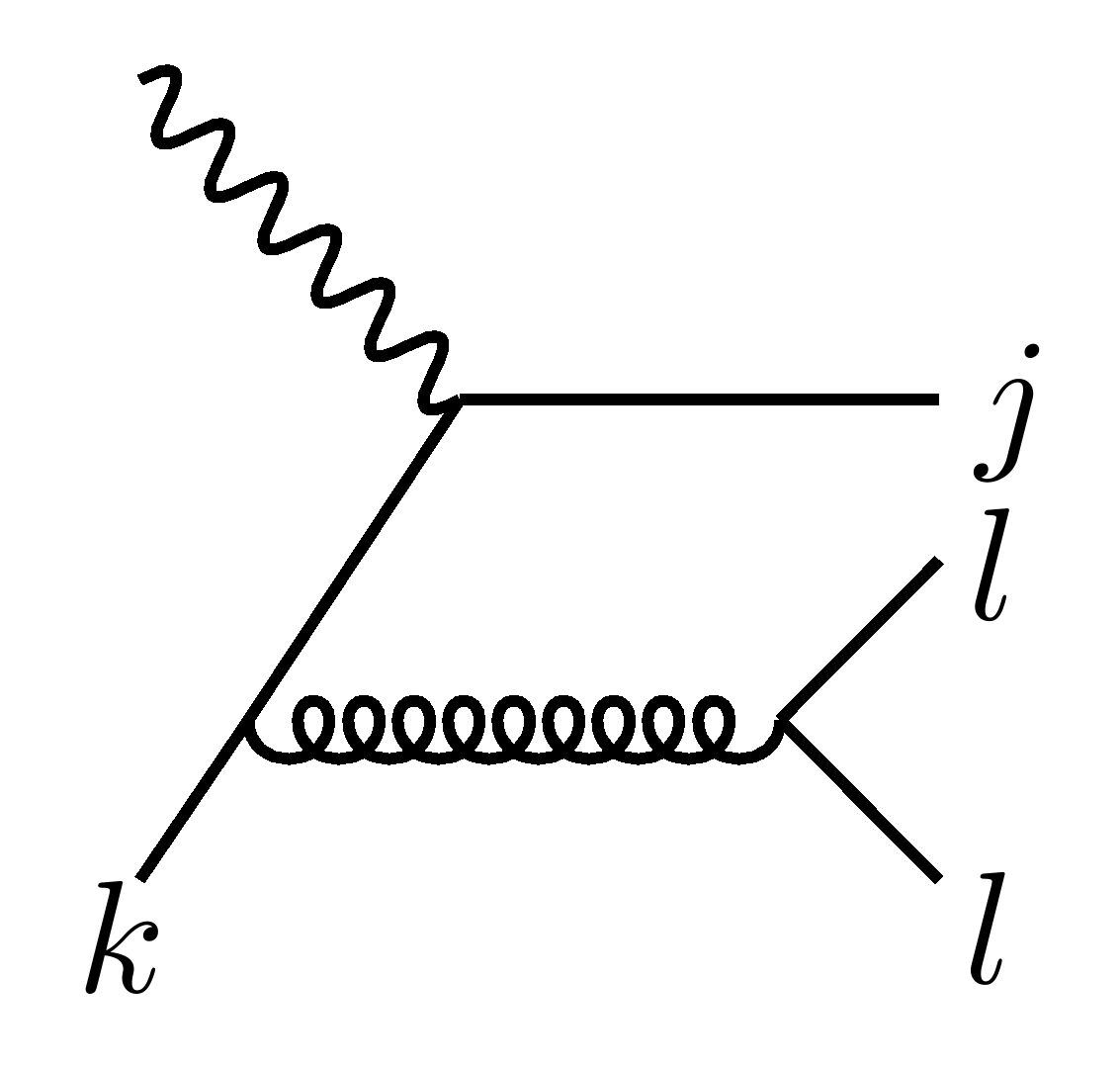}
	}
	\caption{$\order{\alphas^2}$ contributions to $\coef[ns]{\lambda}$. The coefficient can be written as $\coef[ns]{\lambda}=\coef[ns,A]{\lambda} +n_f \coef[ns,B]{\lambda}$, where the graphs contributing to $\coef[ns,A]{\lambda}$ are given in (a) and the graphs contributing to $\coef[ns,B]{\lambda}$ are given in (b).}
	\label{fig:NNLO_ns_CC}
\end{figure}

\begin{figure}[tbh]
	\centering
	\includegraphics[width=0.32\linewidth]{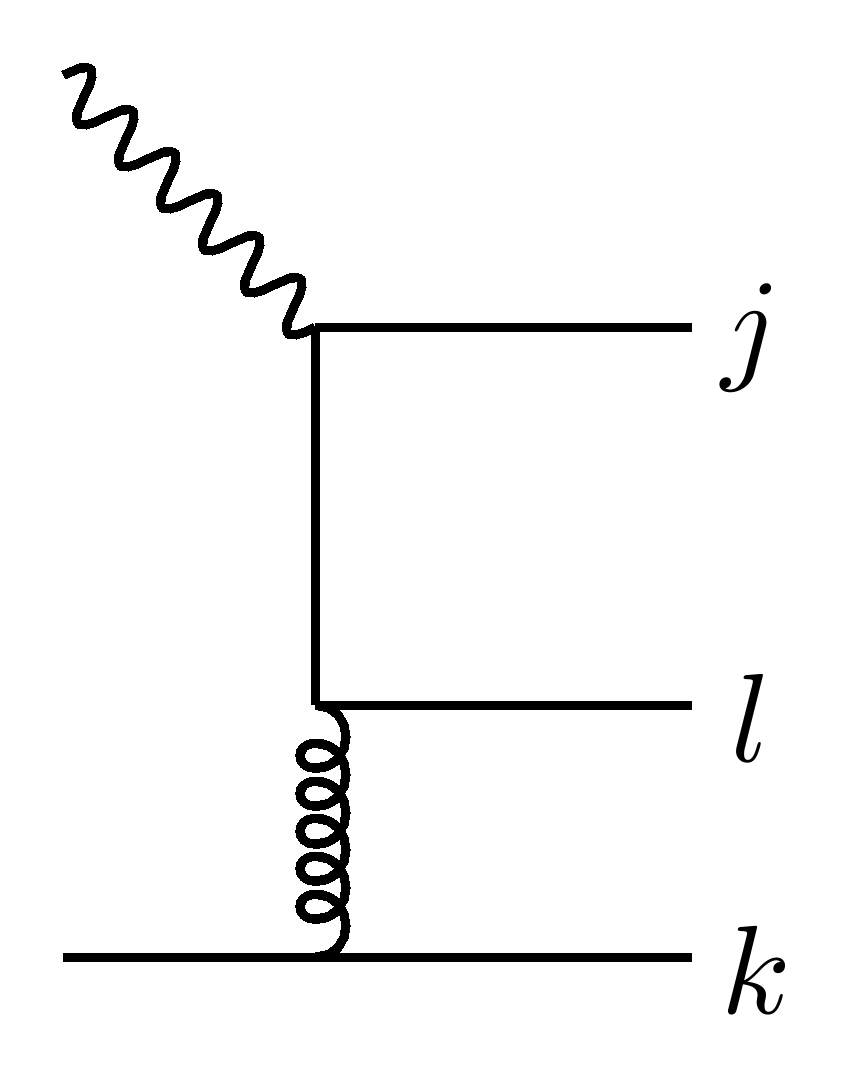}
	\caption{$\order{\alphas^2}$ contribution to $\coef[ps]{\lambda}$. The graph is proportional to $n_f$ and therefore $\coef[ps]{\lambda}$ is as well.}
	\label{fig:NNLO_ps_CC}
\end{figure}

This section displays the relevant Feynman diagrams for charged-current DIS at NNLO with the flavor labeling corresponding to \cref{sec:extension_of_aSACOT-chi_to_CC}. Note that, at this order in $\alphas{}$, graphs with up to three different flavors, namely in \cref{fig:NNLO_ns_CC_nf,fig:NNLO_ps_CC}, are possible for the first time.

\begin{figure}[tbh]
	\centering
	\includegraphics[width=0.32\linewidth]{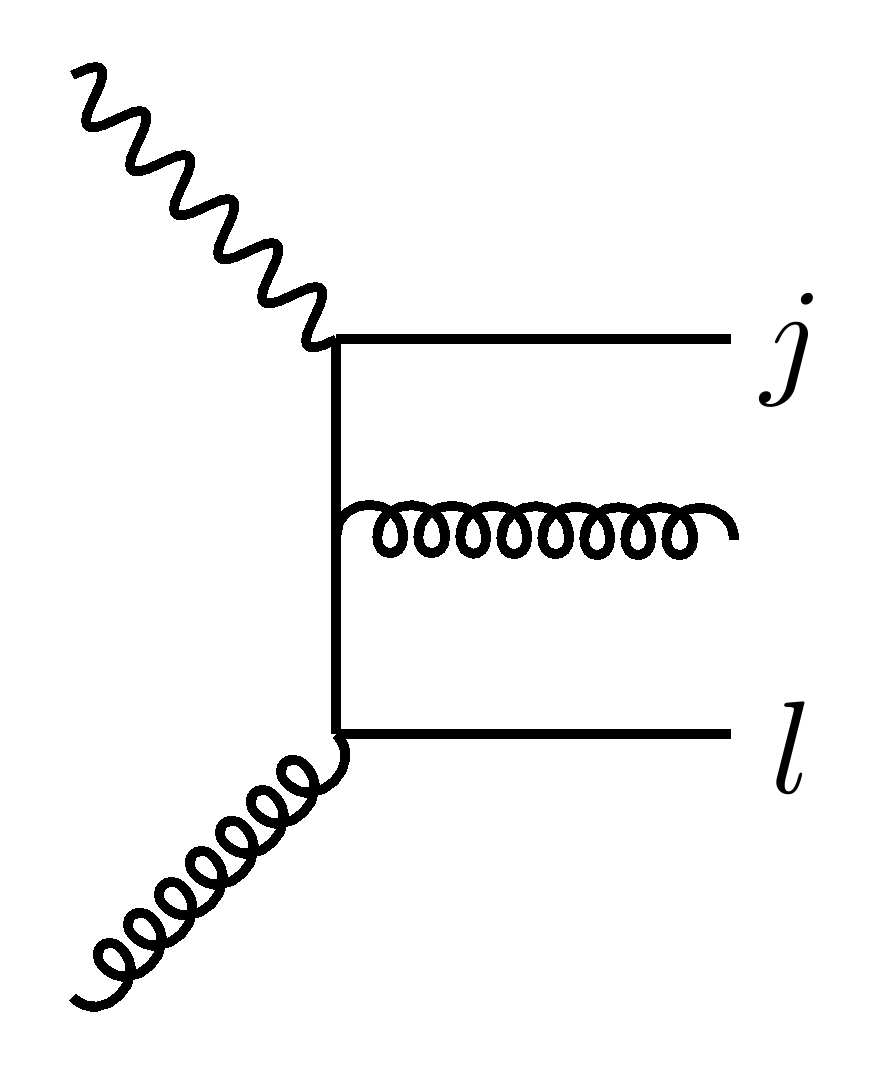}
	\includegraphics[width=0.32\linewidth]{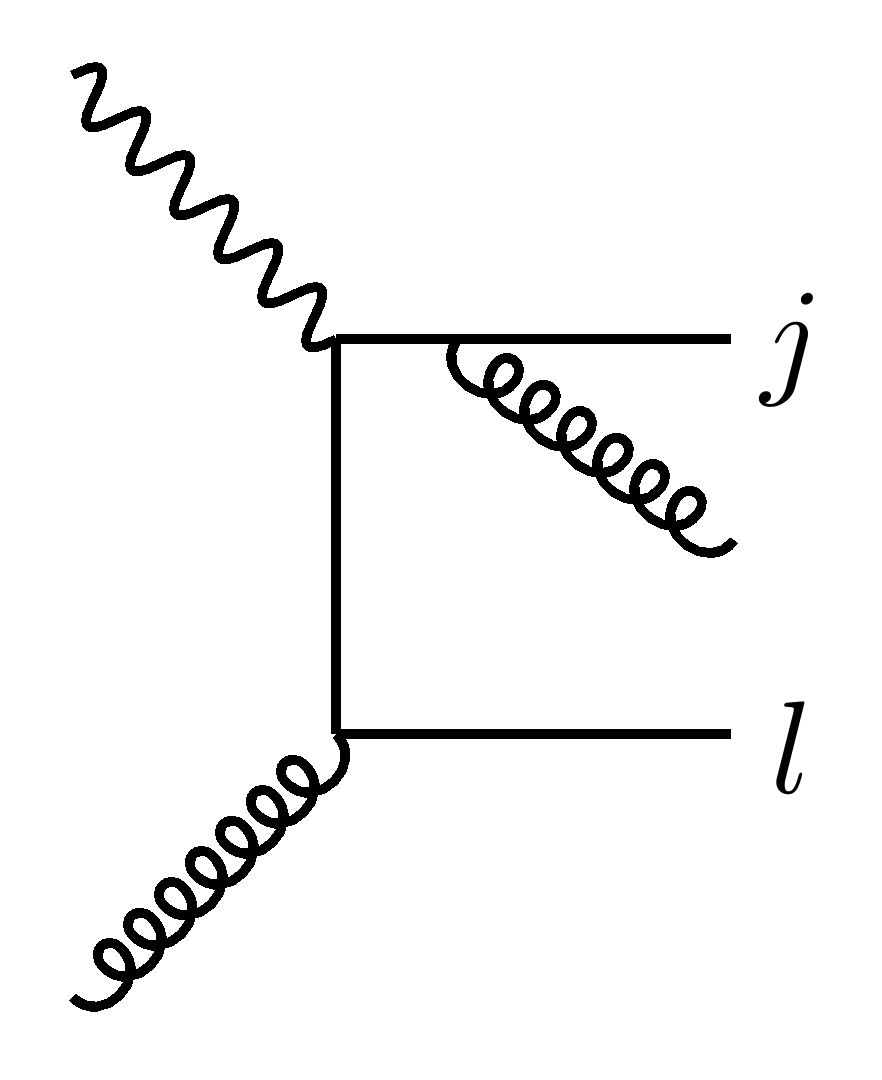}
	
	\includegraphics[width=0.32\linewidth]{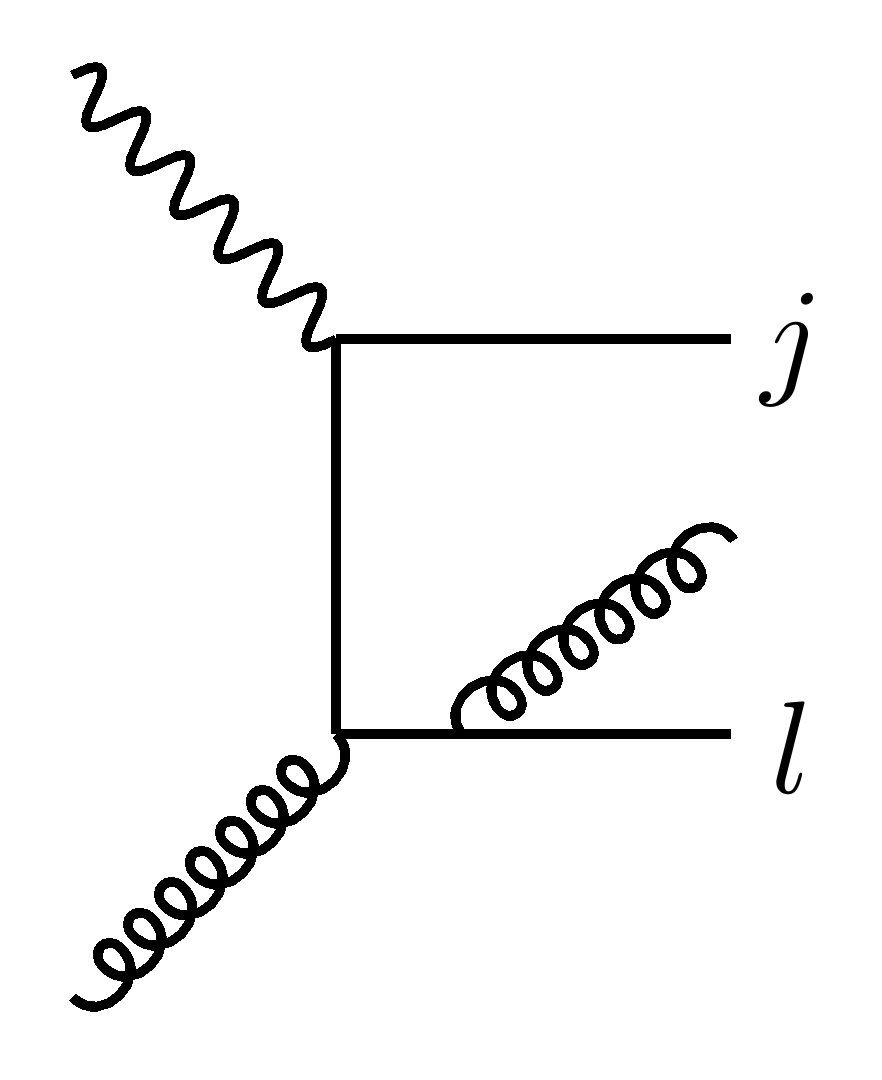}
	\includegraphics[width=0.32\linewidth]{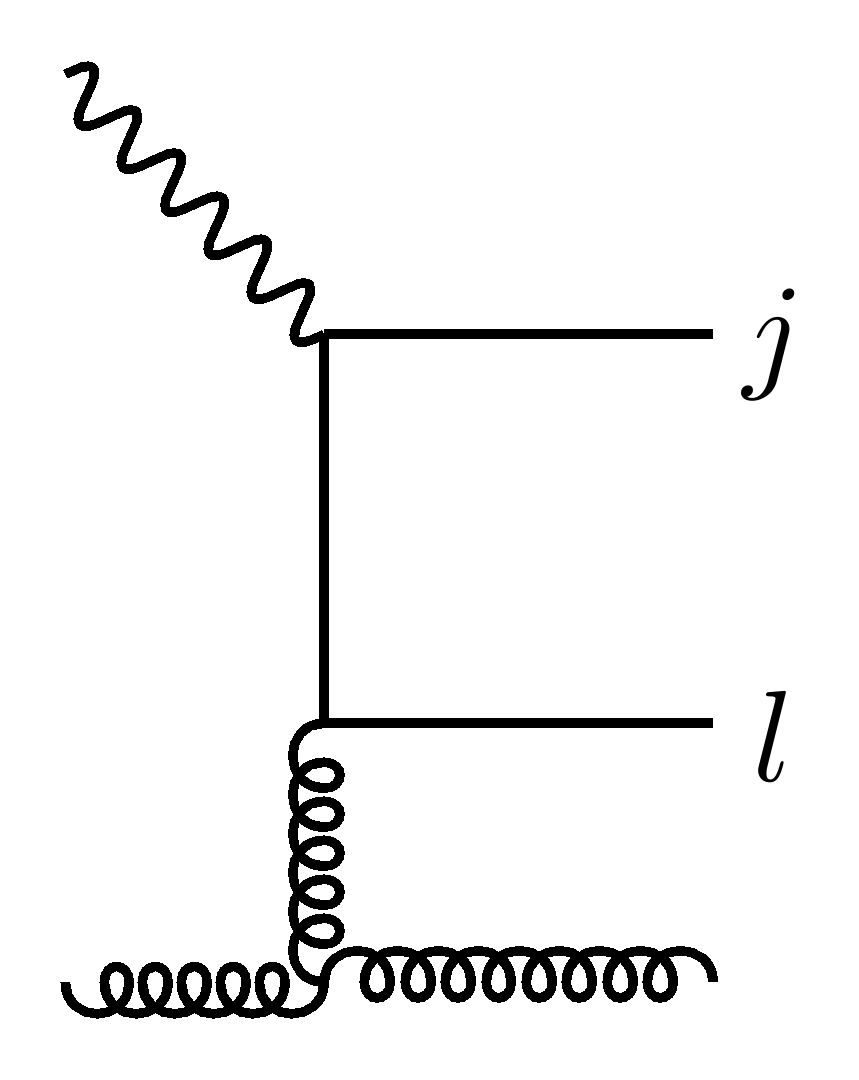}
	\caption{$\order{\alphas^2}$ contributions to $\coef[g]{\lambda}$. All graphs are proportional to $n_f$ and therefore $\coef[g]{\lambda}$ is as well.}
	\label{fig:NNLO_g_CC}
\end{figure}


\begin{figure*}[t]
	\centering
	\includegraphics[width=0.33\textwidth]{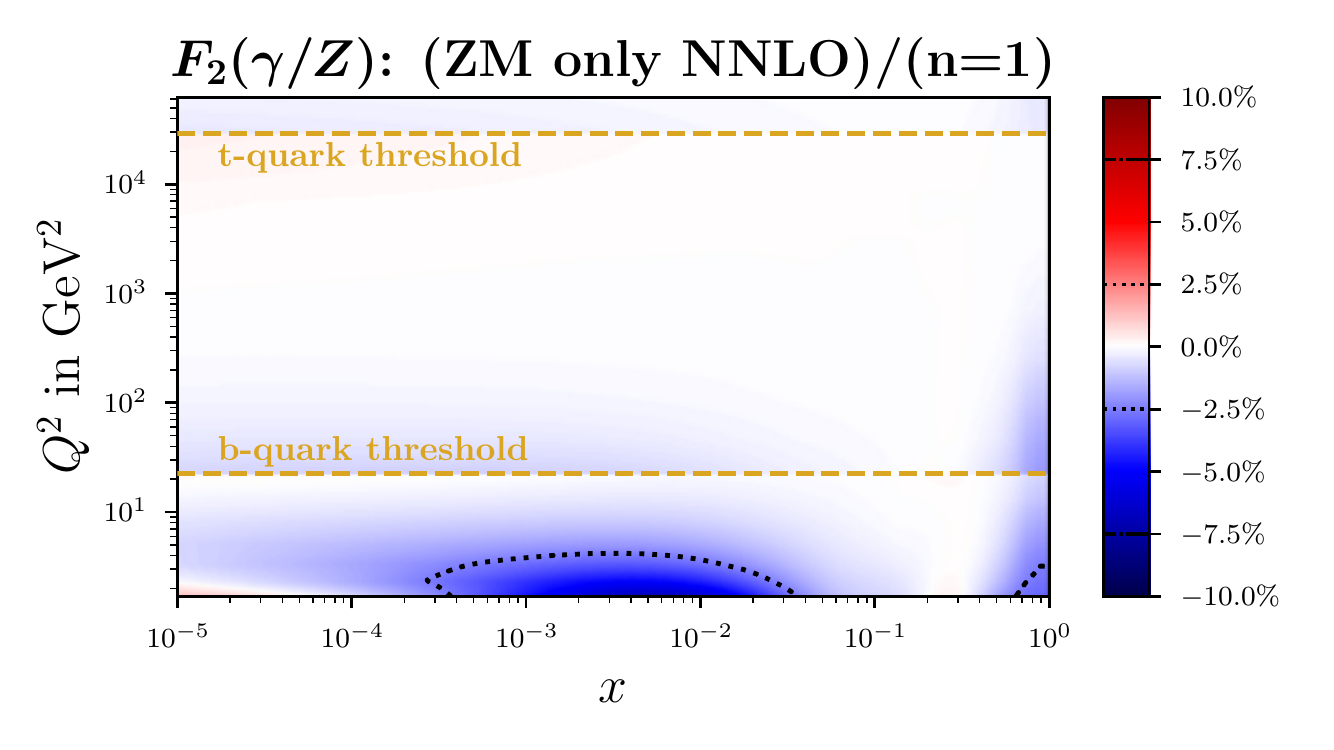}%
    \includegraphics[width=0.33\textwidth]{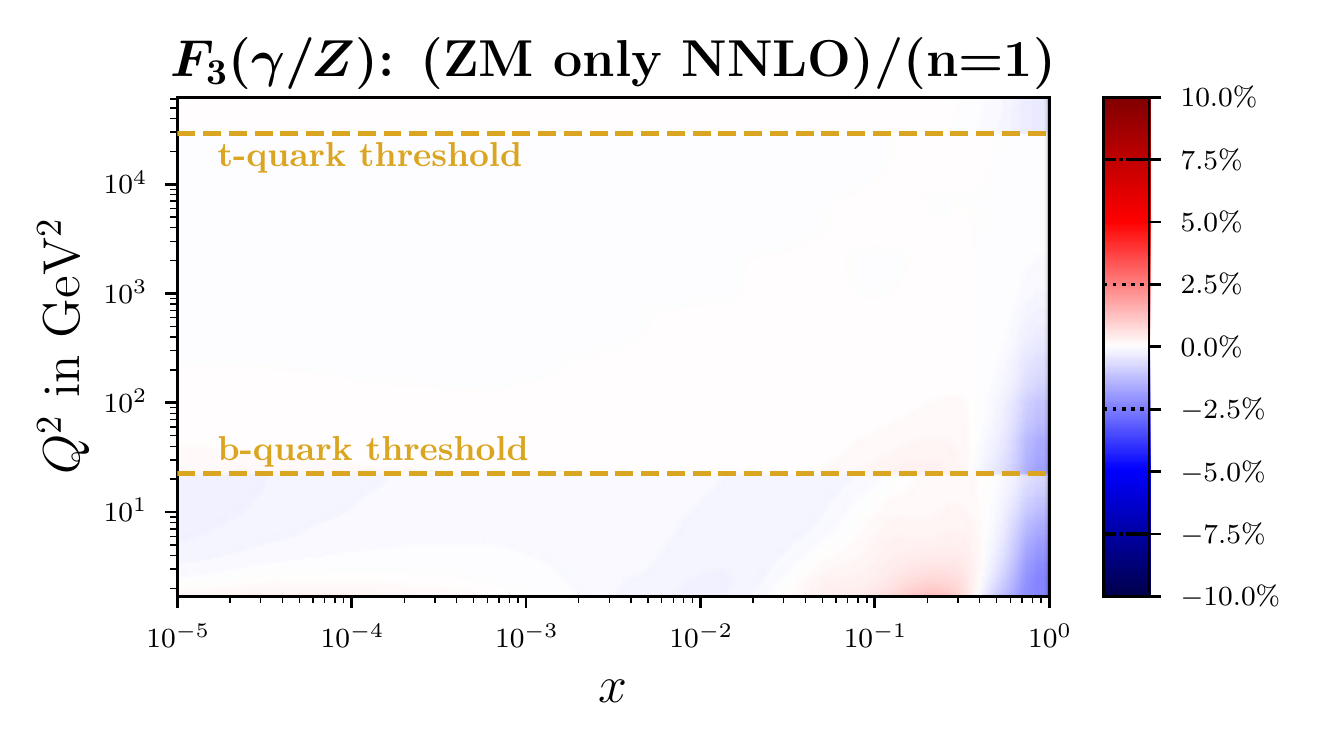}%
    \includegraphics[width=0.33\textwidth]{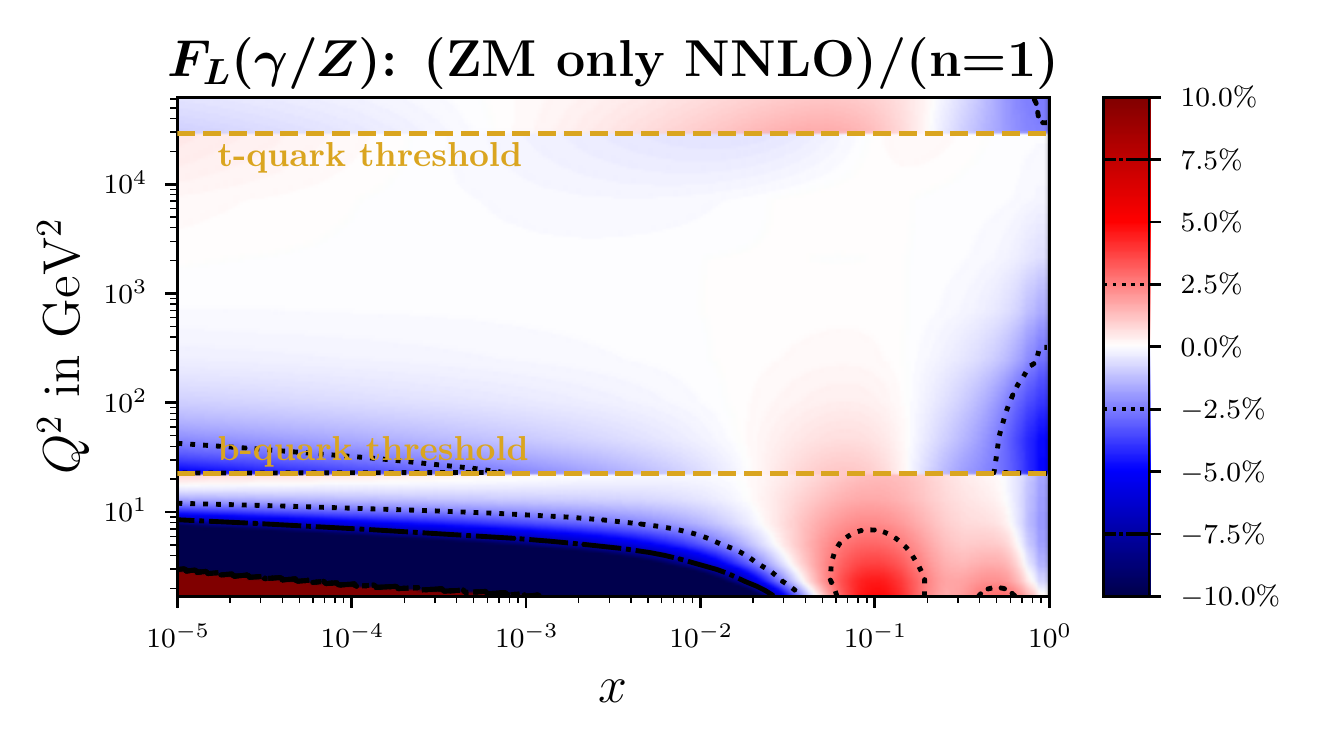}
    \includegraphics[width=0.33\textwidth]{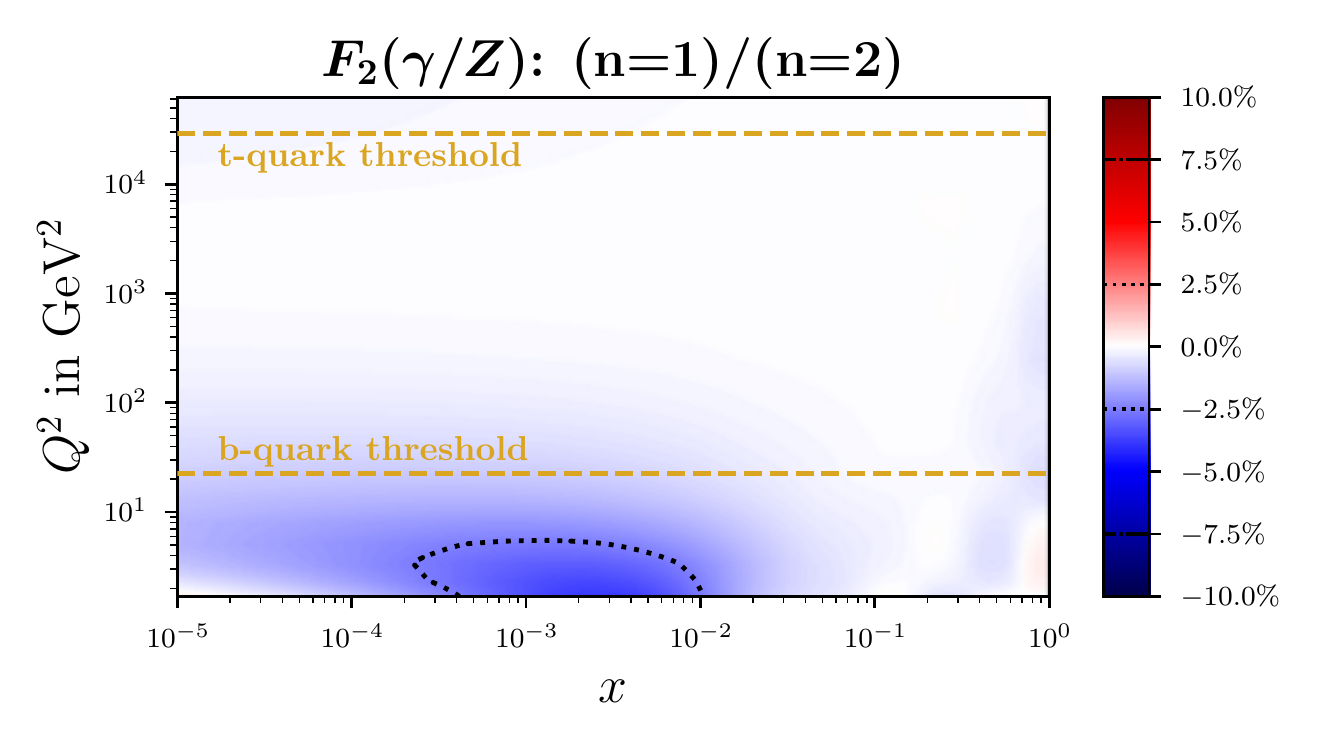}%
    \includegraphics[width=0.33\textwidth]{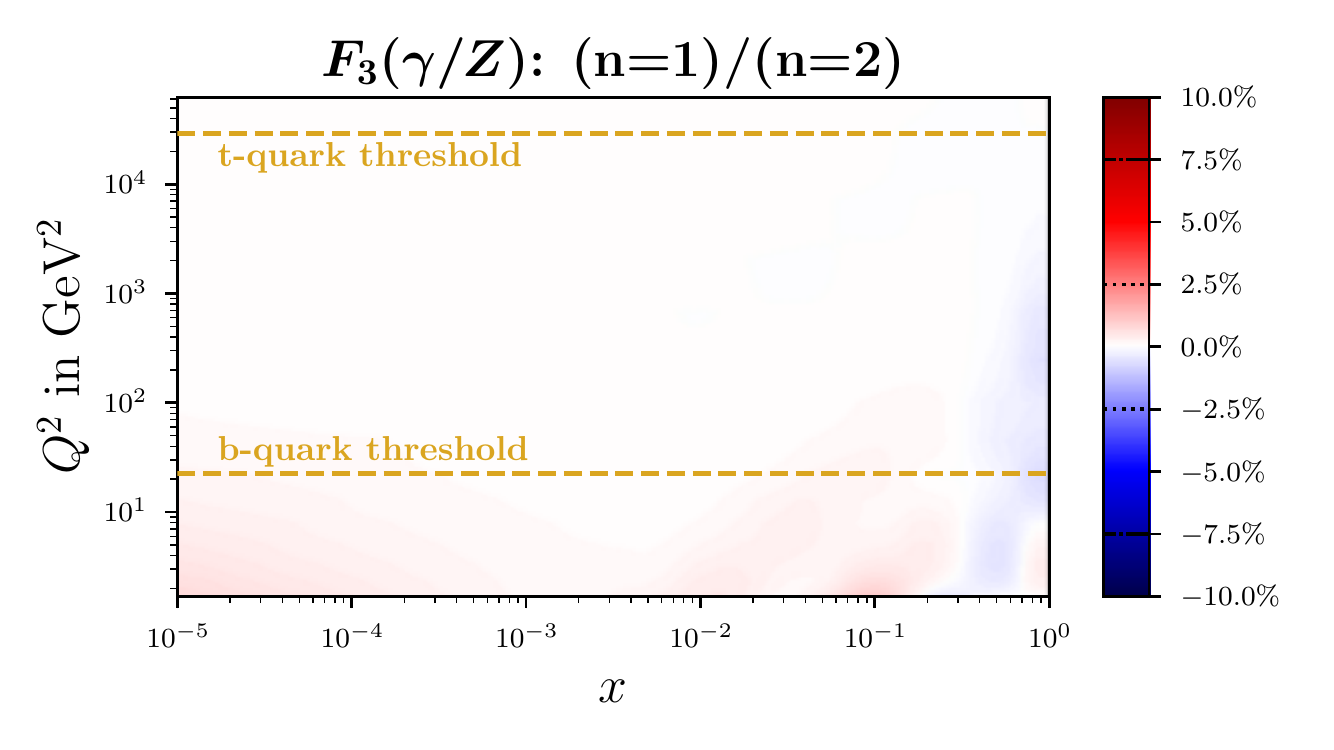}%
    \includegraphics[width=0.33\textwidth]{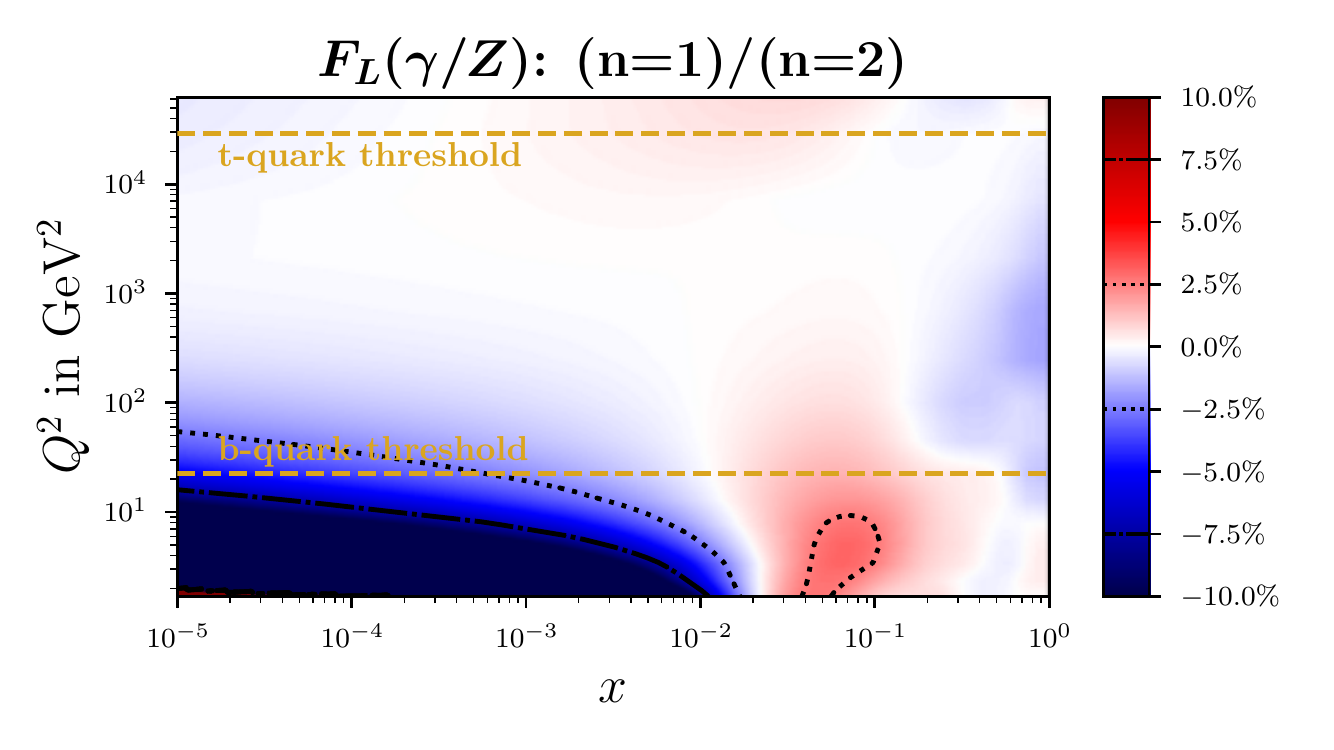}
    \includegraphics[width=0.33\textwidth]{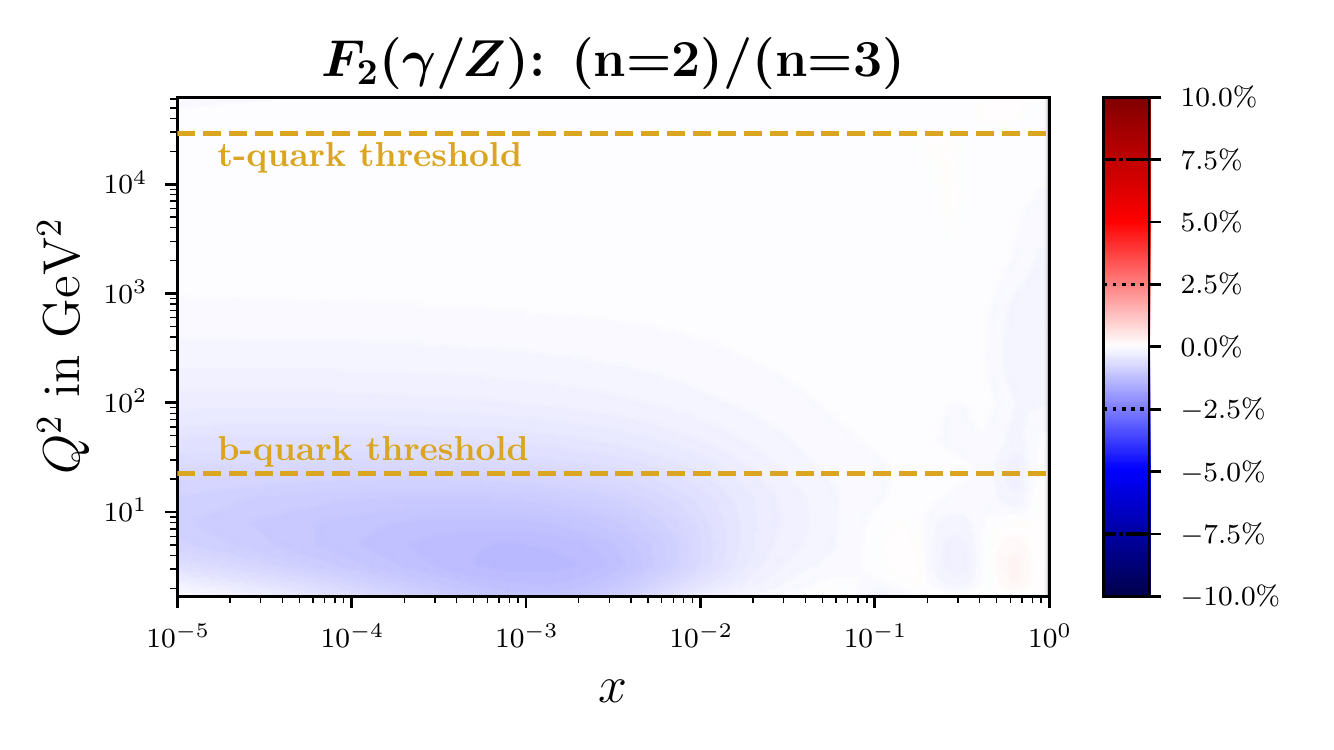}%
    \includegraphics[width=0.33\textwidth]{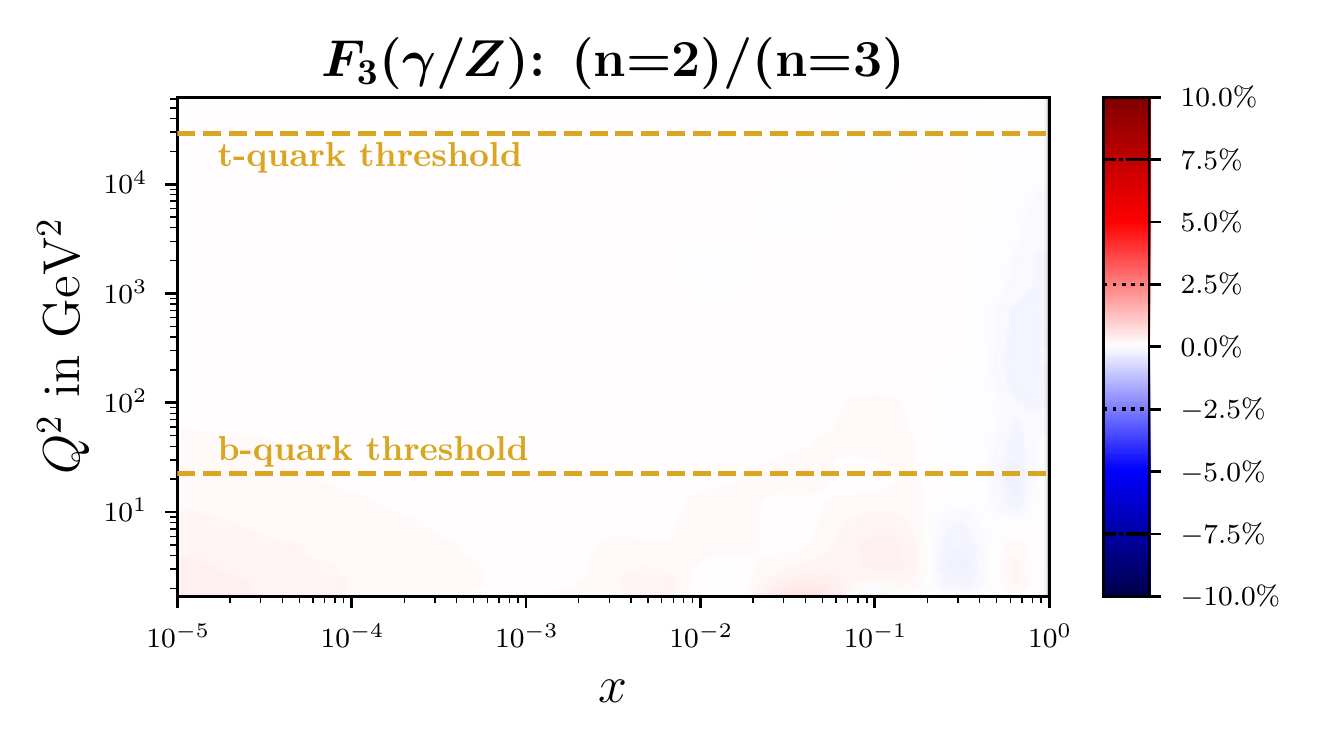}%
    \includegraphics[width=0.33\textwidth]{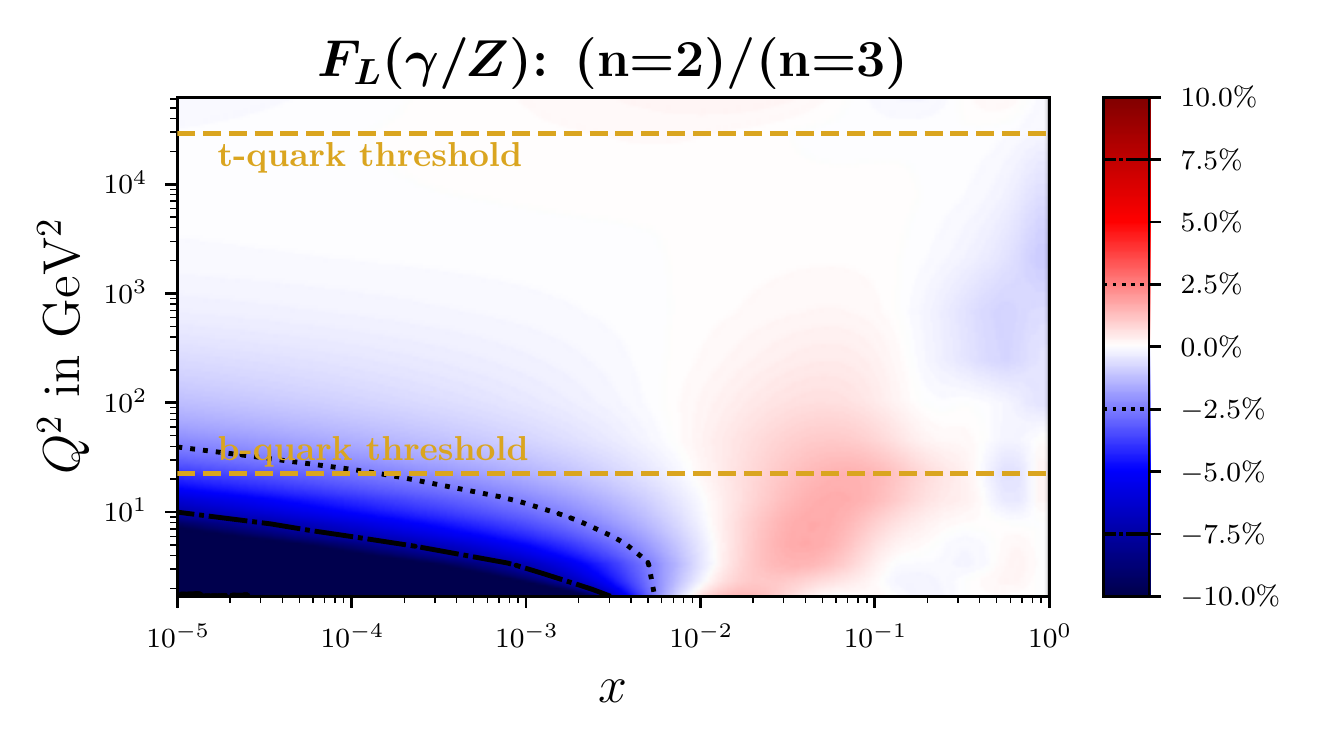}
	\caption{The relative ratio of NNLO neutral current structure functions $F_2,F_3$ and $F_L$ from left to right. From top to bottom, we consider the ratios for ZM/$n=1$, $n=1$/$n=2$, and $n=2$/$n=3$ as the NNLO contributions, respectively. 
    The LO and NLO contributions are always given in the \SACOTchi{} scheme.}
	\label{fig:relative_ratio_F23L_NC}
\end{figure*}
\begin{figure*}[t]
	\centering
	\includegraphics[width=0.33\textwidth]{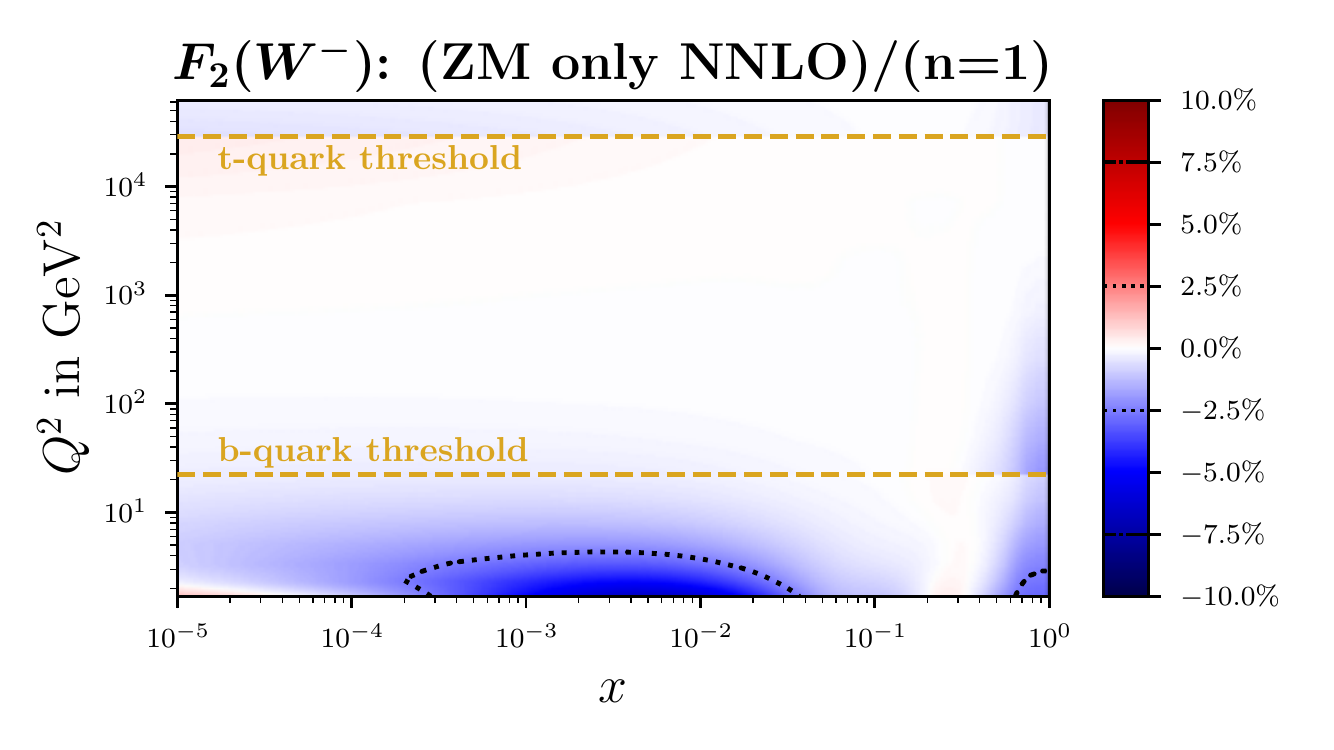}%
    \includegraphics[width=0.33\textwidth]{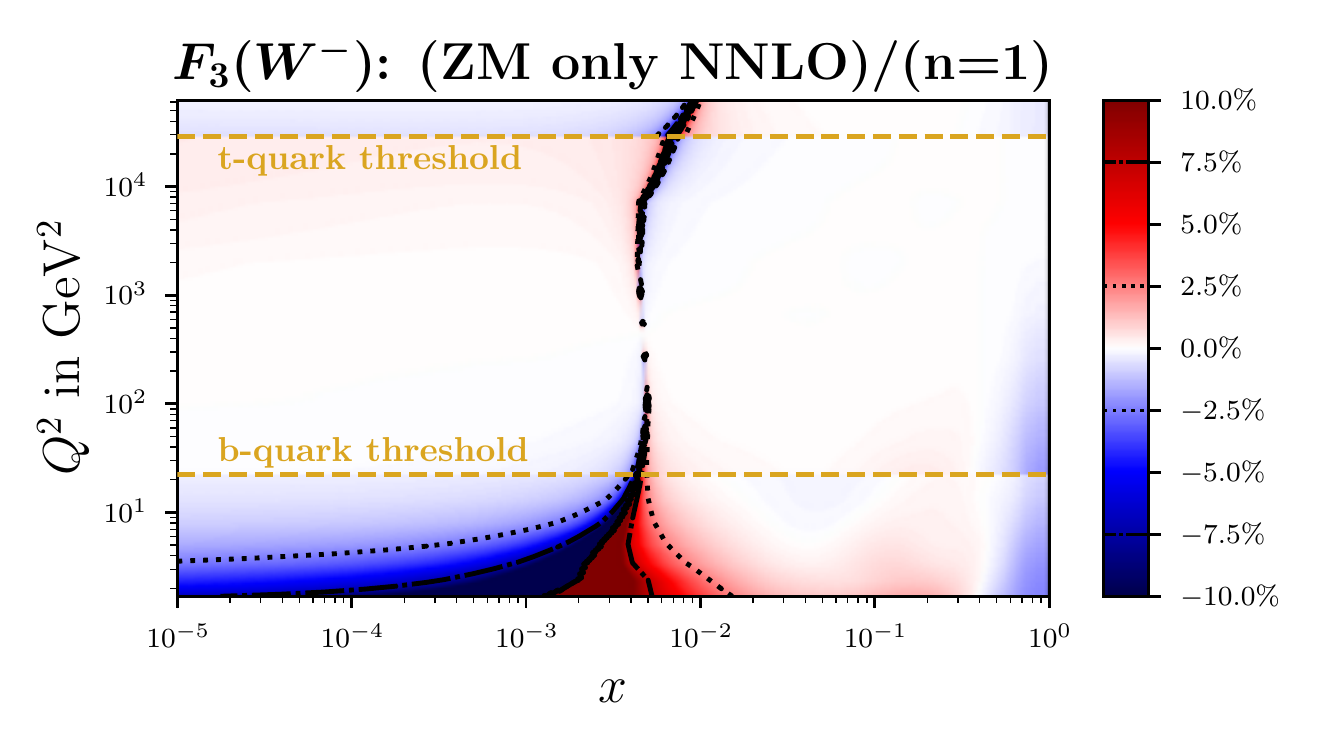}%
    \includegraphics[width=0.33\textwidth]{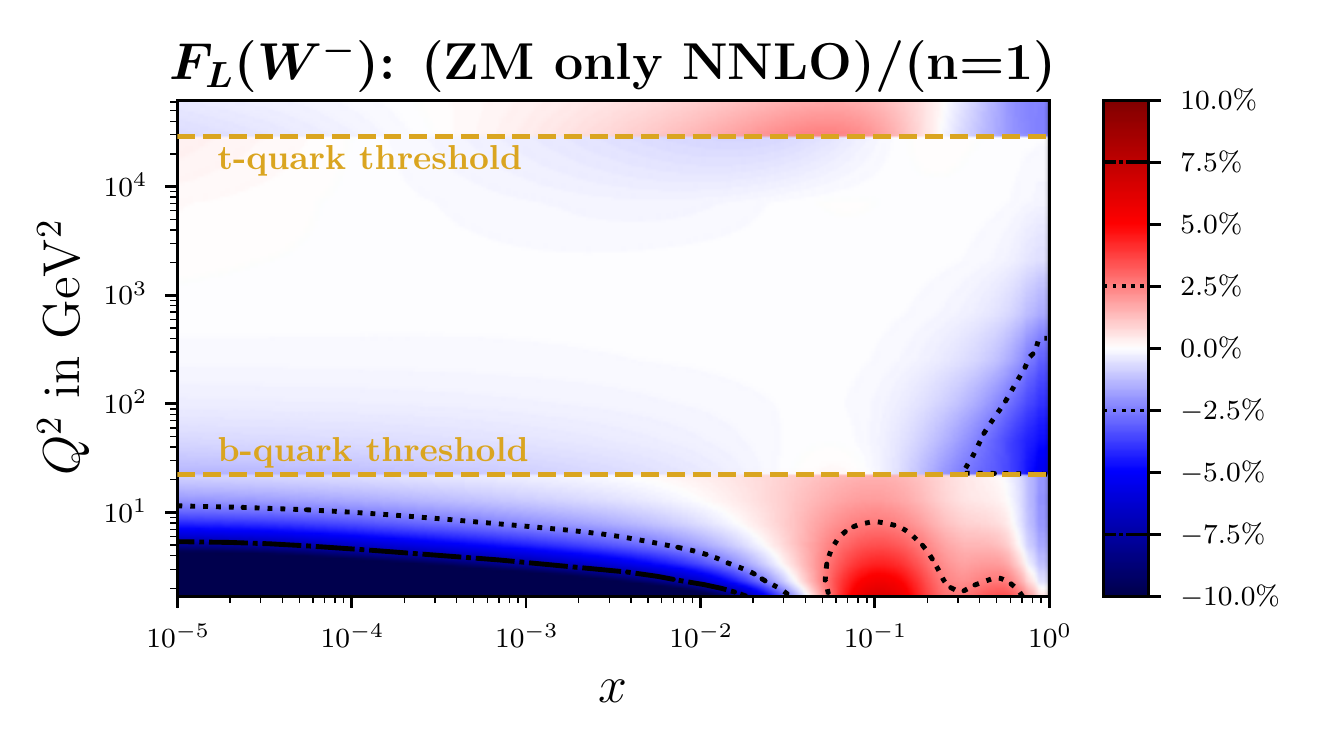}
    \includegraphics[width=0.33\textwidth]{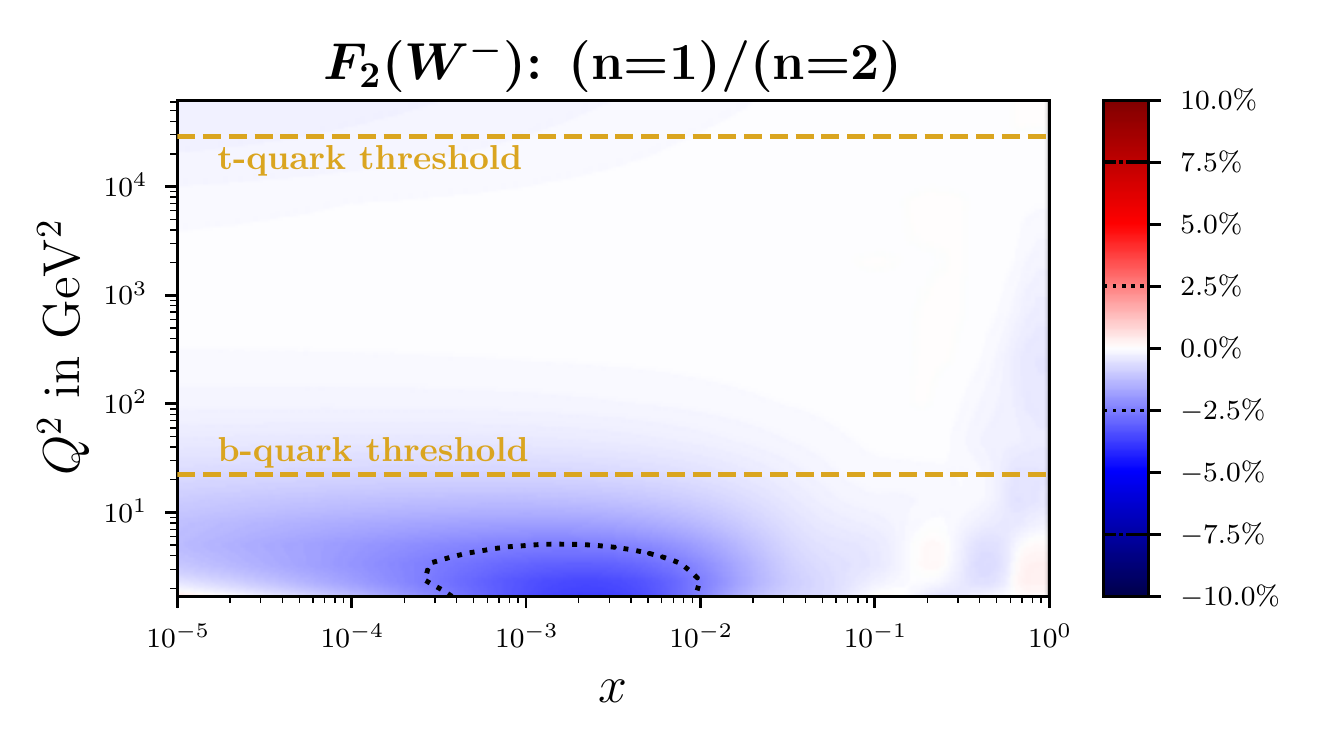}%
    \includegraphics[width=0.33\textwidth]{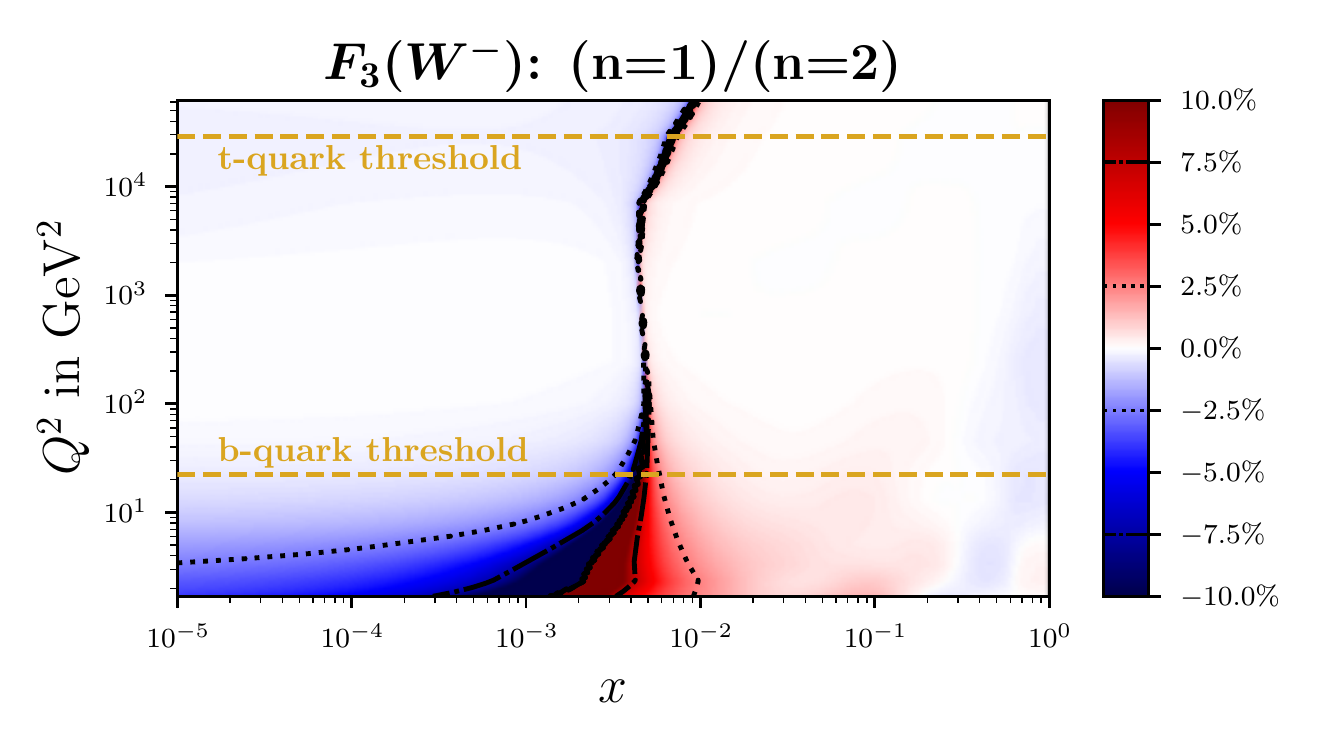}%
    \includegraphics[width=0.33\textwidth]{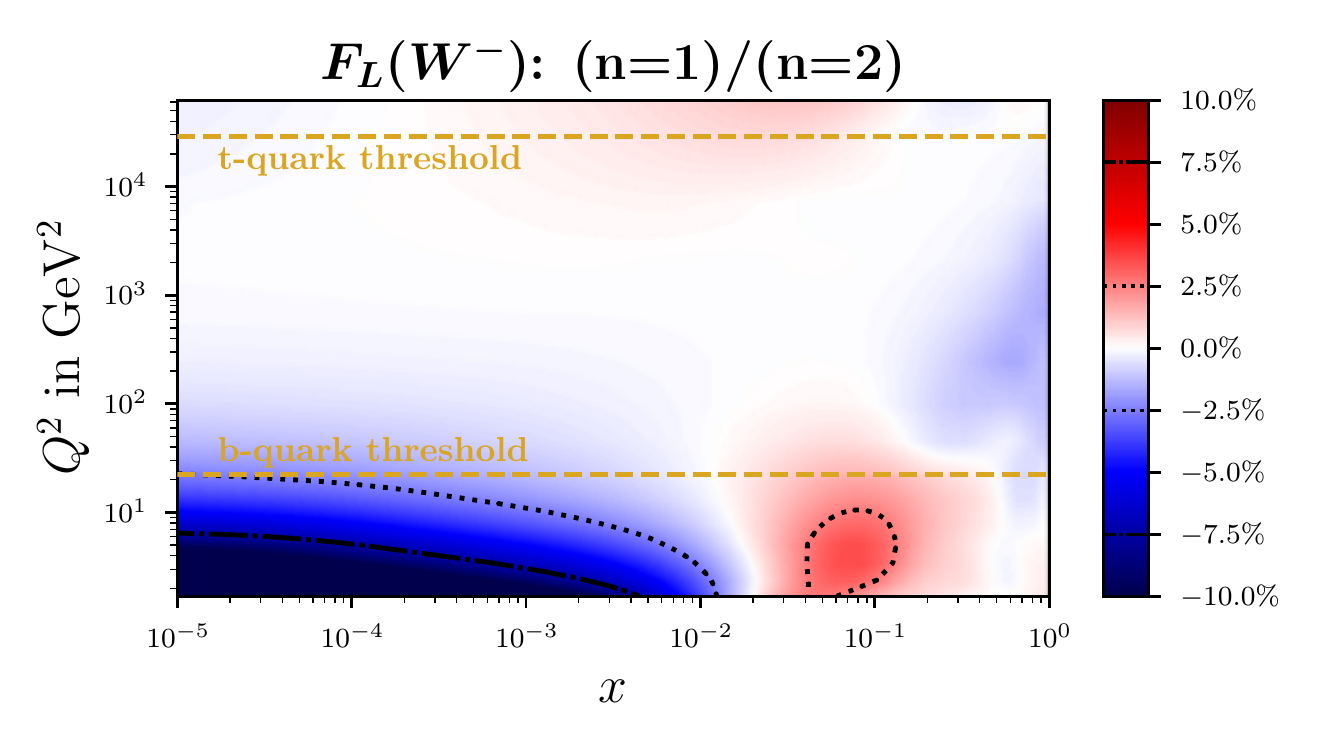}
    \includegraphics[width=0.33\textwidth]{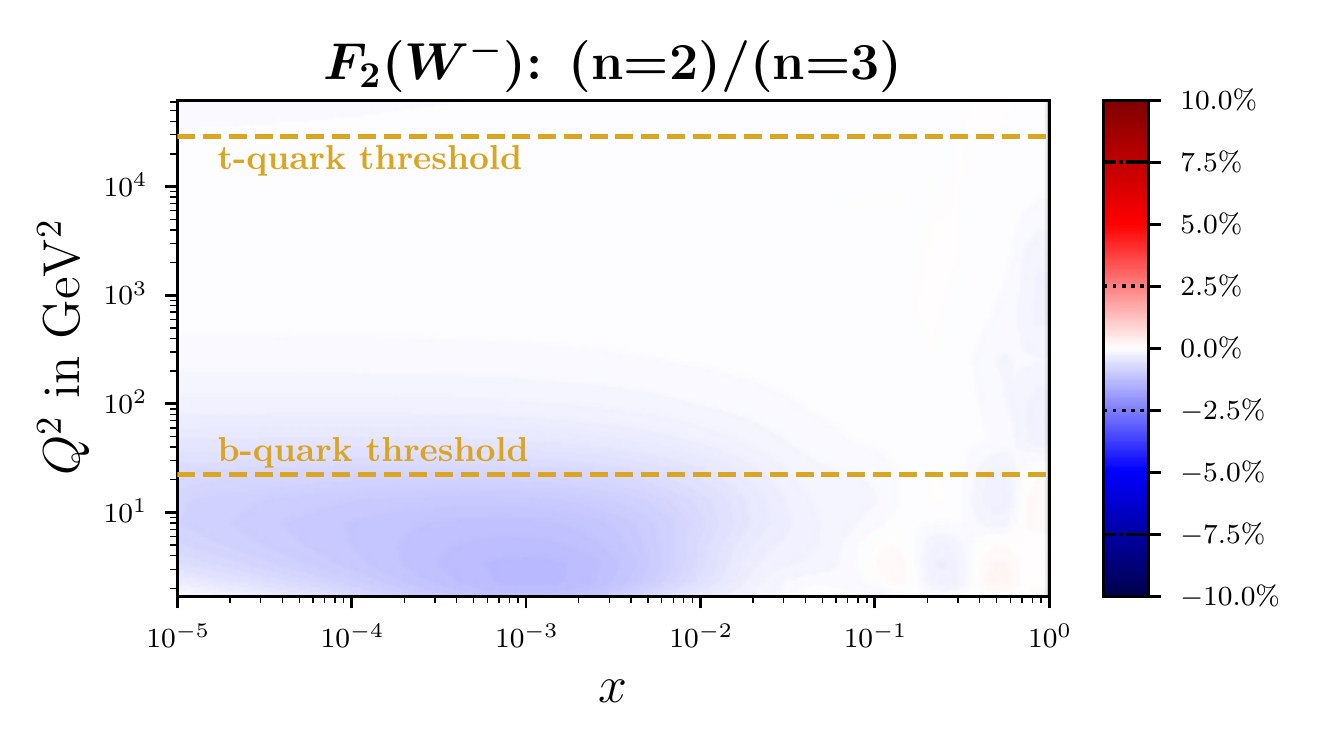}%
    \includegraphics[width=0.33\textwidth]{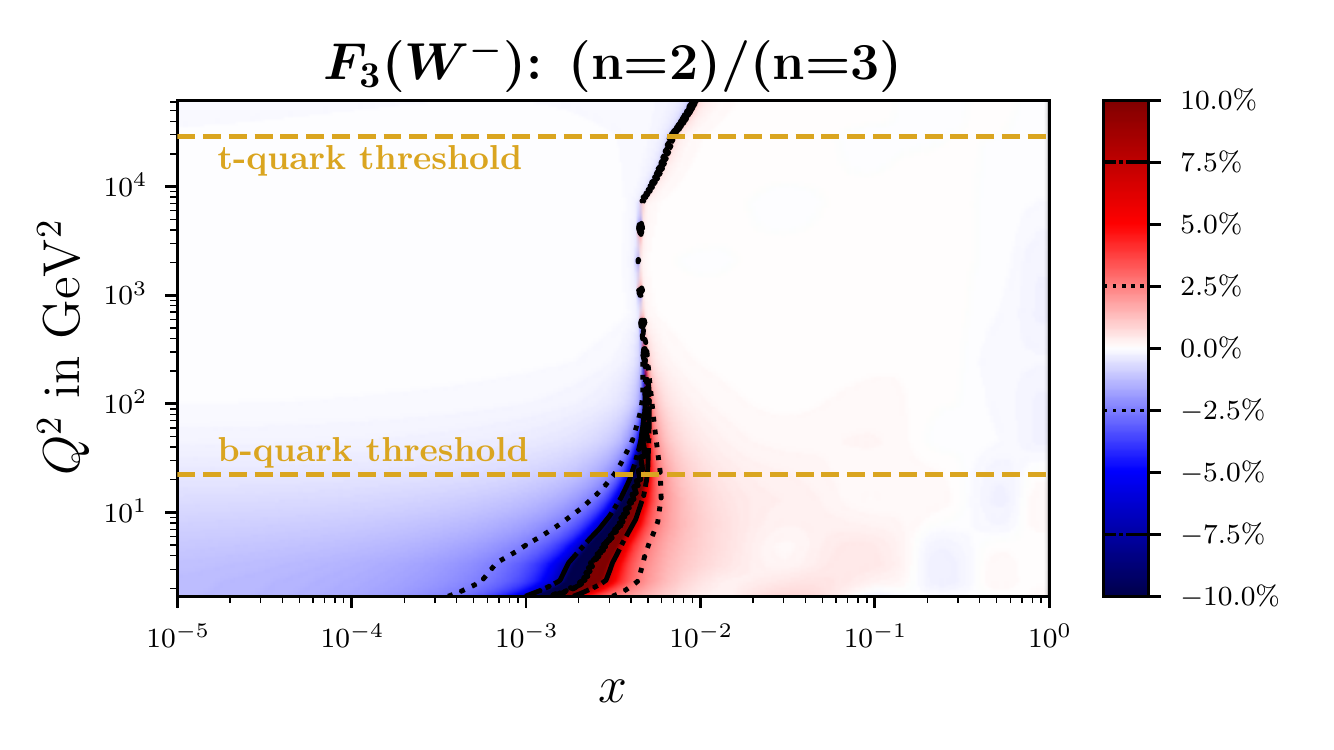}%
    \includegraphics[width=0.33\textwidth]{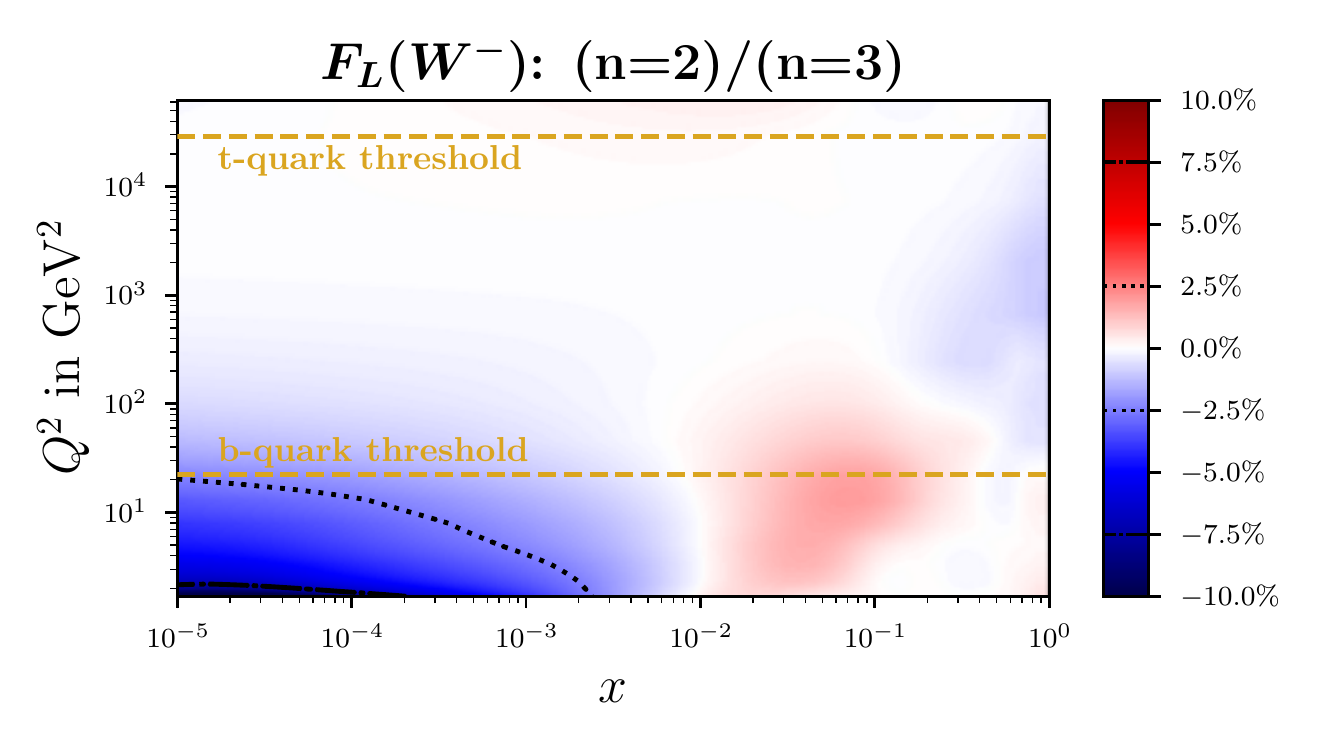}
	\caption{The same as \cref{fig:relative_ratio_F23L_NC} but for charged current with a $W^-$ exchange.}
	\label{fig:relative_ratio_F23L_WM}
\end{figure*}
\begin{figure*}[t]
	\centering
	\includegraphics[width=0.33\textwidth]{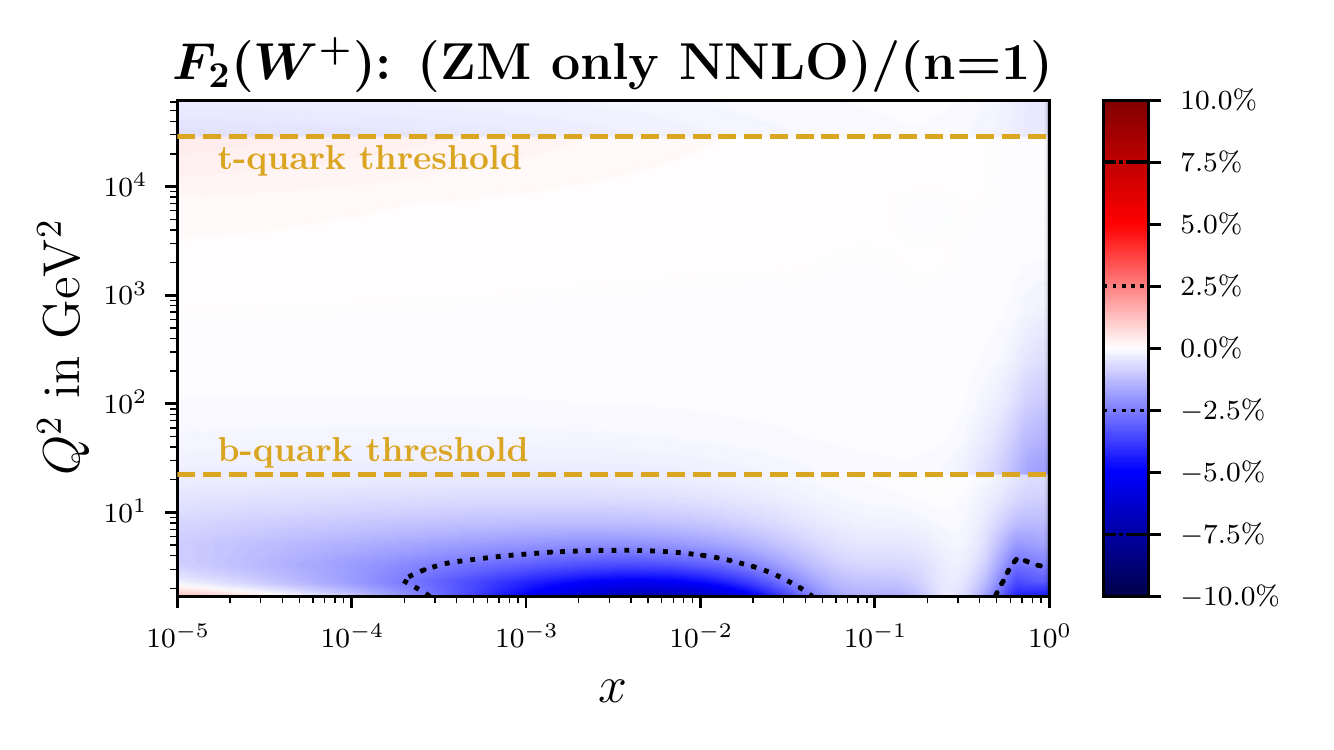}%
    \includegraphics[width=0.33\textwidth]{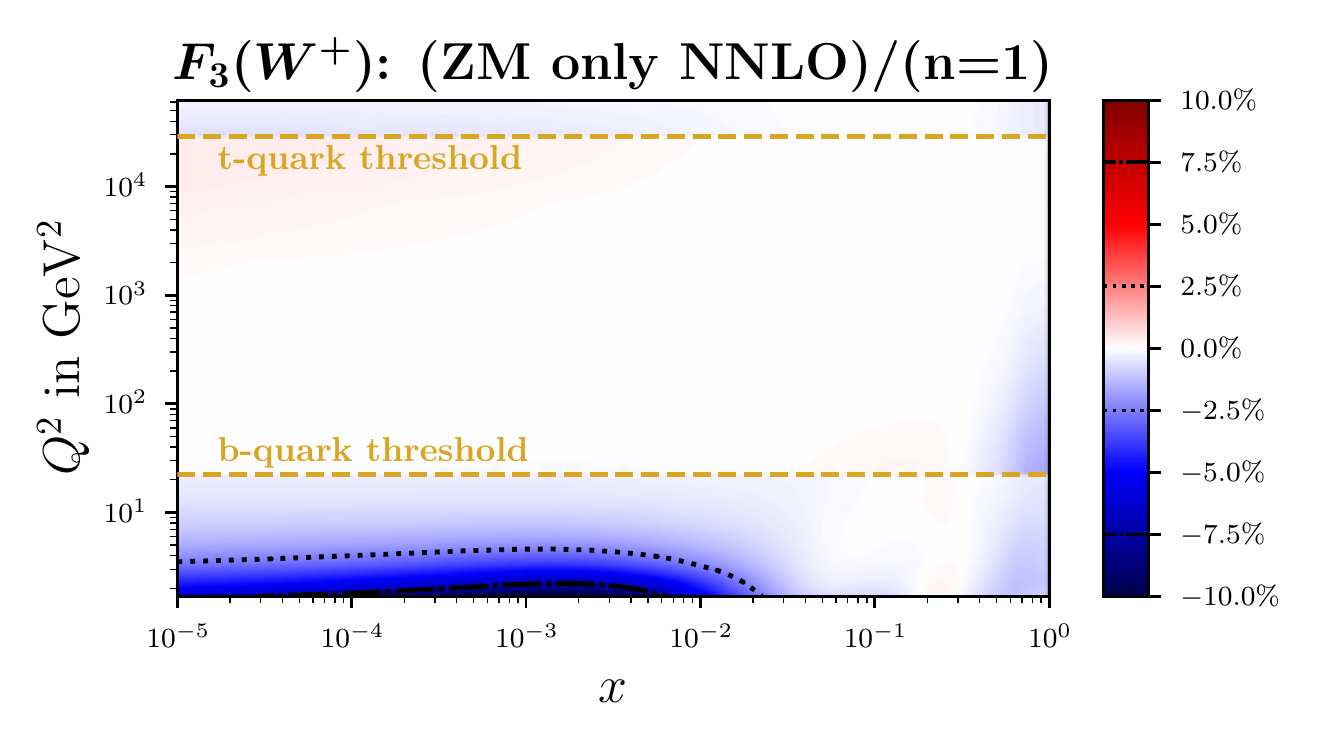}%
    \includegraphics[width=0.33\textwidth]{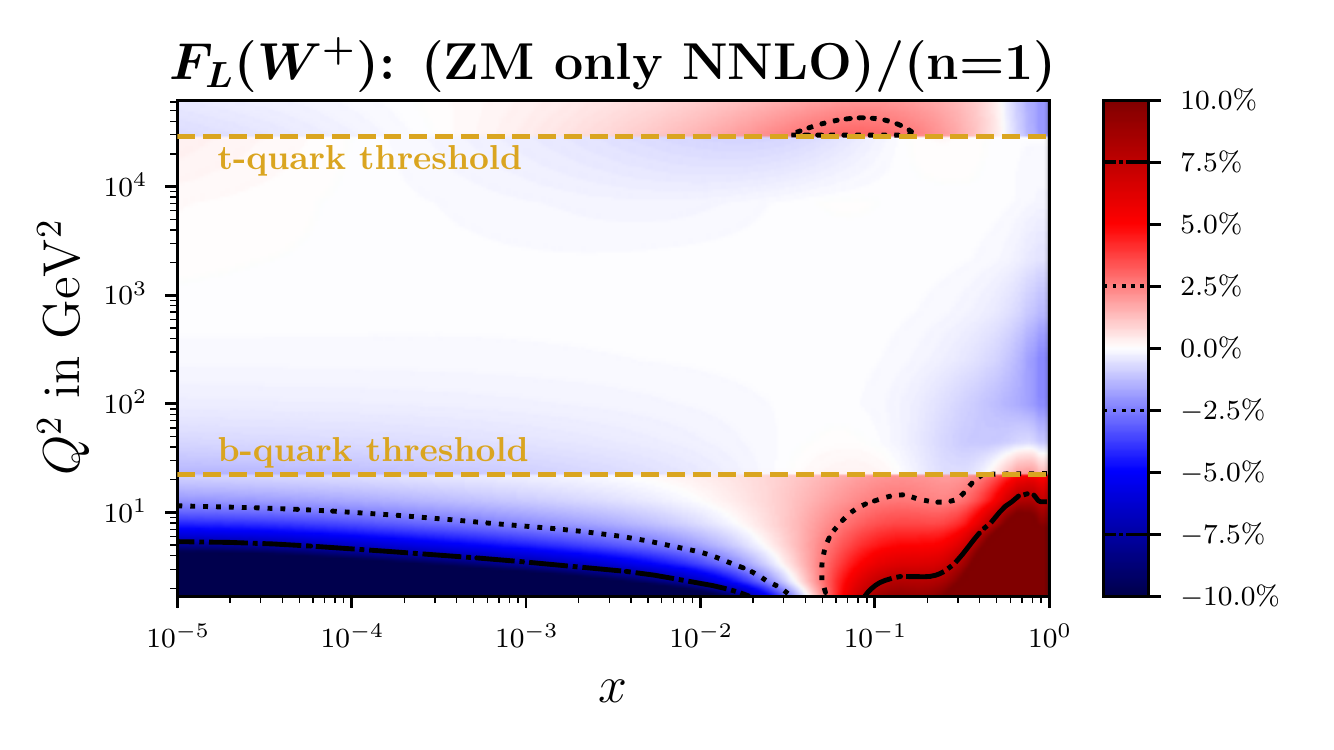}
    \includegraphics[width=0.33\textwidth]{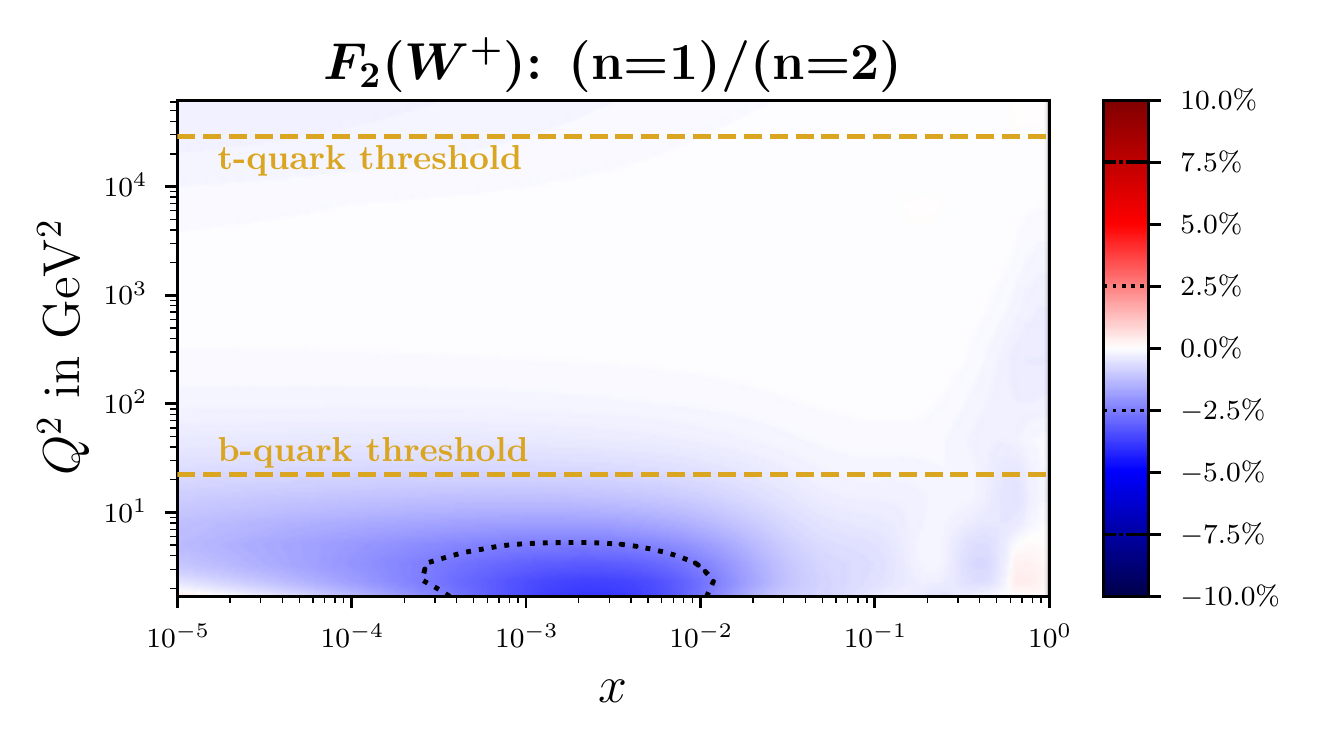}%
    \includegraphics[width=0.33\textwidth]{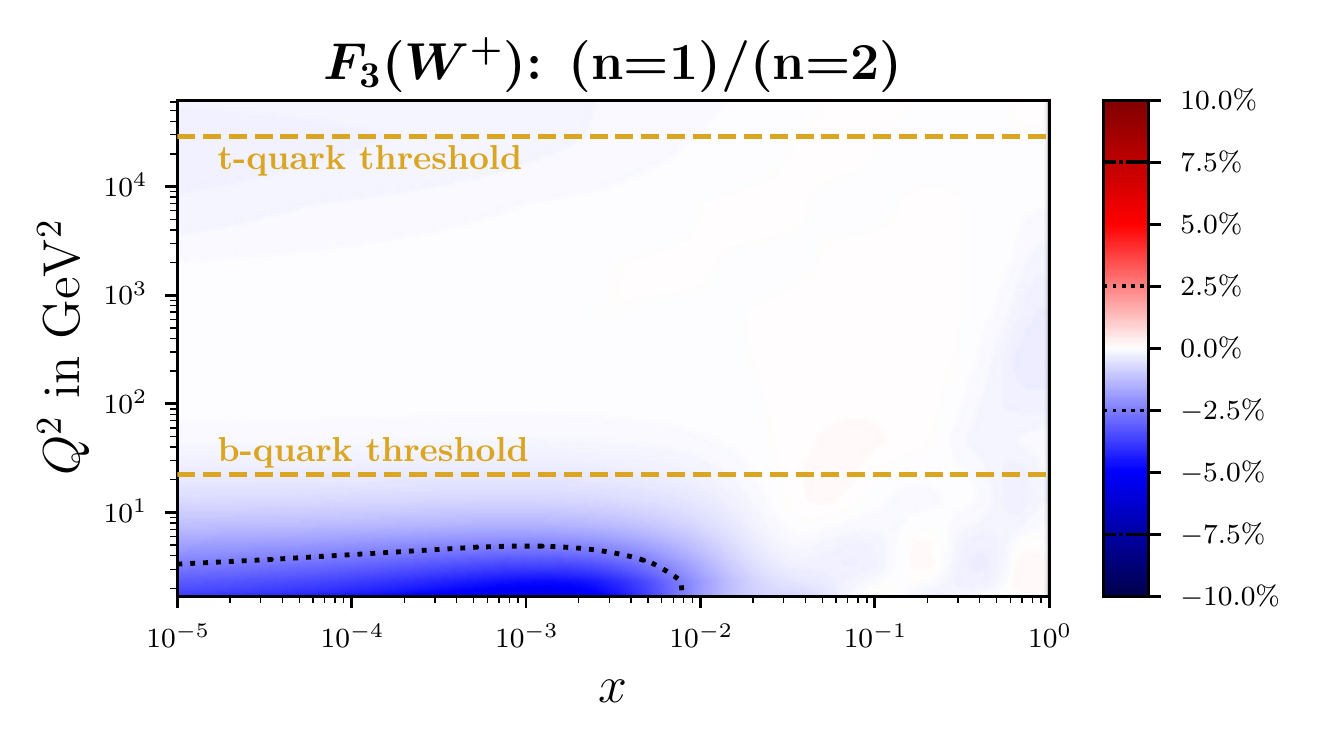}%
    \includegraphics[width=0.33\textwidth]{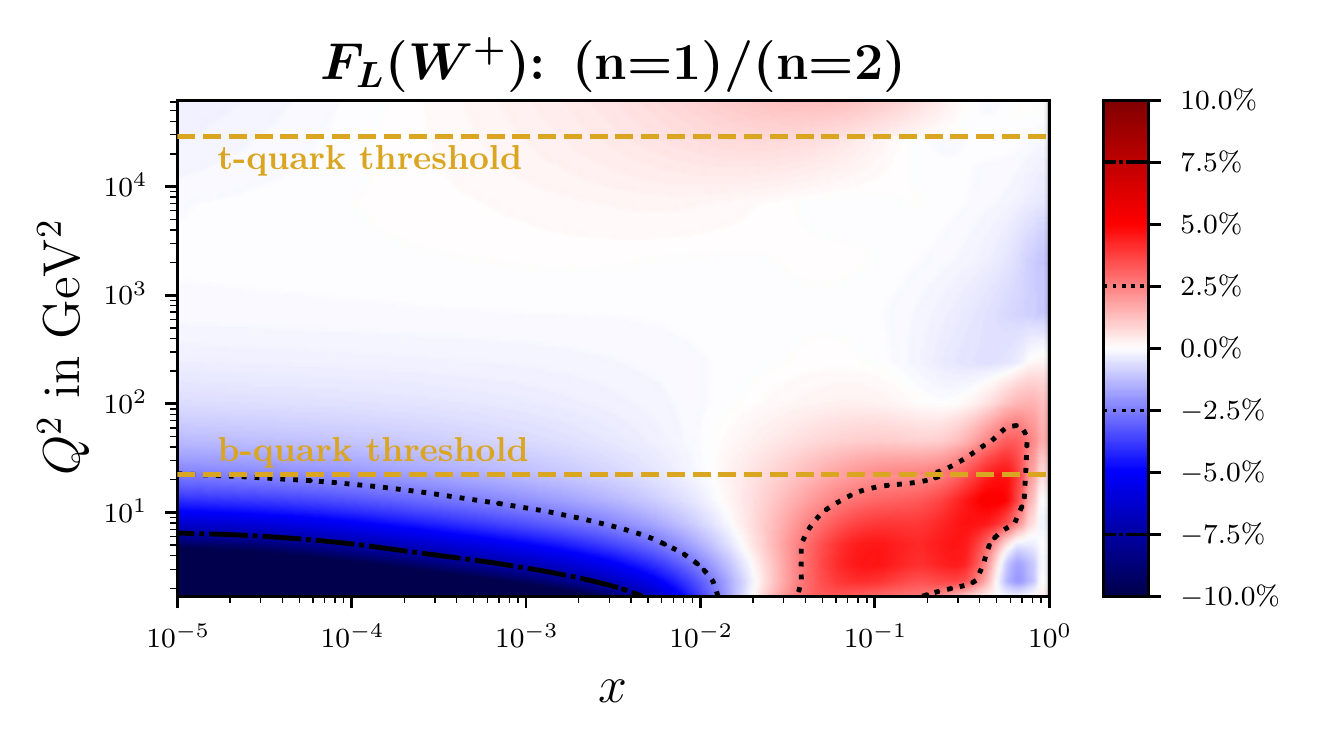}
    \includegraphics[width=0.33\textwidth]{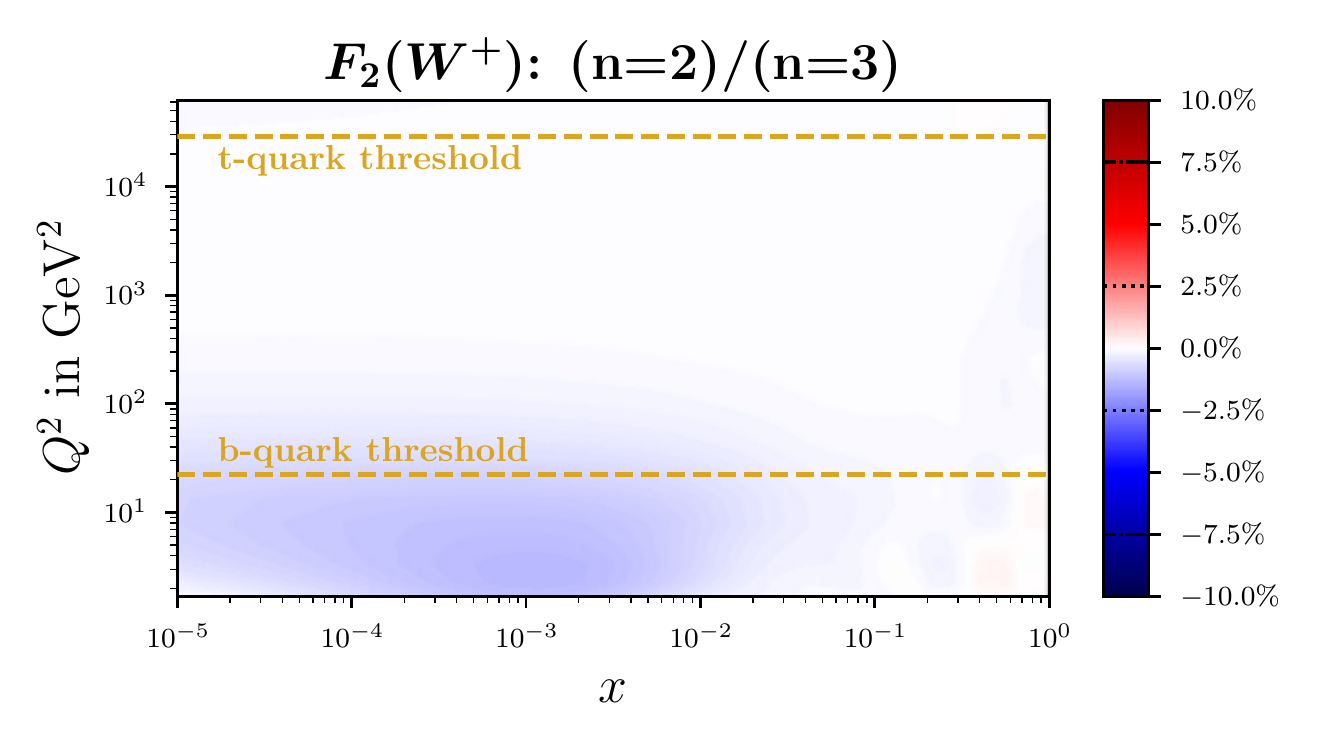}%
    \includegraphics[width=0.33\textwidth]{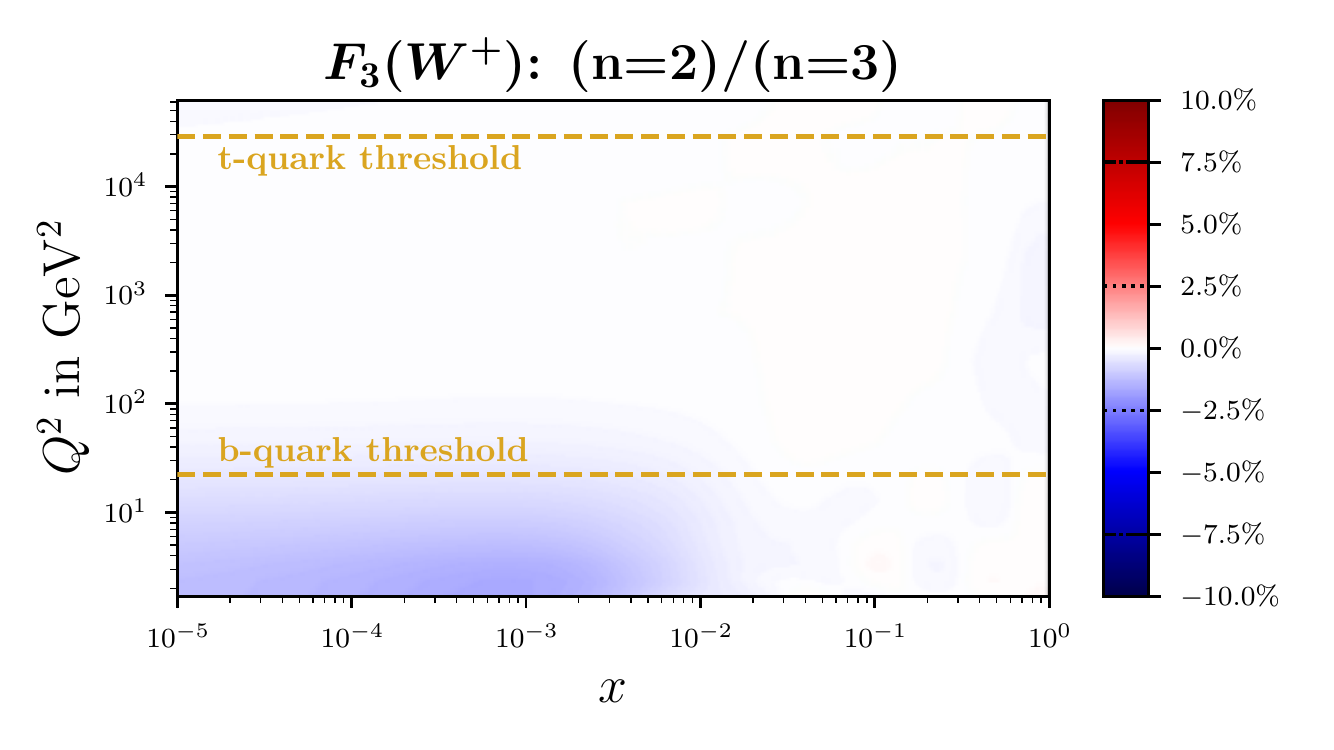}%
    \includegraphics[width=0.33\textwidth]{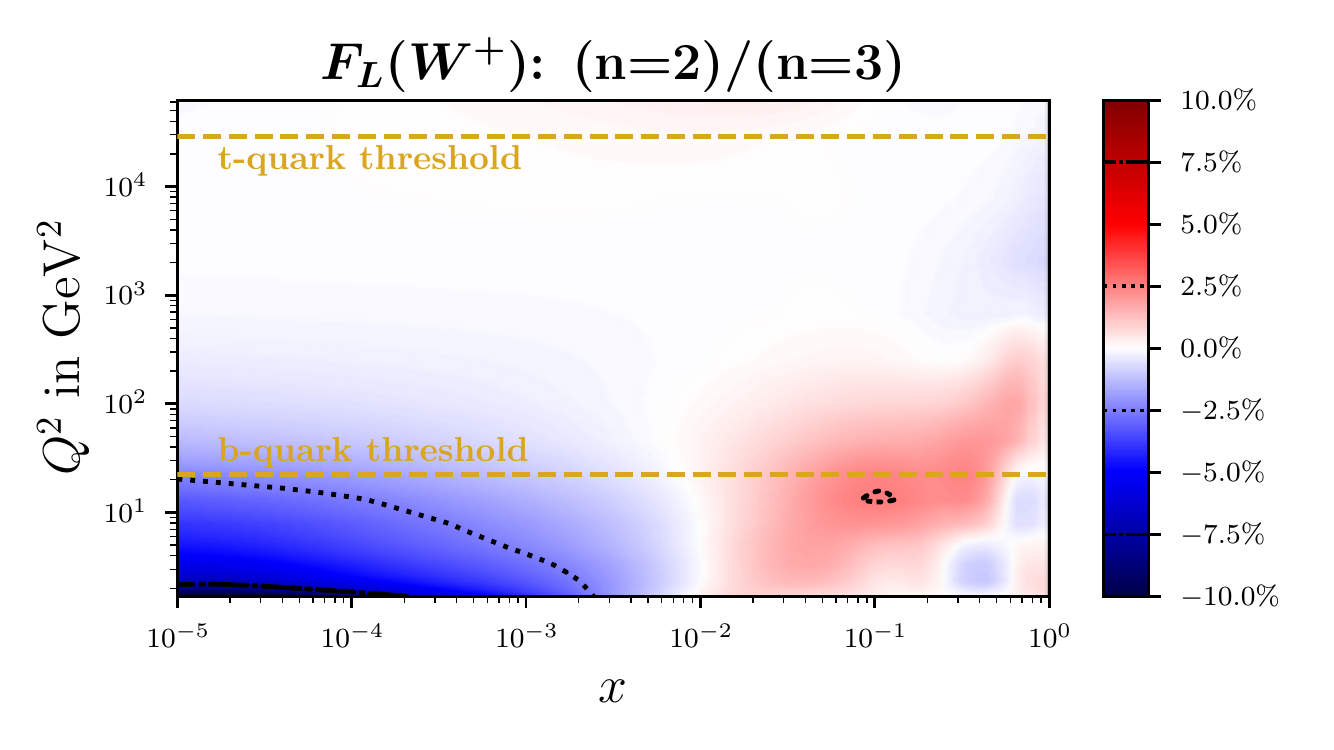}
	\caption{The same as \cref{fig:relative_ratio_F23L_NC} but for charged current with a $W^+$ exchange.}
	\label{fig:relative_ratio_F23L_WP}
\end{figure*}

\section{Approximate higher-order mass contributions}\label{sec:nScaling}

As demonstrated in Ref.~\cite{Stavreva:2012bs}, 
the $n$-rescaling of \cref{eq:n_scaling_variable} yields an approximate mechanism to 
adjust for the reduced phase space available when heavy quarks are produced in the final state.\footnote{%
In particular, see Table~I of Ref.~\cite{Stavreva:2012bs} for an analysis of the NLO phase space factors.}
The $n$-scaling not only provides us with an approximation of the 
proper phase-space suppression, but, by examining the variation of the structure functions with respect to $n$, we can also estimate both the uncertainty of this approximation and the impact of the heavy-quark masses. 
In the following we discuss this aspect in detail.

\subsection{Interpretation of \texorpdfstring{$n$}{n}}\label{subsec:n123}

We start by identifying the correspondence between $n=\{0,1,2,3\}$ and the underlying heavy-quark processes. 
For example, when $n{=}0$, there is no rescaling and this corresponds to the case of a massless quark (ZM).\footnote{Note that in the ZM scheme the gluon coefficient receives contributions from the heavy-quark even below its mass threshold. However, for technical reasons in many implementations (also in \apfelxx{}) a step function of the form $\coeftilde[g]{\lambda}\to\theta\left(Q^2-m^2_H\right)\coeftilde[g]{\lambda}$ is introduced and the heavy quark contribution appears only above the threshold. Our $n=0$ implementation includes the step function in the gluon coefficient to align with the ZM implementation.
However, for $n\neq 0$ the heavy-quark also contributes below the threshold.}
Similarly, $n{=}1$ introduces the typical factor observed in the production of a single heavy quark such as 
the $q\to WQ$ charged-current process~\cite{Aivazis:1993pi}.
For a gluon producing a pair of heavy quarks, 
such as $\gamma g\to Q \bar{Q}$, the $n{=}2$ factor yields the appropriate phase-space suppression. 
Finally, at NNLO we have a new combination as illustrated in~\cref{fig:NNLO_ps_CC} where we can have 3 heavy quarks
in the final state corresponding to $n{=}3$.

Note that $n$ is not required to be an integer and could be an ``effective'' number of heavy quarks instead. In this paper, we restrict the discussion to whole numbers as it reduces the possibilities but still covers the range of physically motivated choices. 
Our numerical implementation allows the user to freely choose the value of $n$.

\subsection{Quality of approximation}\label{subsec:nQuality}

We begin by examining the structure function relative ratio plots of~\cref{fig:relative_ratio_F23L_NC,fig:relative_ratio_F23L_WM,fig:relative_ratio_F23L_WP}.
This will allow us to both i)~gauge the impact of the mass effects
on the $\{F_2, F_3, F_L \}$ structure functions, and 
ii)~estimate the uncertainty due to missing higher-order mass contributions.

\def\fredhead#1{\vspace{4pt}\noindent\textbf{#1}\ }
\fredhead{The \mbox{$\bm{(\textrm{ZM only NNLO})/(n=1)}$} Case:}
We start with the \mbox{$(\textrm{ZM only NNLO})/(n=1)$} comparisons displayed in the top row of the figures. %
Recall that the dotted (dash-dotted) lines indicate 2.5\%
(7.5\%) contours.
We know that the ZM  massless result (effectively $n{=}0$) is not a reasonable estimate since the quark mass effects are essential to consider, especially in the low $Q^2$ region.  
Therefore, these plots show us the minimum effect of the mass terms we are missing when we use the ZM-VFN scheme. 

\begin{addmargin}[\parindent]{0cm}
\fredhead{$\bm{F_2}$:}
For example, the $F_2$ structure functions display a significant change ($\gtrsim 2.5\%$) for low $Q^2$, especially in the intermediate $x$-range ($\sim 10^{-3} \textrm{--} 10^{-2}$). There is also a lesser change ($\lesssim 2.5\%$) for high $x$-values which is especially evident above the $b$-quark threshold. As $F_2$ is often the dominant contribution to the physical cross section, we cannot neglect the mass terms in these regions. 

\fredhead{$\bm{F_3}$:}
For the parity-violating neutral current structure function $F_3$, the mass effects are essentially zero. This is because the only parity-violating contributions come from the $Z$-boson, which is suppressed relative to the photon due to its large mass $M_Z$. 

Conversely, for charged current processes ($W^\pm$), we do observe significant mass effects at low $Q^2$ on a relative scale; however, on an absolute scale, these contributions will be suppressed by a factor $(Q^2/M_W^2)$ due to the large $M_W$ mass
in a manner similar to the neutral current $Z$-contribution above.

Note in~\cref{fig:relative_ratio_F23L_WM} the   $F_3$ 
structure function passes through zero at an intermediate $x$-value (\mbox{$\sim 10^{-2}$}), and yields the observed artifact in the plot. 
In contrast to $F_2$ which is positive definite, there is no constraint on the sign of $F_3$, and we discuss this further in~\cref{subsec:comments_on_the_sign_of_F3}.

\fredhead{$\bm{F_L}$:}
The impact of the mass terms is much more pronounced in the ratios for $F_L$ exceeding  $\gtrsim 7.5\%$ below the bottom threshold, and still exceeding $\gtrsim 2.5\%$ above.
Since the LO contributions to $F_L$ are suppressed relative to the $F_2$ terms,\footnote{%
Contributions to $F_L$ at LO are suppressed due to helicity conservation, and these vanish in the massless limit~\cite{Aivazis:1993kh}.
} 
our calculation for this quantity is effectively one-order lower than for $F_2$; hence the effect of the mass correction is enlarged. Furthermore, we note that $F_L$ is close to zero in the $x<10^{-2}$ regime below the bottom threshold, thus small differences between the schemes are artificially enlarged. We observe large corrections ($\gtrsim 7.5\%$) below the $b$-quark threshold, and these persist above this threshold at both large and small $x$-values. 
\end{addmargin}

\fredhead{The $\bm{(n=2)/(n=3)}$ Case:}
Turning now to examine the \mbox{$n=3$} case, the suppression factor corresponds to an incoming heavy quark $Q$, producing a $Q'\bar{Q'}$ pair via gluon splitting; thus, there are 3 heavy quarks in the final state.  Given that the PDF for a heavy quark is significantly suppressed compared to the gluon and lighter quarks, we expect contributions from this channel to be minimal; therefore, $(n{=}3)$ represents an extreme case which is most likely beyond reasonable expectations for the mass uncertainty.

\begin{addmargin}[\parindent]{0cm}
\fredhead{$\bm{F_2}$:}
For the $F_2$ structure function, 
we examine the bottom row of~\cref{fig:relative_ratio_F23L_NC,fig:relative_ratio_F23L_WM,fig:relative_ratio_F23L_WP}
and observe the relative difference between the $n=2$ and $n=3$ result is minimal ($\lesssim 2.5\%$).
As implied above, this suggests that the heavy-quark contributions are already strongly suppressed in the $n=2$ instance so that the additional suppression for $n=3$ is minimal. 

\fredhead{$\bm{F_3}$:}
For the neutral current $F_3$ structure function, 
the difference between the $n=2$ and $n=3$ result also is minimal 
due to the $(Q^2/M_Z^2)$ suppression of the parity-violating $Z$ contribution. 
Conversely, for the charged-current $F_3$ structure function
we do observe relative differences in the region of low $Q^2$; 
however, the absolute contributions will again be suppressed by the $(Q^2/M_W^2)$ factor due to the $W^\pm$  boson mass. 

\fredhead{$\bm{F_L}$:}
The $F_L$ also displays significant differences due to the fact that this is essentially a lower-order result as compared to $F_2$. Again, this behavior is similar to the $ZM/(n=1)$ instance. 
\end{addmargin}

\fredhead{The $\bm{(n=1)/(n=2)}$ Case:}
The above observations of $n=\{0,3\}$ suggest that our best estimate for the mass uncertainty of our approximation
can be obtained by examining the range between $n=\{1,2\}$. 
Therefore, we examine the middle rows of~\cref{fig:relative_ratio_F23L_NC,fig:relative_ratio_F23L_WM,fig:relative_ratio_F23L_WP}, 
to focus on the $(n=1)/(n=2)$ ratio as an estimate of the neglected higher-order mass contributions.

\begin{addmargin}[\parindent]{0cm}
\fredhead{$\bm{F_2}$:}
For the $F_2$ structure function, as before we see a significant change ($\gtrsim 2.5\%$) for low $Q^2$, especially in the intermediate $x$-range. There is also a lesser change above the $b$-quark threshold; across all vector bosons the difference is less than 1\%. 
As $F_2$ is often the dominant contribution to the physical cross section, the ratios displayed in these plots represent our best estimate of the uncertainty due to our $\aSACOTchi{}$ scheme. 

%
\fredhead{$\bm{F_3}$:}
The neutral current $F_3$ relative ratio of~\cref{fig:relative_ratio_F23L_NC} 
displays minimal dependence on the treatment of the heavy quark mass. 
As before,  the  charged current $F_3$ structure functions (\cref{fig:relative_ratio_F23L_WM,fig:relative_ratio_F23L_WP})
do show relative differences in the region of low $Q^2$; 
however, the absolute contributions will again be suppressed by the $(Q^2/M_W^2)$ factor due to the $W^\pm$  boson mass. 

\fredhead{$\bm{F_L}$:}
As before, the impact of the mass terms is even more pronounced in the ratios for $F_L$ exceeding  $\gtrsim 7.5\%$ below the bottom threshold, and still exceeding $\gtrsim 2.5\%$ above the bottom threshold.
This reflects the fact that the absolute size of the $F_L$ structure function are small compared to $F_2$; 
hence, the relative uncertainty is large.

This is also evident when comparing the experimental uncertainty  of $F_2$ as compared to $F_L$ which is more challenging to measure~\cite{H1:2013ktq}.
While current theoretical uncertainties align with the precision of existing experimental data, $F_L$  stands out as an observable that would significantly benefit from a more refined treatment of heavy-quark masses.
\end{addmargin}

%
\fredhead{Recap:}
In summary, these figures display the kinematic regions where the structure functions are sensitive to the mass dependence. As observed in~\cref{sec:results}, the impact on the charged current HERA and EIC data (which is typically at larger $Q^2$ values) is rather minimal. But the effect on the $\nu$DIS data can be significant, as shown in~\cref{fig:CC_NuTeV_n1_over_all} for NuTeV. 

Additionally, by examining the variation between the $(n=1)$
and $(n=2)$ results, we can also gauge the theoretical uncertainty of our approximations. For the inclusive $F_2$ structure function, these are typically small; for $Q^2>4\,\textrm{GeV}^2$ (a typical cut in global fit analyses), the variation is $\lesssim 2.5\%$.
 
However, for $F_L$, the impact of the mass contributions, and the associated uncertainties, are significantly larger; hence, this quantity can benefit from updated theoretical calculations as the experimental accuracy is improved.

The $\aSACOTchi{}$ scheme and the $\apfelxx{}$ framework are designed with the flexibility to seamlessly incorporate new calculations as they become available. Additionally, the precomputed grid technique enables fast and efficient numerical evaluation, ensuring both adaptability and computational speed in future analyses.

\subsection{The parity-violating \texorpdfstring{$F_3$}{F3}}
\label{subsec:comments_on_the_sign_of_F3}

We observed that the  $F_3$ structure function crosses zero at intermediate $x$-values, and this appears as an artifact in~\cref{fig:relative_ratio_F23L_WM}. 
We briefly examine the physics behind this phenomenon. 

For charged current processes at leading order, we have:
\begin{eqnarray}
    F_3^{W^+} &=& 2\sum_D^{d,s,b}\,\sum_U^{u,c,t} |V_{DU}|^2\big(f_D(x)-f_{\anti{U}}(x)\big)
    \\
    F_3^{W^-} &=&
    2\sum_D^{d,s,b}\,\sum_U^{u,c,t} |V_{DU}|^2\big(f_U(x)-f_{\anti{D}}(x)\big)\,.
\end{eqnarray}
Focusing on the dominant first- and second-generation $\{u,d,s,c\}$ quarks, in the region of small-$x$ (that is, below the valence region), we have effectively:
\begin{equation}
    F_3^{W^+}   \simeq 2 ( d_{sea} - u_{sea} + s - c) \simeq  -F_3^{W^-}
\end{equation}
That is, the dominant contributions to $F_3^{W^+}$ and $F_3^{W^-}$ will have opposite signs; thus, there is no constraint that $F_3$ be positive. 
This is in contrast to $F_2$, which can be written as a sum of (positive definite) longitudinal and transverse cross sections. 

Additionally, we can relate $F_3$ to the left- and right-helicity structure functions (which are positive) via the relation~\cite{Olness:1986mv,Aivazis:1990pe,Aivazis:1993kh,Aivazis:1993pi}:
\begin{equation}
    F_3 \propto  (-F_+ + F_-)
\end{equation}
where $F_\pm$ represents the right/left-helicity structure-function.
Thus, when we change from $W^+/W^-$, we change from a dominantly right/left coupling, and the sign of $F_3$ will flip.  
This is precisely the source of the zero we observe in
$F_3$ for the charged $W^-$ exchange process, 
and explicitly demonstrates the parity-violating nature of $F_3$.

Finally, note that at the level of the cross section the sign in front of $\F{3}$ changes depending on the incoming lepton: ``+'' for a lepton and ``-'' for an anti-lepton. Thus for $\nu$-DIS ($W^+$ exchange) $\F{3}$ contributes with a positive prefactor and for $\anti{\nu}$-DIS ($W^-$ exchange) with a negative prefactor.

\goodbreak
\onecolumngrid
\section{Library documentation}\label{sec:LibraryDocumentation} 

\noindent The code can be downloaded from the git repository under the URL

\parbox{\linewidth}{
\centering\texttt{\url{https://github.com/vbertone/apfelxx}}
}

\noindent The full documentation can be found under this link as well. Currently, the implementation is located in the ``\texttt{ACOT}'' branch; to switch to the branch we use the commands
\begin{lstlisting}[language=bash,style=mystyle]
 cd /location/of/apfelxx
 git checkout ACOT
\end{lstlisting}
The code is installed with the following procedure
\begin{lstlisting}[language=bash,style=mystyle]
 cd build
 cmake -DCMAKE_INSTALL_PREFIX=/your/installation/path ..
 make && make install
\end{lstlisting}
Here, \texttt{/your/installation/path} is the path where the user would like to install \apfelxx{}. Note that the two full stops at the end are important. The installation can be tested with the command 
\begin{lstlisting}[language=bash,style=mystyle]
 make test
\end{lstlisting}

Before tabulating a structure function, a few initializations have to be performed, namely
\begin{lstlisting}[language=C++,style=mystyle]
 // set perturbative order
 const int pto = 2;
 // define a x-grid through a two subgrids that have 100/50 interpolation nodes, 
 // x_min=1e-5/1e-1 and are of degree 3/3
 const apfel::Grid g{{apfel::SubGrid{100,1e-5,3},apfel::SubGrid{50,1e-1,3}}};
 // define a vector of mass thresholds -> needed to not interpolate across a discontinuity
 const std::vector<double> Thresholds = {0,0,0,mc,mb,mt};
 // define a function that returns the effective electro-weak charges as a function of Q
 const auto fEW = [=] (double const& Q) -> std::vector<double> {
    return apfel::ElectroWeakCharges(Q,false);}; // use predefined function within APFEL++
\end{lstlisting}
For the prediction of structure functions, a set of PDFs is required. In the following we assume that we have a function called \lstinline{my_PDF(x,Q)} that returns a \lstinline{std::map<int,double>}, which maps the flavor indices in the LHAPDF-format\footnote{I.e.~we have the integers $\{-6,-5,\dots,-2,-1,1,2,\dots,5,6,21\}$ that map to the flavors $\{\anti{t},\anti{b},\dots,\anti{u},\anti{d},d,u,\dots,b,t,g\}$.}\,\cite{Buckley:2014ana} to the values of the respective distribution at $\{x,Q\}$. The set of PDF needs to be rotated into the QCD evolution basis, which can be done with a helper function part of the library:
\begin{lstlisting}[language=C++,style=mystyle]
 // rotate the PDFs into the QCD evolution basis
 const auto PDFrotated = [&] (double const& x, double const& Q) -> std::map<int,double>{
    return apfel::PhysToQCDEv(my_PDF(x,Q));};
\end{lstlisting}
As the PDF set is usually aligned with a specific implementation of the strong coupling, the library is built such that a user-defined function can be used. For our example, we define the function \lstinline{strongCoupling} with the call structure \lstinline{(double const& mu) -> double}. A built-in implementation of the strong coupling is given through the class \lstinline{AlphaQCD}. 

In the following, we describe how to obtain fast evaluations of the neutral current $\F{2}$ structure function in the \aSACOTchi{} scheme. In \apfelxx{} the structure functions are built in two steps to decouple the expensive integration over the operators from the convolution with the PDFs. 

First, the operators that make up the structure function are precalculated on the $x$- and $Q$-grids. For NC $\F{2}$ in the \aSACOTchi{} scheme this is done using the function \lstinline{InitF2NCsimACOT_NNLO}. Other structure functions or other schemes can be constructed using different functions with a similar naming scheme that can be found in the full documentation. We call
\begin{lstlisting}[language=C++,style=mystyle]
 // Construct the operator grids for the F2 structure function
 const auto F2objects = apfel::InitF2NCsimACOT_NNLO(
    g,Thresholds,IntEps=1e-5,nQ=100,Qmin=2,Qmax=225,intdeg=3,n=1);
\end{lstlisting}
The variables \{\lstinline{nQ,Qmin,Qmax,intdeg}\} define the $Q$-grid, \lstinline{IntEps} is the integration accuracy and the variable \lstinline{n} is the scaling variable from \cref{eq:n_scaling_variable}. This step usually takes a few seconds as we are calculating the expensive integrals over the operators, but has to be done only once.

Second, we build the $\F{2}$ structure function from the operators tables, a process that is very fast. This is done by combining \lstinline{F2objects} with the rotated PDFs, the strong coupling, the EW-charges and selecting the perturbative order.
\begin{lstlisting}[language=C++,style=mystyle]
 // Build the structure function
 const std::map<int,apfel::Observable<>> F2 = apfel::BuildStructureFunctions(
    F2objects,PDFrotated,pto,strongCoupling,fEW);
\end{lstlisting}
This map allows one to access the total $\F{2}$ through
\begin{lstlisting}[language=C++,style=mystyle]
 // Evaluate the total F2 at x=0.1 and Q=10
 F2.at(0).Evaluate(x=0.1,Q=10);
\end{lstlisting}
Note that we have accessed the map at the key ``\lstinline{0}''. In general, the possible keys are \{\lstinline{0,1,2,3,4}\} and result in the total, light, charm, bottom, or top contribution to the structure function, respectively. For the \aSACOTchi{} scheme we implemented only the total structure functions (key ``\lstinline{0}'').

Now, every time the function \lstinline{Evaluate(x,Q)} is called, the operator tables are convolved with the PDFs. A final speed-up can be achieved by interpolating the $Q$-dependence of the final result as well:
\begin{lstlisting}[language=C++,style=mystyle]
 // Interpolate the Q-dependence of the total F2 
 const apfel::TabulateObject<apfel::Distribution> F2total {
    [&] (double const& Q) -> apfel::Distribution{return F2.at(0).Evaluate(Q);},
    nQ=50,Qmin=2,Qmax=225,intdeg=3,Thresholds};
\end{lstlisting}
where we substitute the evaluation of $\F{2}$ as an anonymous function alongside with the definition of the $Q$-grid and the mass thresholds to the interpolation routine. $\F{2}$ can now be evaluated as 
\begin{lstlisting}[language=C++,style=mystyle]
 // Evaluate the Q-interpolated total F2 at x=0.1 and Q=10
 F2total.Evaluate(x=0.1,Q=10);
\end{lstlisting}

\begin{acknowledgments}
The authors would like to thank our nCTEQ collaborators, including 
\mbox{N.~Derakhshanian,}
P.~Duwent\"aster, 
C.~Keppel,
M.~Klasen,
R.~Ruiz,
J.~Wissmann
and
J.~Y.~Yu
for many useful comments and discussions. 

The work of P.R.\ and F.O.~was supported by the U.S.~Department of Energy under Grant \mbox{No.~DE-SC0010129,} 
and by the Office of Science, the Office of Nuclear Physics, within the framework of the Saturated Glue (SURGE) Topical Theory Collaboration.
P.R. thanks the Jefferson Lab for their hospitality. This material is based upon work supported by the U.S. Department of Energy, Office of Science, Office of Nuclear Physics under contract DE-AC05-06OR23177.
The work of FO was performed, in part, at the Aspen Center for Physics, which is supported by National Science Foundation grant PHY-2210452.
The work of V.B. has been supported by l’Agence Nationale
de la Recherche (ANR), project ANR-24-CE31-7061-01.
The work of P.R., T.J.\ and K.K.\ at the university of M{\"u}nster 
was funded by the DFG through the Research Training Group 2149 ``Strong and Weak Interactions - from Hadrons to Dark Matter'' and in the case of T.J. further from the SFB 1225 ``Isoquant,'' {project\nobreakdash-id}~273811115. 
A.K. acknowledges the support of the National Science Centre Poland under the Sonata Bis Grant No.~2019/34/E/ST2/00186.
\end{acknowledgments}

\clearpage
\twocolumngrid
\bibliographystyle{utphys}
\bibliography{ACOT_NNLO}

\end{document}